\documentclass{article}
\usepackage{amssymb}

\usepackage{graphicx}
\usepackage{amsmath}

\input{tcilatex}

\begin{document}

\begin{center}
\bigskip {\Huge Designing new apartment buildings for strings and conformal
field theories. }

{\Huge First steps}

\bigskip
\end{center}

\bigskip

\textbf{Arkady L.Kholodenko\footnote{%
375 H.L. Hunter Laboratories, Clemson University, Clemson, SC 29634-0973,
USA . E-mail: string@clemson.edu}}

\bigskip

\ 

\bigskip

\textbf{\ Abstract} : The concepts of apartments and buildings were
suggested by Tits for description of the Weyl-Coxeter reflection groups. We
use these and many additional facts from the theory of reflection and
pseudo-reflection groups along with results from the algebraic and
symplectic geometry of toric varieties in order to obtain the tachyon-free
Veneziano-like multiparticle scattering amplitudes and the partition
function generating these amplitudes. Although the obtained amplitudes
reproduce the tachyon-free spectra of both open and closed bosonic string,
the generating (partition) function is not that of the traditional bosonic
string. It is argued that it is directly related to the N=2 supersymmetric
quantum mechanical model proposed by Witten in 1982 in connection with his
development of the Morse theory. Such partition function can be
independently obtained with help of the results by Solomon (published in
1963) on invariants of finite (pseudo) reflection groups. Although the
formalism developed in this work is also applicable to conformal field
theories (CFT), it leaves all CFT results unchanged.

\bigskip

\bigskip

\bigskip

\bigskip

\bigskip

\pagebreak

\section{From geometric progression to Weyl character formula}

\subsection{\protect\bigskip General considerations}

Consider finite geometric progression of the type 
\begin{eqnarray}
\mathcal{F(}c,n) &=&\sum\limits_{l=-n}^{n}\exp \{cl\}=\exp
\{-cn\}\sum\limits_{l=0}^{\infty }\exp \{cl\}+\exp
\{cn\}\sum\limits_{l=-\infty }^{0}\exp \{cl\}  \notag \\
&=&\exp \{-cn\}\frac{1}{1-\exp \{c\}}+\exp \{cn\}\frac{1}{1-\exp \{-c\}} 
\notag \\
&=&\exp \{-cn\}\left[ \frac{\exp \{c(2n+1)\}-1}{\exp \{c\}-1}\right] . 
\TCItag{1.1}
\end{eqnarray}
The reason for displaying the intermediate steps will be explained shortly.
But first, we would like to consider the limit : $c\rightarrow 0^{+}$ of $%
\mathcal{F(}c,n)$. Clearly, it is given by $\mathcal{F(}0,n)=2n+1$. The
number $2n+1$ equals to the number of integer points in the segment $[-n,n]$ 
\textit{including} \textit{boundary} points. It is convenient to rewrite the
above result in terms of $x=\exp \{c\}$. So that we shall write formally $%
\mathcal{F(}x,n)$ instead of $\mathcal{F(}c,n).$ Using such notations, let
us consider the related function\footnote{%
Incidentally, such type of relation ( the Erhart-Macdonald reciprocity law )
is characteristic for the Ehrhart polynomial for rational polytopes. Work of
Stanley, Ref.[1], provides many applications of this law.} 
\begin{equation}
\mathcal{\bar{F}(}x,n)=(-1)\mathcal{F(}\frac{1}{x},-n).  \tag{1.2}
\end{equation}
Explicitly, we obtain: 
\begin{equation}
\mathcal{\bar{F}(}x,n)=(-1)\frac{x^{-(-2n+1)}-1}{x^{-1}-1}x^{n}.  \tag{1.3}
\end{equation}
In the limit: $x\rightarrow 1+0^{+}$ we obtain $\mathcal{\bar{F}(}1,n)=2n-1.$
The number 2$n-1$ is equal to the number of integer points \textit{inside}
the segment $[-n,n].$

These seemingly trivial results will be broadly generalized in this work. To
this purpose let us introduce some notations. First, we replace $x^{l}$ by 
\textbf{x}$^{\mathbf{l}}=x_{1}^{l_{1}}x_{2}^{l_{2}}\cdot \cdot \cdot
x_{d}^{l_{d}}$ and, accordingly, we replace the summation sign \ in \ the
left hand side of Eq.(1.1) by the multiple summation, etc. Thus obtained
function $\mathcal{F(}\mathbf{x},n)$ in the limit $x_{i}\rightarrow 1+0^{+},$
$i=1-d,$ produces the anticipated result: 
\begin{equation}
\mathcal{F(}\mathbf{1},n)=(2n+1)^{d}.  \tag{1.4}
\end{equation}
It provides the number of integer points \textit{inside} the $d-$dimensional
cube and at its faces. Analogously, using Eq.(1.2), we obtain 
\begin{equation}
\mathcal{\bar{F}(}\mathbf{1},n)=(2n-1)^{d}.  \tag{1.5}
\end{equation}
This is just the number of integer points \textit{strictly inside} of $d$%
-dimensional cube. In our earlier work [2] Eq.(1.5) (with $2n$ being
replaced by $N$ and $d$ being replaced by $n+1$) was obtained and
interpreted as the Milnor number of the Brieskorn-Pham (B-P) type
singularity, e.g. that of the Fermat-type. The above number counts ''new''
Veneziano-like amplitudes which mathematically are\ interpreted as periods
of the Fermat-type hypersurfaces. The purpose of this work is to make such
identification more accurate. It becomes possible based on detailed Lie
group-theoretic analysis of earlier obtained \ ''new'' amplitudes. As in
known treatments of conformal field theories (CFT) \ whose computational
formalism and its physical interpretation comes directly from that developed
in string theories [3], we demonstrate that the same is also true in the
present case. The same group-theoretic methods which work for the
Veneziano-like amplitudes can be extended to recover CFT results. However,
the physical models leading to these ''new '' amplitudes are noticeably
different from those used traditionally.

This work is made to a large degree self contained and is organized in a
such way that it is expected to be accessible to both mathematicians and
physicists. Such attitude towards our readers had made this paper somewhat
larger than usual. At the same time, in writing this paper we wanted to
expose new elements with sufficient details which can be easily checked and
understood.

In particular, already the rest of Section 1 sets up the tone for the
reminder of this paper. In this section we consider various aspects of the
Weyl character formula from nontraditional point of view. These include:
group-theoretic, combinatorial and dynamical. Some results of this section
are new and as such have independent interest but most of the results are
auxiliary and used in the rest of the paper. Section 2 contains some facts
from the theory of linear algebraic groups. These facts are intertwined with
the description of affine and projective toric varieties. Although some
readers might be familiar with such objects our exposition differs from more
traditional and highly recommended mathematical expositions, e.g. [4],
because it contains various physical applications. Section 3 is built on the
results of section 2 and contains some discussion of the method of coadjoint
orbits and moment map to be used later in Section 5. Although such topics
are usually discussed from the point of view of symplectic geometry [5], the
results of Section 2 allow us to look at these topics from group-theoretic
perspective only. Section 3 also contains few new results and is mainly
auxiliary. The new results appear first in Section 4 devoted to detailed
study of the Veneziano-like amplitudes. If for the reading of Sections 1-3
use of the results of Appendix A is helpful, the reading of Section 4 is
closely interconnected with the results of the Appendix B in which we
discuss the analytical properties of the Veneziano and Veneziano-like
amplitudes. The difference between the Veneziano and \ Veneziano-like
amplitudes lies in the fact that the last ones are tachyon-free by design.
Mathematically, these amplitudes are periods associated with Hodge-De Rham
cohomology basis made of differential forms living on Fermat-type
hypersurfaces. Recently published monograph [6] provides an easily readable
comprehensive mathematical exposition of results related to periods
associated with complex hypersurfaces. Many physically relevant applications
of these mathematical results can be found in our earlier work, Ref.[2].
Section 4 contains also some discussion relating the Veneziano-like
amplitudes to the amplitudes in CFT.\ Roughly speaking, the differences
between these observables are very much the same as \ the differences
between the calculable observables in periodic solids and those in the
vacuum in the absence of periodicity. Such differences are known and treated
accordingly in the theoretical solid state physics. Effectively, CFT
formalism accounts for the periodicity effects. Deeper reasons revealing the
true essence of the affine Kac-Moody algebras are provided in the Appendix A
from which it should be also clear how such results can be extended if
necessary. Section 4 contains as well some solutions to exercises (Chapter5,
paragraph 5, problem set \# 3) from the book by Bourbaki on Lie groups and
Lie algebras [7] while Section 5 contains the rest of solutions to these
exercises. These exercises are based (in part only !) on the fundamental
paper by Solomon [8] on invariants of (pseudo) reflection groups published
in 1963. The results by Solomon allow us to reanalyze group-theoretically
the Veneziano-like amplitudes and to reconstruct the partition function
reproducing these amplitudes. This is accomplished in Section 5 where we
discuss 3 independent ways : group-theoretic, symplectic and supersymmetric
-all leading to the same partition function for the Veneziano-like
amplitudes. We argue in this section that the finite dimensional N=2
supersymmetric quantum mechanical model proposed by Witten in 1982 is
sufficient for reproduction of the Veneziano-like amplitudes and the
associated with them partition function. Nevertheless, using some
group-theoretic considerations inspired by book by Ginzburg [9] and taking
into account the Lefschetz isomorphism theorem [10], we map Witten's
supersymmetric quantum mechanical model into the direct sum of $sl_{2}(%
\mathbf{C})$ Lie algebras. Ultimately, this allows us to reformulate the
associated quantum mechanical problems in the language of semisimple Lie
algebras (since all of them are made of copies of $sl_{2}(\mathbf{C})$
[11]). Based on the book by Kac [12], this allows us to find very precisely
the place at which the existing formalism of CFT and the formalism of \ our
new ''string'' theory come apart. Naturally, by sacrificing mathematical
rigor, this allows, in principle, to recover back the results of ''old''
string theories. In Section 6 we briefly discuss some implications of the
obtained results. These are related to remarkable connections between the
pseudo-reflection groups used in the main text and complex hyperbolic
geometry. Such connections tie together new string theory, hyperbolic space
and quantum mechanics thus making them inseparable. This connection is rigid
in the sense that variations of complex structure (that is ''motions'' in
the moduli space) of complex projective hypersurfaces, e.g. of Fermat type,
do not destroy such connections.

\subsection{\protect\bigskip Connection with integral polytopes}

\bigskip

Let $\mathbf{Z}^{d}$ be a subset of Euclidean space $\mathbf{R}^{d}$
consisting of points with integral coefficients. Select a subset \ $\mathcal{%
P}$ $\mathcal{\subset }$ $\mathbf{R}^{d}$.\ A convex hull of the
intersection\ \ $\mathcal{P\cap }$\textbf{Z}$^{d}$ \ \ is \ called the 
\textit{integral polytope}$.$ The scalar product in Euclidean space $\mathbf{%
R}^{d}$ is defined by 
\begin{equation}
<\mathbf{x},\mathbf{y}>=\sum\limits_{i=1}^{d}x_{i}y_{i}.  \tag{1.6}
\end{equation}
With help of such definition the $d$-dimensional version of Eq.(1.1) is
written below 
\begin{equation}
\sum\limits_{\mathbf{x}\in \mathcal{P\cap }\mathbf{Z}^{d}}\exp \{<\mathbf{c},%
\mathbf{x}>\}=\sum\limits_{\mathbf{v}\in Vert\mathcal{P}}\exp \{<\mathbf{c},%
\mathbf{v}>\}\left[ \prod\limits_{i=1}^{d}(1-\exp \{-c_{i}u_{i}^{v}\})\right]
^{-1}.  \tag{1.7}
\end{equation}
Here $Vert\mathcal{P}$ denotes the vertex set of the integral polytope, in
our case the $d$-dimensional cube. It is made of\ the highest weight
vectors\ of the Weyl-Coxeter reflection group $B_{d}$ appropriate for cubic
symmetry as explained in the Appendix A. The set $\{u_{1}^{v},...,u_{d}^{v}%
\} $ is made of \ vectors\ (not necessarily of unit length) constituting the
orthonormal basis of the $d-$dimensional cube. These vectors are oriented
along the positive semi axes with respect to center of symmetry of the cube.
When parallel translated to the edges ending at particular hypercube vertex $%
v$, they can point either in or out of this vertex. The correctness of
Eq.(1.7) can be readily checked by considering Eq.(1.1) as an example.
Subsequent generalization of this result to a square, a cube, etc. causes no
additional problems.

The above formula was obtained (seemingly independently) in many different
contexts. For instance, in the context of discrete and computational
geometry it is attributed to Brion [13]. In view of \ Eq.(1.2), it can as
well to be attributed to Ehrhart [14] \ and to many others but, actually,
this is just a special case of Weyl's character formula as we are going to
demonstrate shortly below. To prove that this is indeed the case we need to
recall some results of Atiyah and Bott [15] nicely summarized in Ref. [16]
by Bott. It is physically illuminating, however, to make a small digression
in order to discuss the dynamical transfer operators introduced by Ruelle
[17]. This discussion will lead us directly to the results of Atiyah and
Bott [15]. In addition, it enables us to look at these results from the
point of view of thermodynamic formalism developed by Ruelle for description
of evolution of chaotic dynamical systems. Such discussion is helpful for
interpreting the Veneziano-like amplitudes \ in terms of dynamical systems
formalism as suggested in our earlier work, Ref.[2].

\subsection{From \ Ruelle dynamical transfer operator to Atiyah and Bott
Lefschetz-type fixed point formula for elliptic complexes}

\bigskip

To emphasize the physical content of\ Ruelle transfer operator we would like
to reproduce several relevant results from the unpublished book Cvitanovi\v{c%
} \ [18] \footnote{%
Professor Cvitanovi\'{c} had kindly pointed to the author the web site from
which the book was downloaded}. Following [18], any classical dynamical
system can be thought of as the pair $(\mathcal{M},\mathbf{f})$ with $%
\mathbf{f}$ being a map $\mathbf{f}$: $\mathcal{M}\rightarrow \mathcal{M}$ .
More specifically, if the initial conditions at time $t=0$ are given by the
phase vector $\mathbf{\xi }=\mathbf{x}(0)$, then the mapping is just the
phase space trajectory: $\mathbf{x}(t)$=$\mathbf{f}^{t}\mathbf{(\xi )\equiv f%
}^{t}\mathbf{(x(0)).}$ Let $\mathbf{o}=\mathbf{o}(\mathbf{x(}t))$ be some
physically meaningful observable. We associate with it the integrated
observable $\mathbf{O}^{t}(\mathbf{\xi })$ defined by 
\begin{equation}
\mathbf{O}^{t}(\mathbf{\xi })=\int\limits_{0}^{t}d\tau \mathbf{o}(\mathbf{f}%
^{\tau }\mathbf{(\xi )).}  \tag{1.8}
\end{equation}
In terms of these definitions the \textit{evolution operator} $\mathcal{L}%
^{t}$ (the transfer operator) is formally defined as 
\begin{equation}
\mathcal{L}^{t}(\mathbf{y},\mathbf{x})=\delta (\mathbf{y}-\mathbf{f}(\mathbf{%
x}))\exp (\beta \mathbf{O}^{t}(\mathbf{x})).  \tag{1.9}
\end{equation}
In the case when the observable $\mathbf{o}$\textbf{\ }is classical
Hamiltonian the parameter $\beta $ is a sort of the \ inverse dynamical
temperature. More rigorously, the operator $\mathcal{L}^{t}$ is defined by
its action on the bounded scalar function $h(\mathbf{x})$ ''living'' on the
phase space $\mathcal{M}$ : 
\begin{equation}
\mathcal{L}^{t}h(\mathbf{y})=\int\limits_{\mathcal{M}}d\mathbf{x}\delta (%
\mathbf{y}-\mathbf{f}(\mathbf{x}))\exp (\beta \mathbf{O}^{t}(\mathbf{x}))h(%
\mathbf{x}).  \tag{1.10}
\end{equation}
Such operator possess nice semigroup property 
\begin{equation}
\mathcal{L}^{t_{1}+t_{2}}(\mathbf{y},\mathbf{x})=\int\limits_{\mathcal{M}}d%
\mathbf{z}\mathcal{L}^{t_{2}}(\mathbf{y},\mathbf{z})\mathcal{L}^{t_{1}}(%
\mathbf{z},\mathbf{x})  \tag{1.11}
\end{equation}
which should look familiar to every physicist dealing with Feynman's path
integrals. In this work we have no intentions to develop this avenue of
thought however. Instead, assuming that the phase trajectory possesses fixed
point(s), \ i.e. \ if we assume that equation $\mathbf{x}=\mathbf{f}^{t}(%
\mathbf{x})$ (or, more generally, $\mathbf{f}^{n}(\mathbf{x})=\mathbf{x}$ \
with $\mathbf{x}$ being the periodic point of period $n$ ) possesses at
least one fixed point solution, the trace of the evolution operator ( the 
\textit{partition function} in physical language) can be written as follows 
\begin{equation}
tr\mathcal{L}^{n}=\int\limits_{\mathcal{M}}d\mathbf{x}\mathcal{L}^{n}\mathbf{%
(x,x)=}\sum\limits_{\mathbf{x}_{i}\in Fix\mathbf{f}^{n}}\frac{\exp (\beta 
\mathbf{O}^{t}(\mathbf{x}_{i}))}{\left| \det (\mathbf{1}-\mathbf{J}^{n}(%
\mathbf{x}_{i}))\right| }  \tag{1.12}
\end{equation}
where the set $Fix\mathbf{f}^{n}$ is defined by $Fix\mathbf{f}^{n}=\{\mathbf{%
x}:\mathbf{f}^{n}(\mathbf{x})=\mathbf{x}\}$ and $\mathbf{J}^{n}(\mathbf{x}%
_{i})$ denoting the Jacobian matrix to be defined more accurately shortly
below. This result is useful to compare with Eq.(1.7). Clearly, \ when the
denominator of the right hand side of Eq.(1.7) is rewritten in terms of
determinants both expressions become identical.

To make connections with the work of Atiyah and Bott [15], we would like to
rewrite just presented intuitive physical results in a more mathematically
systematic way. Following Ruelle[17], let us consider a map $f$: $\mathcal{M}%
\rightarrow \mathcal{M}$ and a scalar function $g$: $\mathcal{M}\rightarrow 
\mathbf{C.}$Based on these data, the transfer operator $\mathcal{L}$ is
defined now by 
\begin{equation}
\mathcal{L}\Phi (x)=\sum\limits_{y:fy=x}g(y)\Phi (y).  \tag{1.13}
\end{equation}
By analogy with Eq.(1.11), if \ $\mathcal{L}_{1}$ and $\mathcal{L}_{2}$ are
transfer operators associated with maps $f_{1}$ , $f_{2}$ : $\mathcal{M}%
\rightarrow \mathcal{M}$ and weights $g_{1}$ and $g_{2}$ such that, as
before, $\mathcal{M}\rightarrow \mathbf{C}$ then, 
\begin{equation}
(\mathcal{L}_{1}\mathcal{L}_{2}\Phi
)(x)=\sum\limits_{y:f_{2}f_{1y}=x}g_{2}(f_{1}y)g_{1}(y)\Phi (y).  \tag{1.14}
\end{equation}
Finally, by analogy with Eq.(1.12), \ it is possible to obtain 
\begin{equation}
tr\mathcal{L}=\sum\limits_{x\in Fixf}\frac{g(x)}{\left| \det
(1-D_{x}f^{-1}(x))\right| }  \tag{1.15}
\end{equation}
with $D_{x}f$ being derivative of $f$ acting in the tangent space $T_{x}%
\mathcal{M}$ and the graph of $f$ is required to be transversal to the
diagonal $\Delta \subset \mathcal{M}\times \mathcal{M}$ . Eq.(1.15)
coincides with that obtained \ in the work by Atiyah and Bott [15]\footnote{%
They use $D_{x}f$ instead of $D_{x}f^{-1}$ which makes no difference for
fixed points and invertible functions. The \ important (for chaotic
dynamics) non invertible case is discussed by Ruelle also but the results
are not much different.}. These authors make several additional important\
(for the purposes of this work) observations to be discussed now. \ Ruelle,
Ref. [17], uses essentially the same type of arguments. These are as
follows. Define the \textit{local} Lefschetz index $\mathcal{L}_{x}(f)$ by 
\begin{equation}
\mathcal{L}_{x}(f)=\frac{\det (1-D_{x}f(x))}{\left| \det
(1-D_{x}f(x))\right| }  \tag{1.16}
\end{equation}
where $x\in Fixf$ , then, as in our previous work, Ref.[2], the \textit{%
global} Lefschetz index $\mathcal{L(}\mathit{f)}$ is defined by 
\begin{equation}
\mathcal{L(}\mathit{f)=}\sum\limits_{f(x)=x}\mathcal{L}_{x}(f).  \tag{1.17}
\end{equation}
Taking into account that $\det (1-D_{x}f(x))$=$\prod\limits_{i=1}^{d}(1-%
\alpha _{i}),$ where $\alpha _{i\text{ }}$are the eigenvalues of the
Jacobian matrix, the determinant can be rewritten in the following useful
form [19], page 133, 
\begin{equation}
\det (1-D_{x}f(x))=\prod\limits_{i=1}^{d}(1-\alpha
_{i})=\sum\limits_{k=1}^{d}(-1)^{k}e_{k}(\alpha _{1},...,\alpha _{d}) 
\tag{1.18}
\end{equation}
where the elementary symmetric polynomial $e_{k}(\alpha _{1},...,\alpha
_{d}) $ \ is defined by 
\begin{equation}
e_{k}(\alpha _{1},...,\alpha _{d})=\sum\limits_{1\leq \text{ }i_{1}<\cdot
\cdot \cdot <\text{ }i_{k}\leq d}\alpha _{i_{1}}\cdot \cdot \cdot \alpha
_{i_{k}}  \tag{1.19}
\end{equation}
with $e_{k=0}=1.$With help of these results the local Lefschetz index,
Eq.(1.16), can be rewritten \ alternatively as follows 
\begin{equation}
\mathcal{L}_{x}(f)=\frac{\sum\limits_{k=1}^{d}(-1)^{k}e_{k}(\alpha
_{1},...,\alpha _{d})}{\left| \det (1-D_{x}f(x))\right| }\equiv \frac{%
\sum\limits_{k=1}^{d}(-1)^{k}tr(\wedge ^{k}D_{x}f(x))}{\left| \det
(1-D_{x}f(x))\right| }  \tag{1.20}
\end{equation}
with $\wedge ^{k}$denoting \ the $k$-th power of the exterior product. With
help of these results, following Ruelle [17], the transfer operator $%
\mathcal{L}^{(k)}$ can be defined analogously to $\mathcal{L}$ in Eq.(1.15),
i.e. 
\begin{equation}
tr\mathcal{L}^{(k)}=\sum\limits_{x\in Fixf}\frac{g(x)tr(\wedge ^{k}D_{x}f(x))%
}{\left| \det (1-D_{x}f^{-1}(x))\right| }  \tag{1.21}
\end{equation}
and, accordingly, \ in view of \ Eq.s(1.16)-(1.21), we obtain as well 
\begin{equation}
\sum\limits_{k=0}^{d}(-1)^{k}tr\mathcal{L}^{(k)}=\sum\limits_{x\in Fixf}g(x)%
\mathcal{L}_{x}(f).  \tag{1.22}
\end{equation}
Finally, if in the above formulas we replace \ $Fixf$ by $Fixf^{n}$\ \ we
have to replace $tr\mathcal{L}^{(k)}$ by $tr\mathcal{L}_{n}^{(k)}$. Since 
\begin{equation}
\exp (\sum\limits_{n=1}^{\infty }\frac{tr(\mathbf{A}^{n})}{n}t^{n})=\left[
\det (\mathbf{1}-t\mathbf{A})\right] ^{-1}  \tag{1.23}
\end{equation}
it is convenient to combine this result with Eq.(1.21) in order to obtain
the following zeta function: 
\begin{eqnarray}
Z(t) &=&\exp (\sum\limits_{n=1}^{\infty }\frac{t^{n}}{n}\left\{
\sum\limits_{k=0}^{d}(-1)^{k}tr\mathcal{L}_{n}^{(k)}\right\} )  \notag \\
&=&\prod\limits_{k=0}^{d}\left[ \exp (\sum\limits_{n=1}^{\infty }\frac{tr%
\mathcal{L}_{n}^{(k)}}{n}t^{n})\right] ^{(-1)^{k}}  \notag \\
&=&\prod\limits_{k=0}^{d}\left[ \det (\mathbf{1-}t\mathcal{L}^{(k)})\right]
^{(-1)^{k+1}}.  \TCItag{1.24}
\end{eqnarray}
The final result coincides with that obtained by Ruelle as required. Thus
obtained zeta function possess dynamical, number-theoretic and
algebro-geometric interpretation. Such type zeta function was discussed in
our earlier work, Ref.[2], in connection with dynamical/number-theoretic
interpretation of the $p$-adic Veneziano amplitudes. The discussion above
provides support \ in favor of \ earlier obtained dynamical interpretation
of $p$-adic Veneziano amplitudes. The crucial Eq.(5.29a) of Section 5
provides additional support in favor of such interpretation.

In the meantime, based on the paper by Atiyah and Bott [15], we would like
to demonstrate that Eq.(1.15) is actually the Weyl's character formula [16].
To prove that this is the case is not entirely trivial. The Appendix A
provides some general background on Weyl-Coxeter reflection groups needed
for understanding of \ the arguments presented below.

\subsection{From Atiyah-Bott-Lefschetz fixed point formula to character
formula by Weyl}

We begin with observation that earlier obtained Eq.s (1.12) and (1.15) are
essentially the same. In addition, Eq.s(1.12) and (1.7) are equivalent.
Because of this, it is sufficient to demonstrate that the r.h.s of Eq.(1.7)
\ indeed coincides with the Weyl's character formula. Although Eq.(1.7)
(and, especially, Eq.s(1.15),(1.21)) looks similar to that obtained in the
paper by Atiyah and Bott [15], part I, page 379, leading to the Weyl
character formula ( Eq.(5.12), Ref. [15], part II ) neither Eq.(1.7) nor
Eq.(5.11) of Atiyah and Bott paper, Ref. [15], part II, provide immediate
connection with their Eq.(5.12). Hence, the task now is to restore some
missing links.

To this purpose, following Bourbaki [7], consider a finite set of formal
symbols $e(\mu )$ possessing the same multiplication properties as the usual
exponents\footnote{%
In the case of \textbf{usual} exponents it is being assumed that all the
properties of formal exponents are transferable to the usual ones.}, i.e. 
\begin{equation}
e(\mu )e(\nu )=e(\mu +\nu ),\text{ }\left[ e(\mu )\right] ^{-1}=e(-\mu )%
\text{ and }e(0)=1.  \tag{1.25}
\end{equation}
Such defined set of formal exponents is making a free $\mathbf{Z}$ module
with the basis $e(\mu ).$ Subsequently we shall require that $\mu \in \Delta 
$ with $\Delta $ defined in the Appendix A. Suppose also that we are having
a polynomial ring A[\textbf{X}] made of all linear combinations of terms $%
\mathbf{X}^{\mathbf{n}}\equiv X_{1}^{n_{1}}\cdot \cdot \cdot X_{d}^{n_{d}}$
with $n_{i}\in \mathbf{Z}$, then one can construct another ring A[P] \ made
of linear combinations of elements $e(\mathbf{p}\cdot \mathbf{n})$ with $%
\mathbf{p}\cdot \mathbf{n=}p_{1}n_{1}+\cdot \cdot \cdot p_{d}n_{d}$.
Clearly, the above rings are isomorphic. Let $x=\sum\limits_{p\in
P}x_{p}e(p)\in $A[P] \ with P=$\{p_{1},...,p_{d}\},$ then using Eq.(1.25) we
obtain, 
\begin{eqnarray}
x\cdot y &=&\sum\limits_{s\in P}x_{s}e(s)\sum\limits_{r\in
P}y_{r}e(r)=\sum\limits_{t\in P}z_{t}e(t)\text{ with}  \notag \\
z_{t} &=&\sum\limits_{s+r=t}x_{s}y_{r}\text{ and, accordingly,}  \notag \\
x^{m} &=&\sum\limits_{t\in P}z_{t}e(t)\text{ with }z_{t}=\sum\limits_{s+%
\cdot \cdot \cdot +r=t}x_{s}\cdot \cdot \cdot y_{r}\text{, }m\in \mathbf{N} 
\TCItag{1.26}
\end{eqnarray}
with $\mathbf{N}$ being some non negative integer. Introduce now the
determinant of $w\in W$ via rule: 
\begin{equation}
\det (w)\equiv \varepsilon (w)=(-1)^{\mathit{l}(w)},  \tag{1.27}
\end{equation}
where all notations are taken from the Appendix A. If, in addition, we would
require 
\begin{equation}
w(e(p))=e(w(p))  \tag{1.28}
\end{equation}
then, all elements of the ring A[P] are subdivided into two classes defined
by 
\begin{equation}
w(x)=x\text{ (invariance)}  \tag{1.29a}
\end{equation}
and 
\begin{equation}
w(x)=\varepsilon (x)\cdot x\text{ (anti invariance).}  \tag{1.29b}
\end{equation}
These classes are very much like subdivision into bosons and fermions in
quantum mechanics. All anti invariant elements can be built from the basic
anti invariant element $J(x)$ defined by 
\begin{equation}
J(x)=\sum\limits_{w\in W}\varepsilon (w)\cdot w(x).  \tag{1.30}
\end{equation}
From the definition of P and from Appendix A it should be clear that the set
P can be identified with the set of reflection elements $w$ of the Weyl
group $W$. Therefore, for all $x\in $ A[P] and $w\in W$ we obtain the
following chain of equalities: 
\begin{equation}
w(J(x))=\sum\limits_{v\in W}\varepsilon (v)\cdot w(v(x))=\varepsilon
(w)\sum\limits_{v\in W}\varepsilon (v)\cdot v(x)=\varepsilon (w)J(x) 
\tag{1.31}
\end{equation}
as required. Accordingly, \textit{any} anti invariant element $x$ can be
written as $x=$ $\sum\limits_{l\in P}x_{p}J(\exp (p)).$ \ The denominator of
Eq.(1.7), when \ properly interpreted with help of the results of the
Appendix A, can be associated with $J(x)$. Indeed, without loss of
generality let us choose the constant $\mathbf{c}$ as $\mathbf{c}%
=\{1,...,1\} $. Then, for fixed $v$ the denominator of Eq.(1.7) can be
rewritten as follows: 
\begin{eqnarray}
\prod\limits_{i=1}^{d}(1-\exp \{-u_{i}^{v}\}) &\equiv &\prod\limits_{\alpha
\in \Delta ^{+}}(1-\exp (-\alpha ))\equiv \tilde{d}\exp (-\rho ),\text{ } 
\TCItag{1.32} \\
\text{where }\rho &=&\frac{1}{2}\sum\limits_{\alpha \in \Delta ^{+}}\alpha 
\text{ and }  \notag \\
\tilde{d} &=&\prod\limits_{\alpha \in \Delta ^{+}}(\exp (\frac{\alpha }{2}%
)-\exp (-\frac{\alpha }{2})).  \TCItag{1.33}
\end{eqnarray}

To prove that $\ $thus defined $\tilde{d}$ belongs to the anti invariant
subset of A[P] is not difficult. Indeed, consider the action of a reflection 
$r_{i}$ on $\tilde{d}$. Taking into account that $r_{i}(\alpha _{i})=-\alpha
_{i}$ we obtain 
\begin{eqnarray}
r_{i}(\tilde{d}) &=&(\exp (-\frac{\alpha _{i}}{2})-\exp (\frac{\alpha _{i}}{2%
}))\prod\limits_{\substack{ \alpha \neq \alpha _{i}  \\ \alpha \in \Delta
^{+} }}(\exp (\frac{\alpha }{2})-\exp (-\frac{\alpha }{2}))  \notag \\
&=&-\tilde{d}\equiv \varepsilon (r_{i})\tilde{d}  \TCItag{1.34}
\end{eqnarray}
Hence, clearly, 
\begin{equation}
\tilde{d}=\sum\limits_{p\in P}x_{p}J(\exp (p)).  \tag{1.35}
\end{equation}
Moreover, it can be shown [7] that, actually, $\tilde{d}=J(\exp (\rho ))$
which, in view of Eq.s(1.32), (1.33), produces identity originally obtained
by Weyl : 
\begin{equation}
\hat{d}\exp (-\rho )=\prod\limits_{\alpha \in \Delta ^{+}}(1-\exp (-\alpha
)).  \tag{1.36}
\end{equation}
Applying reflection $w$ to the above identity while taking into account
Eq.s(1.28),(1.34) produces: 
\begin{equation}
\prod\limits_{\alpha \in \Delta ^{+}}(1-\exp (-w(\alpha )))=\exp (-w(\rho ))%
\text{ }\varepsilon (w)\text{ }\hat{d}.  \tag{1.37}
\end{equation}
The result just obtained is of central importance for the proof of Weyl's
formula. Indeed, in view of Eq.s (1.28) and (1.37), inserting \ the identity
: $1=\dfrac{w}{w}$ $\ $\ into the sum over the vertices on the r.h.s.of
Eq.(1.7) \ and taking into account that: a) $\varepsilon (w)=\pm 1$ so that $%
\left[ \varepsilon (w)\right] ^{-1}=\varepsilon (w)$ ; b) actually, the sum
over the vertices is the same thing as \ the sum \ over the members of the
Weyl-Coxeter group (since all vertices of the \ integral polytope can be
obtained by use of the appropriate reflections applied to the highest weight
vector pointing to chosen vertex), we obtain the Weyl's character formula: 
\begin{equation}
tr\mathcal{L(\lambda )=}\frac{1}{\hat{d}}\sum\limits_{w\in \Delta
}\varepsilon (w)\text{ exp}\{w(\lambda +\rho )\}.  \tag{1.38}
\end{equation}
It was obtained with help of the results of Appendix A, Eq.s(1.15),(1.28)
and (1.37). Looking at the l.h.s. of Eq.(1.7) we can, of course, replace $tr%
\mathcal{L(\lambda )}$ by the sum in the l.h.s. of Eq.(1.7) if we choose the
constant $\mathbf{c}$ as before. This is not too illuminating however as we
shall explain now.

Indeed, since $J(x)$ in Eq.(1.30) is by construction the basic anti
invariant element and the r.h.s of Eq.(1.38) is by design manifestly
invariant element of A[P], it is only natural to look for the basic
invariant element analogue of $J(x).$ Then, clearly, $tr\mathcal{L(\lambda
)\equiv \chi (\lambda )}$ should be expressible as follows 
\begin{equation}
\mathcal{\chi (\lambda )=}\sum\limits_{w\in W}n_{w}(\lambda )\text{ }e(w). 
\tag{1.39}
\end{equation}
This result admits clear physical interpretation. Indeed, the partition
function $\Xi $ of any quantum mechanical system can be represented as 
\begin{equation}
\Xi =\sum\limits_{n}g_{n}\exp \{-\beta E_{n}\}\equiv tr(\exp (-\beta \hat{H}%
))  \tag{1.40}
\end{equation}
where $\hat{H}$ is \ the quantum Hamiltonian of the system, $\beta $ is the
inverse temperature and $g_{n}$ is the degeneracy factor. If we introduce
the density of states $\rho (E)$ via 
\begin{equation}
\rho (E)=\sum\limits_{n}\delta (E-E_{n})  \tag{1.41}
\end{equation}
then, the partition function $\Xi $ can be written as the Laplace transform 
\begin{equation}
\Xi (\beta )=\int\limits_{0}^{\infty }dE\rho (E)\exp \{-\beta E\}. 
\tag{1.42}
\end{equation}
Clearly, Eq.(1.39) is just the discrete analogue of Eq.(1.42) so that it
does have a statistical/quantum mechanical interpretation as partition
function. From condensed matter physics it is known that all important
statistical information is contained in the density of states. Its
calculation is of primary interest in physics. Evidently, the same should be
true in the present case as well. Indeed, the results of Section 5
illustrate this point extensively. The density of states $n_{w}(\lambda )$
is known in group theory as Kostant's multiplicity formula [20]. Cartier
[21] had simplified the original derivation by Kostant to the extent that it
is worth displaying it here since it supplements nicely what was presented
already and to be discussed further in the rest of the paper. Cartier
noticed that the denominator of the Weyl character formula, Eq.(1.38), can
be formally expanded with help of Eq.(1.36) as follows: 
\begin{equation}
\left[ \exp (\rho )\prod\limits_{\alpha \in \Delta ^{+}}(1-\exp (-\alpha ))%
\right] ^{-1}=\sum\limits_{w^{\prime }\in W}P(w^{\prime })e(-\rho -w^{\prime
})  \tag{1.43}
\end{equation}
By combining Eq.s(1.38),(1.39) and (1.43) we obtain 
\begin{equation}
\sum\limits_{w\in W}n_{w}(\lambda )\text{ }e(w)=\sum\limits_{w\in
W}\varepsilon (w)\text{ exp}\{w(\lambda +\rho )\}\sum\limits_{w^{\prime }\in
W}P(w^{\prime })e(-\rho -w^{\prime }).  \tag{1.44}
\end{equation}
Comparing the left side with the right we obtain finally the Kostant
multiplicity formula 
\begin{equation}
n_{w}(\lambda )=\sum\limits_{w^{\prime }\in W}\varepsilon (w^{\prime
})P(w^{\prime }(\lambda +\rho )-(\rho +w)).  \tag{1.45}
\end{equation}
The obtained formula allows us to determine the density of states $%
n_{w}(\lambda )$ provided that we can obtain the function $P$ explicitly.
This will be discussed in Section 5.

Next, the obtained results allow us to clear up yet another problem: when
compared with the r.h.s. of Eq.(5.11) of Atiyah and Bott paper, Ref.[15],
part II, the r.h.s. of our Eq.(1.7) is not looking the same. We would like
to explain that, actually, these expressions are equivalent. To this
purpose, let us reproduce Eq.(5.11) of Atiyah and Bott first. Actually, for
this purpose it is more convenient to use the paper by Bott, Ref.[16],
Eq.(28). In terms of notations taken from this reference we have: 
\begin{equation}
trace\text{ }T_{g}=\sum\limits_{\substack{ w\in W  \\ \alpha <0}}\left[ 
\frac{\lambda }{\prod \left( 1-\alpha \right) }\right] ^{w}.  \tag{1.46}
\end{equation}
Comparing with the r.h.s. of our Eq.(1.7) and taking into account Eq.(1.28),
the combination $\lambda ^{w}$ in the numerator of Eq.(1.46) is the same
thing as $\exp \{w\lambda \}$ in Eq.(1.38). As for the denominator, Bott
uses the same Eq.(1.36) as we do so that it remains to demonstrate that 
\begin{equation}
\left[ \prod\limits_{\alpha \in \Delta ^{+}}(1-\exp (-\alpha ))\right]
^{w}=\exp (-w(\rho ))\text{ }\varepsilon (w)\text{ }\hat{d}.  \tag{1.47}
\end{equation}
In view of Eq.(1.36), we need to demonstrate that 
\begin{equation}
\left[ \hat{d}\exp (-\rho )\right] ^{w}=\exp (-w(\rho ))\text{ }\varepsilon
(w)\text{ }\hat{d},  \tag{1.48}
\end{equation}
i.e. that $\left[ \hat{d}\right] ^{w}=\varepsilon (w)$ $\hat{d}.$ In view of
Eq.(1.34), this requires us to assume that $\left[ \hat{d}\right] ^{w}=w\hat{%
d}.$ But, in view of Eq.s(1.28), (1.30) and (1.35), we conclude that this is
indeed the case. This proves the fact that Eq.(1.46), that is Eq.(5.11) of
Ref.[15], is indeed the same thing as the Weyl's character formula,
Eq.(5.12) of Ref.[15], or Eq.(1.38) above. According to Kac, Ref.[12], page
174, the classical Weyl character formula, Eq.(1.38), is formally valid for
both finite dimensional semisimple Lie algebras and infinite dimensional
affine Kac-Moody algebras. This circumstance and the Proposition A.1 of
Appendix A play important motivating role in developments presented in this
work. Our treatment so far had been limited only to the $d-$ dimensional
hypercube. This deficiency can be easily corrected with help of the concept
of \textit{zonotope }which we would like to explain now.

\section{ From zonotopes to fans and toric varieties}

\subsection{\protect\bigskip From zonotopes to fans}

The concept of zonotope is actually not new. According to Coxeter [22] it
belongs to 19th century Russian crystallographer Fedorov. Nevertheless, this
concept has been truly appreciated only relatively recently in connection
with oriented matroids . For our purposes it is sufficient to consider only
the most elementary properties of zonotopes. To this purpose, following
Ref.[23] let us consider a $p-$dimensional cube $C_{p}$ defined by 
\begin{equation}
C_{p}=\{\mathbf{x}\in \mathbf{R}^{p},-1\leq x_{i}\leq 1\text{ , }i=1-p\} 
\tag{2.1}
\end{equation}
and the \ surjective map $\pi :\mathbf{R}^{p}\rightarrow \mathbf{R}^{d}$
given by $\pi :$ $\mathbf{x}\rightarrow V\mathbf{x}+\mathbf{z}$ \ with $V$
being $d\times p$ matrix written in the vector form as $V=\{\mathbf{v}%
_{1},...,\mathbf{v}_{p}\}$ so that, actually, $\mathbf{x}^{\prime }=\mathbf{z%
}+\sum\limits_{i=1}^{p}x_{i}\mathbf{v}_{i}$ with $\mathbf{x}^{\prime }\in 
\mathbf{R}^{d}$ and $-1\leq x_{i}\leq 1.$ A \textit{zonotope} $\ Z(V)$ is
the image of a $p-$cube under affine projection $\pi ,$ i.e. $Z(V)$=$VC_{p}+%
\mathbf{z}$ \ It \ can be shown [23] that such created zonotope is a\
centrally symmetric $d-$polytope. Because of central symmetry, it is
sometimes convenient to associate with such $d$-polytope its fan. This
concept can be easily understood with help of pictures. In particular, the
simplest zonotope construction is displayed in Fig.1.

\FRAME{ftbphFU}{1.817in}{1.7867in}{0pt}{\Qcb{Two dimensional zonotope Z from
three dimensional cube C$_{3}$}}{\Qlb{zonotope}}{figure1.gpj}{\special%
{language "Scientific Word";type "GRAPHIC";maintain-aspect-ratio
TRUE;display "PICT";valid_file "F";width 1.817in;height 1.7867in;depth
0pt;original-width 8.9794in;original-height 8.6983in;cropleft "0";croptop
"0.9629";cropright "0.9489";cropbottom "0";filename
'zonotope3.jpg';file-properties "XNPEU";}}

\bigskip

The fan associated with such planar zonotope is schematically displayed in
Fig.2. As Fig.2 suggests, such fan is a collection of cones. These cones can
be constructed by drawing semi- infinite strips whose semi- infinite edges
are perpendicular to the corresponding edges of zonotope and then, by
identifying the \ set of cones with the complement of zonotope Z (together
with these strips) in \textbf{R}$^{2}$. Extension of such procedure to
dimensions $d\geq 2$ is straightforward. Thus formed \textit{normal} fan
[23] is \textit{complete} if upon parallel translation of apexes of the
cones to one point \ in \textbf{R}$^{2}($ or \textbf{R}$^{d}$ in general
case) the resulting picture spans \textbf{R}$^{2}$ (or \textbf{R}$^{d})$ as
depicted in Fig.2.

\FRAME{ftbphFU}{1.964in}{1.0136in}{0pt}{\Qcb{Construction of the complete
fan associated with two dimensional zonotope Z}}{\Qlb{fan}}{figure2.jpg}{%
\special{language "Scientific Word";type "GRAPHIC";maintain-aspect-ratio
TRUE;display "USEDEF";valid_file "F";width 1.964in;height 1.0136in;depth
0pt;original-width 13.5836in;original-height 6.9168in;cropleft "0";croptop
"1";cropright "1";cropbottom "0";filename 'fan.jpg';file-properties "XNPEU";}%
}

For zonotopes, in view of their central symmetry, the fans thus constructed
are always complete. For the purposes of development, we would like to give
more formal \ mathematical definitions\ now.

\textbf{Definition}.\textbf{1}. Let $\mathcal{R}$\ $\subset $ \textbf{R}$%
^{d} $ \ be a subset of \textbf{R}$^{d}$ made of finite set of vectors \ $%
\mathbf{y}_{1},...,\mathbf{y}_{k}$. Then, a \textit{convex polyhedral} cone $%
\sigma $ is a set 
\begin{equation}
\sigma =\sum\limits_{i=1}^{k}r_{i}\mathbf{y}_{i}\text{ }\in \mathcal{R\ }%
\text{, }r_{i}\geq 0.  \tag{2.2}
\end{equation}
The scalar product, Eq.(1.6), allows \ us to introduce

\textbf{Definition 2}. The \textit{dual} cone $\sigma ^{\vee }$ is defined
as 
\begin{equation}
\sigma ^{\vee }==\{\mathbf{u}\in \mathcal{R}^{\ast }:\text{ }<\mathbf{u},%
\mathbf{y>}\text{ }\mathbf{\geq }\text{ }0\text{ for all }\mathbf{y}\in
\sigma \}.  \tag{2.3}
\end{equation}
Finally, we have as well

\textbf{Definition 3}. A \textit{face} $\tau $ of $\sigma $ as an
intersection of $\sigma $ with any \textit{supporting} hyperplane H$_{\alpha
}$ defined by $<\mathbf{u}_{\alpha },\mathbf{y}>=0$ , i.e. 
\begin{equation}
\tau =\sigma \cap \mathbf{u}_{\alpha }=\{\mathbf{y}\in \sigma :\text{ }<%
\mathbf{u}_{\alpha },\mathbf{y}>=0\}  \tag{2.4}
\end{equation}
for some $\mathbf{u}_{\alpha }$ in $\sigma ^{\vee }.$

By analogy with Eq.(1.6), if vectors \ $\mathbf{y}_{1},...,\mathbf{y}_{k}\in 
\mathbf{Z}^{d}$ the cone is called \textit{rational} \textit{polyhedral}. We
shall assume that this is the case from now on.

In view of Eq.(2.4), there is a correspondence between the faces and the
supporting hyperplanes. Fig.2 helps us to realize that assembly of rational
polyhedral cones is made out of complements of these supporting hyperplanes.
It can be proven, e.g. [24], page 144, or Appendix A (part c)), that these
hyperplanes intersect each other just in one point (apex) which can be
chosen as origin in \textbf{Z}$^{d}$. Fig.2 provides an intuitive support to
this claim. Clearly, as results of Appendix A suggest, \ the polyhedral
cones are just the chambers and the supporting hyperplanes are the
reflecting hyperplanes. Fig.2 helps to visualize the interrelationship
between the polytopes and the chamber system. All this, surely, can be made
absolutely rigorous so that we refer our readers to literature [7,24,25]. In
this work, we would like to discuss only the results of immediate relevance.
In particular, the interrelationships between the affine and the projective
toric varieties and the rational polyhedral cones.

We begin with a couple of definitions.

\textbf{Definition 4.} \textit{A semi-group S\ }that is a non-empty set with
associative\textit{\ }operation\textit{\ }is called \textit{monoid\ }if it
is commutative, satisfies cancellation law\textit{\ (i.e.s+x=t+x }implies%
\textit{\ s=t for all s,t,x}$\in S)$ and has zero element $(i.e.s+0=s,s\in
S).$

\textbf{Definition 5}. \textit{A monoid S is} \textit{finitely generated }if
exist set\textit{\ }$a_{1},...,a_{k}\in S$\textit{, }called $generators$,
such that 
\begin{equation}
S=Z_{\geq 0}a_{1}+\cdot \cdot \cdot +Z_{\geq 0}a_{k}.  \tag{2.5}
\end{equation}
Taking into account these definitions, it is clear that the monoid $\mathbf{S%
}_{\sigma }$ =$\sigma \cap \mathbf{Z}^{d}$ \ for the rational polyhedral
cone is finitely generated.

Next, we consider a polynomial 
\begin{equation}
f(\mathbf{z})=f(z_{1},...,z_{n})=\sum\limits_{\mathbf{i}}\lambda _{\mathbf{i}%
}\mathbf{z}^{\mathbf{i}}=\sum\limits_{\mathbf{i}}\lambda
_{i_{1}....i_{n}}z_{1}^{i_{1}}\cdot \cdot \cdot z_{n}^{i_{n}},\text{ (}%
\lambda _{\mathbf{i}},\text{ }z_{m}^{i_{m}}\in \mathbf{C,}1\leq m\leq n). 
\tag{2.6}
\end{equation}
It belongs to the polynomial ring K[$\mathbf{z}$] closed under ordinary
addition and multiplication.

\textbf{Definition 6}.\ An\textit{\ affine} algebraic variety $V\in \mathbf{C%
}^{n}$ is the set of zeros of \ collection of polynomials from the above
ring.

Collection of such polynomials is \textbf{finite} according to famous
Hilbert's Nullstellensatz and forms the set \ $I(\mathbf{z}):=\{f\in $ K[$%
\mathbf{z}$]$,$ $f(\mathbf{z})=0\}$ of maximal ideals usually denoted \emph{%
spec}K[\textbf{z}].

\textbf{Definition 7}. Zero set of a \textit{single} function belonging to $%
I(\mathbf{z})$ is called \textit{algebraic hypersurface }so that the set\ $I(%
\mathbf{z})$ corresponds to the \textit{intersection} of a finite number of
hypersurfaces.

\subsection{From fans to affine toric varieties}

\ To connect these results \ with those introduced earlier, let us consider
now the set of Laurent monomials of the type $\lambda z^{\mathbf{\alpha }%
}\equiv \lambda z_{1}^{\alpha _{1}}\cdot \cdot \cdot z_{n}^{\alpha _{n}}$ .
We shall be particularly interested in \textit{monic} monomials for which $%
\lambda =1$ because such monomials form a \ closed polynomial subring with
respect to usual multiplication and addition. The crucial step forward is to
assume that the exponent $\mathbf{\alpha \in S}_{\sigma }.$ This allows us
to define the following mapping 
\begin{equation}
u_{i}:=z^{a_{i}}  \tag{2.7}
\end{equation}
with $a_{i}$ \ being \ one of generators of the monoid $\mathbf{S}_{\sigma }$
and$\ z\in \mathbf{C}.$ Such mapping provides an isomorphism between the
additive group of exponents $a_{i}$ and the multiplicative group of monic
Laurent polynomials. Let us recall now that the function $\phi $ is
considered to be \textit{quasi homogenous} of degree $d$ with exponents 
\textit{l}$_{1},...,l_{n}$ if 
\begin{equation}
\phi (\lambda ^{\mathit{l}_{1}}x_{1},...,\lambda ^{\mathit{l}%
_{n}}x_{n})=\lambda ^{d}\phi (x_{1},...,x_{n}),  \tag{2.8}
\end{equation}
provided that $\lambda \in \mathbf{C}^{\ast }.$ Applying this result to $z^{%
\mathbf{a}}\equiv z_{1}^{a_{1}}\cdot \cdot \cdot z_{n}^{a_{n}}$ we reobtain
equation for the cone 
\begin{equation}
\sum\limits_{j}\left( l_{j}\right) _{i}a_{j}=d_{i}  \tag{2.9}
\end{equation}
which belongs to the monoid $\mathbf{S}_{\sigma }.$ Here the index $i$ is
numbering different monomials. Clearly, the same result can be achieved if
instead we would consider products of the type $u_{1}^{l_{1}}\cdot \cdot
\cdot u_{n}^{l_{n}}$ and rescale all $z_{i}^{\prime }s$ by the same factor $%
\lambda .$ Eq(2.9) is actually a scalar product with $\left( l_{j}\right)
_{i}$ living in the space \textit{dual} to $a_{j}^{\prime }s$ . So that, in
view of Eq.(2.3), the set of $\left( l_{j}\right) _{i}^{\prime }s$ can be
considered as the set of generators for the dual cone $\sigma ^{\vee }$ .
Next, in view of Eq.(2.6) let us consider polynomials of the type 
\begin{equation}
f(\mathbf{z})=\sum\limits_{\mathbf{a\in S}_{\sigma }}\lambda _{\mathbf{a}}%
\mathbf{z}^{\mathbf{a}}=\sum\limits_{\mathbf{l}}\lambda _{\mathbf{l}}\mathbf{%
u}^{\mathbf{l}}.  \tag{2.10}
\end{equation}
They form a polynomial ring as before. The ideal for this ring can be
constructed based on the observation that for the fixed\ $d_{i}$ \ and the
assigned set of \ cone generators $a_{i}^{\prime }$ there is more than one
set of generators for the dual cone. This redundancy produces relations of
the type 
\begin{equation}
u_{1}^{l_{1}}\cdot \cdot \cdot u_{k}^{l_{k}}=u_{1}^{\tilde{l}_{1}}\cdot
\cdot \cdot u_{k}^{\tilde{l}_{k}}.  \tag{2.11}
\end{equation}
If now we require $u_{i}^{{}}\in \mathbf{C}_{i},$ then it is clear that the
above equation belongs to the ideal $I(\mathbf{z})$ of the above polynomial
ring and that Eq.(2.11) represents the hypersurface. \ As before, $I(\mathbf{%
z})$ represents the intersection of these hypersurfaces thus forming the
affine \textit{toric} variety $X_{\sigma ^{\vee }}.$ The generators \{$%
u_{1},...,u_{k}\}\in \mathbf{C}^{k}$ are coordinates for $X_{\sigma ^{\vee
}} $. They represent the same point in $X_{\sigma ^{\vee }}$ if and only if 
\textbf{u}$^{\mathbf{l}}=$\textbf{u}$^{\mathbf{\tilde{l}}}.$ \ Thus formed
toric variety corresponds to just one (dual) cone.

\subsection{Building toric varieties from affine toric varieties}

As Appendix A suggests, there is one- to- one correspondence between cones
and chambers. From chambers one can construct a gallery and, hence, a
building. Accordingly, information leading to the design of particular
building can be used for construction of toric variety from the set of
affine toric varieties. To do so one only needs the set of \textit{gluing
maps }$\mathit{\{}$ $\Psi _{\sigma ^{\vee }\breve{\sigma}^{\vee }}\}$. Thus,
we obtain the following

\textbf{Definition 8. }Let $\Sigma $ be the complete fan and $%
\coprod\nolimits_{\sigma ^{\vee }\in \Sigma }X_{\sigma ^{\vee }}$ be the
disjoint union of affine toric varieties. Then, using the set of gluing maps 
$\mathit{\{}$ $\Psi _{\sigma ^{\vee }\breve{\sigma}^{\vee }}\}$ such that
each of them identifies two points $x\in X_{\sigma ^{\vee }}$ and $\breve{x}%
\in X_{\breve{\sigma}^{\vee }}$ on respective affine varieties, one obtains
the toric variety $X_{\Sigma }$ determined by fan $\Sigma .$

\textit{Remark }1.Thus constructed variety $X_{\Sigma }$ may contain
singularities which should be obvious just by looking at Eq.(2.11). There is
a procedure of desingularization described, for example in Ref.[26]. In this
work (for the sake of space) we are going to bypass this circumstance. This
is permissible because in the end the results are going to come up the same
as for nonsingular varieties.

\subsection{Torus action and its invariants}

\bigskip

Based on the results just presented we can accomplish more. We begin with

\textbf{Definition 9}. The set $T:=(\mathbf{C}\backslash 0)^{n}=:(\mathbf{C}%
^{\ast })^{n}$ is called complex algebraic torus.

Since each $z\in \mathbf{C}^{\ast }$ can be written as $z=r\exp (i\theta )$
so that for each $r>0$ the fiber: \{$z\in \mathbf{C}^{\ast }\mid \left|
z\right| =r\}$ is a circle of radius $r$, we can represent $T$ as the
product ($R_{>0})^{n}\times \left( S^{1}\right) ^{n}$. The product of $n$
circles $\left( S^{1}\right) ^{n}$ which is the deformation retract of $T$
is indeed a topological torus. Following Fulton [26], we are going to call
it a \textit{compact} \textit{torus} $S_{n}$. So that the algebraic torus is
a product of a compact torus and a vector space. This circumstance is
helpful since whatever we can prove for the deformation retract can be
extended to the whole torus $T$. This explains the name ''algebraic torus''.
Let now $G$ be the group acting (multiplicatively) on the set $X$ via
mapping $G\times X\rightarrow X,$ i.e. $(g,x)\rightarrow gx$ provided that
for all $g,h\in G$, $g(hx)=(gh)x$ and $ex=x$ for some unit element $e$ of $G$%
. The subset $Gx:=\{gx\mid g\in X\}$ of $X$ is called the \textit{orbit} of $%
x$. Denote by $H$ the subgroup of $G$ that fixes $x$, i.e. $H:=\{gx=x\mid
g\in X\}$ the $isotropy$ group. Surely, there could be more than one fixed
point for equation $gx=x$ and all of them are conjugate to each other. A 
\textit{homogenous} space for $G$ is the subspace of $X$ on which $G$ acts
without fixed points. The crucial step forward can be made by introducing a
concept of an \textit{algebraic} group [27].

\textbf{Definition 10.} A \textit{linear algebraic group} $G$ \ is a) an
affine algebraic variety and b) a group in the sense given above, i.e. 
\begin{eqnarray}
\mu &:&G\times G\rightarrow G\text{ };\text{ }\mu (x,y)=xy  \TCItag{2.12a} \\
i &:&G\rightarrow G\text{ };\iota (x)=x^{-1}.  \TCItag{2.12b}
\end{eqnarray}

\textit{Remark} 2. It can be shown [28], page150, that $G$ as linear
algebraic group is isomorphic (as an algebraic group) to a closed subgroup
of $GL_{n}($K$)$ for some $n\geq 1$ and any closed field K such as $\mathbf{C%
}$ or $p-$adic.

This fact plays a crucial role in the whole development given below.
Moreover, another no less important direction of development comes from

\textit{Remark} 3. Let G be a linear algebraic group, $A=$K$[G]$ its affine
algebra, then, the set of rules given by Eq.s(2.12) can be replaced by the
following set 
\begin{eqnarray}
\mu ^{\ast } &:&A\rightarrow A\otimes _{K}A\text{ \ comultiplication} 
\TCItag{2.13a} \\
\iota ^{\ast } &:&A\rightarrow A\text{ taking the antipode of }A. 
\TCItag{2.13b}
\end{eqnarray}
Thus, study of properties of linear algebraic groups is equivalent to study
of coassociative Hopf algebras leading to quantum groups [27,28], etc.

In this work we are not going to develop this line of thought however,
except for few comments made in Section 6. Rather, we would like to connect
previous discussion with Remark 2. To this purpose, in accord with earlier
results, we need to introduce the following

\textbf{Definition} \textbf{11}. The \textit{torus action} is a continuous
map : $T\times X_{\Sigma }\rightarrow X_{\Sigma }$ such that for each affine
variety corresponding to the dual cone it is given by 
\begin{equation}
T\times X_{\sigma ^{\vee }}\rightarrow X_{\sigma ^{\vee }}\text{ , }%
(t,x)\mapsto tx:=(t^{a_{1}}x_{1},...,t^{a_{k}}x_{k}).  \tag{2.14}
\end{equation}
Naturally, such action should be compatible with gluing maps thus extending
it from one cone (chamber) to the entire variety $X_{\Sigma }($building).
The compatibility is easy to enforce since for each of Eq.s (2.11)
multiplication by $t-$factors will not affect the solutions set. This can be
formally stated as follows. Let $\Psi :$ $X_{\Sigma }\rightarrow X_{\tilde{%
\Sigma}}$ be a map and $\alpha :T\rightarrow T^{\prime }$ a homomorphism,
then the map $\Psi $ is called \textit{equivariant} if it obeys the
following identity 
\begin{equation}
\Psi (cx)=\alpha (c)\Psi (x)\text{ for all }c\in T.  \tag{2.15}
\end{equation}
Naturally, $\alpha (c)$ is character of the algebraic torus group.\footnote{%
This fact is known as Borel-Weil theorem [29]. As such it belongs to the
theory of induced group representations [30].} For physical applications it
is more advantageous, following Stanley [31], to consider $invariants$ given
by 
\begin{equation}
\Psi (cx)=\Psi (x)  \tag{2.16}
\end{equation}
rather than $\alpha -$invariants (equivariants) in Stanley's terminology. To
obtain invariants we need to study the orbits of the torus action first. To
this purpose, in view of Eq.(2.14), we need to consider the following fixed
point equation 
\begin{equation}
t^{a}x=x.  \tag{2.17}
\end{equation}
Apart from trivial solutions$:x=0$ and $x=\infty $, there is a nontrivial
solution $t^{a}=1$ for any $x$. For integer $a^{\prime }s$ this is a
cyclotomic equation whose nontrivial $a-1$ solutions all lie on the circle $%
S^{^{1}}.$ In view of this circumstance, it is possible to construct
invariants for this case as we are going to explain now. First, such an
invariant can be built as a \textit{ratio} of two equivariant mappings of
the type given by Eq.(2.15). By construction, such ratio is the\textbf{\ }%
\textit{projective} toric variety. Unlike the affine case, such varieties
are \textbf{not} represented by functions of homogenous coordinates in 
\textbf{CP}$^{n}.$ Instead, they are just constants associated with points
in \textbf{CP}$^{n}$ which they represent. Second option, is to restrict the
algebraic torus to compact torus acting on the circle. These two options are
interrelated in important way as we would like to explain now.

To this purpose we notice that Eq.(2.14) still holds if some of $t-$factors
are replaced by 1's. This means that one should take into account all
situations when one, two, etc. $t$-factors in Eq.(2.14) are replaced by 1's
and account all permutations involving such cases. This leads to torus
actions on toric subvarieties. It is very important that different orbits
belong to different subvarieties which do not overlap. Thus, by design, $%
X_{\Sigma }$ is the disjoint union of finite number of orbits which are
identified with \ subvarieties of $X_{\Sigma }.$ To make all this more
interesting, let us consider instead of Eq.(2.14) its transpose. Then, it
can be viewed as part of the eigenvalue problem 
\begin{equation}
\left( 
\begin{array}{ccccc}
t^{a_{1}} &  &  &  &  \\ 
& \cdot &  &  &  \\ 
&  & \cdot &  &  \\ 
&  &  & \cdot &  \\ 
&  &  &  & t^{a_{k}}
\end{array}
\right) \left( 
\begin{array}{c}
x_{1} \\ 
\cdot \\ 
\cdot \\ 
\cdot \\ 
x_{k}
\end{array}
\right) =\left( 
\begin{array}{c}
t^{a_{1}}x_{1} \\ 
\cdot \\ 
\cdot \\ 
\cdot \\ 
t^{a_{k}}x_{k}
\end{array}
\right) .  \tag{2.18}
\end{equation}
Under such conditions the\ vector $\left( x_{1},...,x_{k}\right) $ forms a
basis of \ $k-$dimensional vector space $V$ so that the vector $\left(
x_{1},...,x_{i}\right) $ , $i\leq k,$ forms a basis of subspace $V_{i}$ .
This allows us to introduce a complete flag $f_{0}$ of subspaces in $V$
(just like in our previous work, Ref.[32]) 
\begin{equation}
f_{0}:0=V_{0}\subset V_{1}\subset ...\subset V_{k}=V.  \tag{2.19}
\end{equation}
Consider now action of $G$ on $f_{0}$ . Taking into account the Remark 2,
effectively, $G$=$GL_{n}(K)$. The matrix representation of this group
possess remarkable property. To formulate this property requires several
definitions. They are provided below. To make our exposition less formal, we
would like to explain their rationale now so that their importance cannot be
underestimated.

The eigenvalue problem for some matrix \textbf{A }is reduced to finding
eigenvectors and eigenvalues of the equation \textbf{Ax}=$\mathbf{\lambda x.}
$ Eq.(2.18) fits trivially to this category of problems. Suppose now that
the matrix \textbf{A }can be presented as\textbf{\ A=}UTU$^{-1}$ with U
being some matrix which we need to find, then, of course, the above
eigenvalue equation can be written as TU$^{-1}\mathbf{x}=$U$^{-1}\mathbf{x}$
and, as usual, if we replace U$^{-1}\mathbf{x}$ by \textbf{y}, we reobtain
back Eq.(2.18). This standard exercise can be made more entertaining by
making an assumption (to be justified below) that equation U$^{-1}\mathbf{x=x%
}$ or, more generally, equation U$^{-1}\mathbf{x=\pi x}$ \ can be solved,
where $\mathbf{\pi }$ is some permutation of indices of $\mathbf{x}$. If
this is possible, then, instead of the torus action defined by Eq.(2.14),
one can consider much more general action $G\times X_{\sigma ^{\vee }}.$
Moreover, by making a quotient $G\times X_{\sigma ^{\vee }}/H$ with $H$ made
of U$^{-1}\mathbf{x=x}$ we shall obtain transitive action of $G$ on $%
X_{\sigma ^{\vee }}$ which permutes the ''roots'', i.e. the set $%
\{t^{a_{1}},...,t^{a_{n}}\}.$ That is it acts as the Weyl reflection group
and, indeed, it is the Weyl group (to be defined and explained below) for
this case. In the language of linear algebraic groups, taking the quotient
of the action of such group on $X_{\sigma ^{\vee }}$ by $H$ causes $G$ to
act transitively not on affine but on projective (quasi-projective to be
exact [28], page 160) toric variety. From the definition of such varieties
(e.g. read discussion next to Eq.(2.17)) it follows that such transitive
action is associated with ''motion'' on the fixed point set of such
projective variety. This fact has important physical consequences affecting
the rest of this paper, especially, Section 5. In the meantime, let us
return to the promised definitions. We have

\textbf{Definition 12.} Given that $\ $the set $GL_{n}(K)=\{x\in M_{n}(K$)$%
\mid \det x\neq 0\}$ with $M_{n}(K$) being \ $n\times n$ matrix with entries 
$x_{i,j}\in K$, forms the general linear group, the matrix $x\in M_{n}($K) is

a) $semisimple$ \ ($x=x_{s}$), if it is diagonalizable, that is $\exists
g\in GL_{n}(K)$ such that $gxg^{-1}$ is a diagonal matrix;

b) $nilpotent$ ($x=x_{n}$ ) if $x^{m}=0,$that is for some positive integer $%
m $ all eigenvalues of matrix $x^{m}$ are zero;

c) $unipotent$ ($x=x_{u}$), if $x-1_{n}$ is nilpotent, i.e. $x$ is the
matrix whose only eigenvalues are 1's.

\bigskip

Just like with the odd and even numbers the above matrices, \textit{if they
exist,} form closed disjoint subsets of $GL_{n}(K)$, e.g. all $x,y\in
M_{n}(K)$ commute; if $x,y$ are semisimple so is their sum and product, etc.
Most important for us is the following fact:

\bigskip

\textbf{Proposition 1}.\textit{Let }$x\in GL_{n}($\textit{K}$)$\textit{,
then }$\exists $\textit{\ }$x_{u}$\textit{\ and }$x_{s}$\textit{\ such that }%
$x=x_{s}x_{u}=x_{u}x_{s}$\textit{\ Both x}$_{s}$\textit{and x}$_{u}$\textit{%
\ are determined by the above conditions uniquely.}

\textit{\bigskip }The proof can be found in Ref.[33], page96.

This proposition is in fact a corollary of the Lie-Kolchin theorem \ which
is of central importance for us. To formulate this theorem \ we need to
introduce yet another couple of definitions. In particular, \ if $A$ and $B$
are \ closed (finite) subgroups of an algebraic group $G$ one can construct
the group $(A,B)$ made of commutators $xyx^{-1}y^{-1}$, $x\in A,$ $y\in B$ .
With help of such commutators the following definition can be made

\textbf{Definition 13.} The group $G$ is \textit{solvable} if its derived
series terminates in the unit element $e$. The derived series is being
defined inductively by $\mathcal{D}^{\left( 0\right) }G=G,\mathcal{D}%
^{\left( i+1\right) }G=(\mathcal{D}^{\left( i\right) }G,\mathcal{D}^{\left(
i\right) }G),i\geq 0.$

Such definition implies that an algebraic group $G$ is solvable if and only
if there is exist a chain $G=G^{\left( 0\right) }\supset G^{(1)}\supset
\cdot \cdot \cdot G^{\left( n\right) }=e$ for which $(G^{\left( i\right)
},G^{\left( i\right) })\subset G^{i+1}$ $(0\leq i\leq n)$, Ref.$[33]$, page
111$.$ Finally,

\textbf{Definition 14}.The group is called \textit{nilpotent} if $\mathcal{E}%
^{(n)}\mathit{G=e}$ for some $n$, where $\mathcal{E}^{\left( 0\right) }=G,%
\mathcal{E}^{\left( i+1\right) }=(G,\mathcal{E}^{\left( i\right) }G).$

Such group is represented by the nilpotent matrices. Based on this it is
possible to prove [112] that Every nilpotent group is solvable [33],page
112.These results lead to the Lie-Kolchin theorem of major importance

\textbf{Theorem 1.} \textit{Let }$G$\textit{\ be connected solvable \
algebraic group acting on a projective variety X. Then }$G$\textit{\ has a
fixed point in }$X$\textit{.}

( for proofs e.g .see Ref.[33], page113)

In view of the Remark 2, we know that \ such $G$ is a subgroup of $GL_{n}($%
\textit{K}$).$ Moreover, $GL_{n}($\textit{K}$)$ has at least another
subgroup, called $semisimple$, for which Theorem 1 does not hold.

\textbf{Definition 15}. The group $G$ is \textit{semisimple} if it has no
closed connected commutative normal subgroups other than $e$.

Such group is represented by semisimple, i.e. diagonal (or torus) matrices
while the members of the unipotent group are represented by the upper
triangular matrices with all diagonal entries being equal to 1. In view of
the Theorem 1, the unipotent group is also solvable and, accordingly, there
must be an element \ $B$ of such group which fixes the flag $\ f_{0}$ given
by Eq.(2.19), i.e. $Bf_{0}=f_{0}.$ Let now $g\in GL_{n}($\textit{K}$)$.
Then, naturally, $gf_{0}=f$ where $f\neq f_{0}.$ From here we obtain, $%
f_{0}=g^{-1}f$. Next, we get as well $Bg^{-1}f=g^{-1}f$ and, finally, $%
gBg^{-1}f=f$ . From here, it follows that $gBg^{-1}=\tilde{B}$ is also an
element \ of $GL_{n}($\textit{K}$)$ which fixes flag $\ f$, etc. This means
that all such elements are conjugate to each other and form the $\emph{Borel}
$ $\emph{subgroup}$.We shall denote all elements of this sort by $B\footnote{%
These are made of upper triangular matrices belonging to $GL_{n}($\textit{K}$%
).$ Surely, such matrices satisfy Proposition 1.}$. Clearly, the quotient
group $G/B$ will act transitively on $X$. Since this quotient is an
algebraic linear group, it is also a projective variety called \textit{flag
variety\footnote{%
Flag variety is directly connected with \textit{Schubert variety} [34 ],page
124. Schubert varieties were considered in our work, Ref. [32], in
connection with exact combinatorial solution of the Kontsevich-Witten model.
This observation naturally leads to combinatorial treatment of the whole
circle of problems in this work }}, Ref.[28], page 176. It should be clear
by now that the group $G$ is made out of at least two subgroups: $B$ just
described, and $N$. The \textit{maximal torus} $T$ subgroup can be defined
now as $T=B\cap N$ . This allows to define the Weyl group $W=N/T$. Although
this group has the same name as that discussed in the Appendix A, its true
meaning in the present case requires some explanations to be provided below.
This is done in several steps.

First, following Appendix A, we notice that the ''true'' Weyl group is made
of reflections, i.e. involutions of order 2. Following Tits [7], we
introduce a quadruple $(G,B,N,S)$ \ (the Tits system) where $S$ is subgroup
of $W$ \ made of elements such that $S=S^{-1}$ and 1 $\notin S\footnote{%
Such soubroup always exist for compact Lie groups considered as symmetric
spaces .}.$ Then, it can be shown that $G=BWB$ (\textit{Bruhat decomposition}%
) \ and, moreover, that the Tits system is isomorphic to the Coxeter system,
i.e. to the Coxeter reflection group of Appendix A. The full proof can be
found in Bourbaki [7], Chr.6, paragraph 2.4.

Second, since $W=N/T$ it is of interest to see the connection (if any)
between $W$ and the quotient $G/B=BWB/B$ $=\left[ B\left( N/T\right) B\right]
/B.$ Suppose that $N$ commutes with $B$, then evidently we would have $%
G/B\simeq \left( N/T\right) B$ and since $B$ fixes the flag $f$ we are left
with action of $N$ on the flag. Looking at Eq.(2.18) and noticing that the
diagonal matrix $T$ (the centralizer) can be chosen as a reference
(identity) transformation, so that the commuting matrix $N$ (the normalizer)
should permute $t^{a_{i}}$ 's . To prove that this is indeed the case
requires few additional steps. To begin, using Eq.s(2.7)-(2.9),(2.11),(2.14)
and (2.15) consider the map $\Psi $ for monomial $\mathbf{u}^{\mathbf{l}%
}=u_{1}^{l_{1}}\cdot \cdot \cdot u_{n}^{l_{n}}\equiv z_{1}^{l_{1}a_{1}}\cdot
\cdot \cdot z_{n}^{l_{n1}a_{n}}$ . For such map the character $c(t)$ is
given by 
\begin{equation}
c(t)=t^{<\mathbf{l}\cdot \mathbf{a>}}  \tag{2.20}
\end{equation}
where $<\mathbf{l}\cdot \mathbf{a}>=\sum\nolimits_{i}l_{i}a_{i}$ and both $%
l_{i}$ and $a_{i}$ are being integers. Following Ref.[35], consider limit $%
t\rightarrow 0$ in the above expression. Clearly, we obtain: 
\begin{equation}
c(t)=\left\{ 
\begin{array}{c}
1\text{ if }<\mathbf{l}\cdot \mathbf{a}>=0 \\ 
0\text{ if }<\mathbf{l}\cdot \mathbf{a}>\neq 0
\end{array}
\right. .  \tag{2.21}
\end{equation}
Evidently, the equation $<\mathbf{l}\cdot \mathbf{a}>=0$ describes a
hyperplane or, better, a set of hyperplanes for given vector $\mathbf{a}$
(Appendix A). Based on previous discussion such set forms at least one
chamber. To be more accurate, following the same reference, we would like to
complicate matters by introducing the subset $I$ $\subset $ $\{1,...,n\}$
such that say only those $l_{i}^{\prime }s$ which belong to this subset
satisfy $\ <\mathbf{l}\cdot \mathbf{a}>=0$ then, naturally, one obtains one-
to- one correspondence between such subsets and earlier defined flags.
Clearly, the set of thus constructed monomials forms invariant of torus
group action. In Section 1 we had discussed invariance with respect to the
Coxeter-Weyl reflection group, e.g. see Eq.(1.29a). It is of interest to
discuss now if such set of monomials is also invariant with respect to
action of the reflection group. More on this will be discussed in Section%
\textbf{\ }5\textbf{.}This can be achieved if we demonstrate that the Weyl
group $W=N/T$ permutes $a_{i}$'s thus transitively ''visiting'' different
hyperplanes. This will be demonstrated momentarily. Before doing this, we
would like to change the rules of the game slightly. To this purpose, we
shall replace the limiting $t\rightarrow 0$ procedure by requiring $t=\xi $ $%
($e.g. see Eq.(2.17) and discussion following this equation) where $\xi $ is
nontrivial $n$-th root of unity. After such replacement we formally entering
the domain of pseudo-reflection groups (Appendix A ). The results which
follow can be obtained with help of both real and pseudo reflection groups
as it will become clear upon reading. Replacing $t$ by $\xi $ causes us to
change the rule, Eq.(2.21), as follows 
\begin{equation}
c(\xi )=\left\{ 
\begin{array}{c}
1\text{ if }<\mathbf{l}\cdot \mathbf{a}>=0\text{ }\func{mod}\text{\textit{n}}
\\ 
0\text{ if }<\mathbf{l}\cdot \mathbf{a}>\neq 0
\end{array}
\right.  \tag{2.22}
\end{equation}
For reasons which will become clear in Section 4.4.2. we shall call equation 
$<\mathbf{l}\cdot \mathbf{a}>=0$ (or, more relaxed, $<\mathbf{l}\cdot 
\mathbf{a}>=n)$ the \textit{Veneziano condition} while $\mathit{<}\mathbf{l}%
\mathit{\cdot }\mathbf{a}\mathit{>=0}$ $\func{mod}\mathit{n}$ the \textit{Kac%
} \textit{condition\footnote{%
Actually, this condition should be called Kac--Moody-Bloch-Bragg (K-M-B-B).
The name Bloch comes from the Bloch equation/condition for the
Schr\"{o}dinger wave function of a single electron in perfect crystals [36].
It is essentially of the same nature as Eq.s (2.15),(2.16). The name Bragg
comes from the Bragg condition in X-ray crystallography. The analogy with
results from condensed matter physics should not be totally unexpected in
view of results of Appendix A (part c)) and Eq.(1.42).We shall encounter it
later in Sections 4.4.2. and 4.4.3. }}. The results of the Appendix A (part
c)) indicate that the first option is characteristic for the standard
Weyl-Coxeter reflection groups while the second is characteristic for the
affine Weyl-Coxeter groups thus leading to Kac-Moody affine algebras.

At this moment we are ready to demonstrate that $W=N/T$ is indeed the Weyl
reflection group.\footnote{%
Since $G/B\simeq N/T$ and since $G/B$ is the projective flag variety
(footnote 8). The same should be true for $N/T$ .This is indeed the case as
was demonstrated in Ref. [37] and, using different metods, by Danilov [2]
and Ewald [24]. Homology and cohomology calculations for such varieties are
rather sophisticated [2,24]. Fortunately, the major results (needed for
physical applications) can be obtained much simpler. This is explained below
and in Section 5.} Although we had mentioned earlier that such proof can be
found in Bourbaki [7], still, it is instructive to provide a qualitative
arguments exhibiting the essence of the problem. These arguments also will
be of some help for the next section. Let us begin with assembly of $\left(
d+1\right) \times \left( d+1\right) $ matrices with complex coefficients.
They belong to the group $GL_{d+1}(\mathbf{C}).$ Consider a subset of all
diagonal matrices and, having in mind the results to be presented in
Sections.4 and 5, let us assume that the diagonal entries are made of $n-th$
roots of unity \ $\xi .$ Taking into account the results of Appendix A (part
d)) on pseudo-reflection groups, each diagonal entry can be represented by $%
\xi ^{k}$ with $1\leq k\leq n-1$ so that there are $\left( n-1\right) ^{d+1}$
different diagonal matrices- all commuting with each other. Among these
commuting matrices we would like to single out those which have all \ $\xi
^{k}$ $^{\prime }s$ the same. Evidently, there are $n-1$ of them. They are
effectively the unit matrices and they are forming the centralizer of $W$.
The rest belongs to normalizer.\footnote{%
As with Eq.(2.20) one can complicate matters by considering matrices which
have several diagonal entries which are the same. Then, as before, one
should consider the flag system where in each subsystem the entries are all
different. The arguments \ applied to such subsystems will proceed the same
way as in the main text.} The number $\left( n-1\right) ^{d+1}$ $/(n-1)=$ $%
\left( n-1\right) ^{d}$ was obtained earlier, e.g. see Eq.(1.5) (with $n$
being replaced by $2n$). This is not just a mere coincidence. In Section 5
we shall provide some refinements of this result motivated by relevant
physics. It should be clear already that we are discussing only the simplest
possibility for the sake of illustration of general principles.

Next, let us consider just one of the diagonal matrices \ $\tilde{T}$ whose
entries are all different and made of powers of $\xi $. Let $g\in GL_{d+1}(%
\mathbf{C})$ and consider the automorphism: $\mathcal{F}(\tilde{T}):=g\tilde{%
T}g^{-1}$. Along with it we would like to consider an orbit $O(\tilde{T}):=g%
\tilde{T}C$ where $C$ is any of diagonal matrices which belong to earlier
discussed centralizer.\footnote{%
The presence of $C$ factor undescores the fact that we are considering the
orbit of the factorgroup $W=N/T$.} Clearly, $O(\tilde{T})=g\tilde{T}g^{-1}gC=%
\mathcal{F}(\tilde{T})gC=\mathcal{F}(\tilde{T})C.$ Denote now $\tilde{T}$ =$%
\tilde{T}$ $_{1}$ and consider another matrix $\tilde{T}$ $_{2}$ which
belongs to the same set and suppose that there is such matrix $g_{12}$ that $%
\tilde{T}$ $_{2}C=\mathcal{F}(\tilde{T})C.$ If such matrix exist, it should
belong to the normalizer and, naturally, the same arguments can be used to $%
\tilde{T}_{3}$, etc. Hence, the following conclusions can be drawn. If we
had started with some element \ $\tilde{T}_{1}$ of maximal torus, the orbit
of this element will return back and intersect the maximal torus in \textit{%
finite} number of points ( in our case the number of points is exactly $%
\left( n-1\right) ^{d}).$ By analogy with the theory of dynamical systems we
can consider these intersection points of the orbit $O(\tilde{T})$ with the
\ $T$-plane as Poincare$^{\prime }$ crossections. Hence, as it is done in
dynamical systems, we have to study the transition map between these
crossections. The orbit associated with such map is precisely the orbit of
the Weyl group $W$. It acts on these points transitively [33],page 147.
Provided that such set of fixed point exists, such arguments justify
dynamical interpretation of Weyl's character formula presented in Section 1.
The fact that such fixed point set exist is guaranteed by the Theorem 10.6
by Borel [27]. Its proof relies heavily on the Lie-Kolchin theorem\ (our
Theorem 1).

Mathematical results presented thus far can be made physically relevant if
we close the circle of ideas by considering transition back to zonotopes.
This is discussed in the next section.

\section{From toric varieties back to zonotopes}

\subsection{Coadjoint orbits}

\bigskip

So far we were working with Lie groups. To move forward we need to use the
Lie algebras associated with these groups. In what follows, we expect
familiarity with basic relevant facts about Lie groups which can be found in
the books by Serre [38], Humphreys [11] and Kac [12]. First, we notice that
the Lie group matrices $h_{i}$ associated with the Lie group maximal tori $%
T_{i}$ (that is with all diagonal matrices considered earlier) are commuting
with each other thus forming the Cartan subalgebra, i.e. 
\begin{equation}
\lbrack h_{i},h_{j}]=0.  \tag{3.1}
\end{equation}
The matrices belonging to the normalizer are made of two types $x_{i}$ and $%
y_{i}$ corresponding to the root systems $\Delta ^{+}$ and $\Delta ^{-}$ .
The fixed point analysis described at the end of previous section is
translated into the following set of commutators 
\begin{eqnarray}
\left[ x_{i},y_{j}\right] &=&\left\{ 
\begin{array}{c}
h_{i}\text{ if }i=j \\ 
0\text{ if }i\neq j
\end{array}
\right.  \TCItag{3.2a} \\
\left[ h_{i},x_{j}\right] &=&<\alpha _{i}^{\vee },\alpha _{j}>x_{j} 
\TCItag{3.2b} \\
\left[ h_{i},y_{j}\right] &=&-<\alpha _{i}^{\vee },\alpha _{j}>y_{j} 
\TCItag{3.2c}
\end{eqnarray}
$i=1,...,n$. To insure that the matrices(operators) $x_{i}^{\prime }s$ and $%
y_{i}^{\prime }s$ are nilpotent (that is their Lie group ancestors belong to
the Borel subgroup) one must impose two additional constraints. According to
Serre [38] these are: 
\begin{eqnarray}
\left( \text{ad}x_{i}\right) ^{-<\alpha _{i}^{\vee },\alpha _{j}>+1}(x_{j})
&=&0,\text{ }i\neq j  \TCItag{3.2d} \\
\left( \text{ad}y_{i}\right) ^{-<\alpha _{i}^{\vee },\alpha _{j}>+1}(y_{j})
&=&0,i\neq j.  \TCItag{3.2e}
\end{eqnarray}
where ad$_{X}$ $Y=[X,Y].$\ \ \ From the book by Kac [12] one finds that 
\textit{exactly the same} relations characterize the Kac-Moody affine Lie
algebra. This fact is very much in accord with general results presented in
the Appendix A. It is important for our purposes to realize that for each $i$
Eq.s(3.2a-c) can be brought to form (upon rescaling) which coincides with
the Lie algebra $sl_{2}$(\textbf{C})\footnote{%
This fact is known as Jacobson-Morozov theorem [9]} and, if we replace 
\textbf{C} with any closed \ number field $\mathbf{F}$, then all semisimple
Lie algebras are made of copies of $sl_{2}$(\textbf{F}) [11], page 25.The
Lie algebra $sl_{2}$(\textbf{C}) is isomorphic to the algebra of operators
acting on differential forms living on Hodge-type complex manifolds [10].
This observation is essential for physical applications presented in Section
5.4. The connection with Hodge theory can be also established through the
method of coadjoint orbits which we would like to discuss now. Surely, there
are other reasons to discuss this method as it will become clear upon
reading the rest of this section.

We begin by considering the orbit in Lie group. \ It is given by the Ad
operator, i.e. $O(X)=$Ad$_{g}X=gXg^{-1}$ where $g\in G$ and $X\in $\textsf{g}
with $G$ being the Lie group and \textsf{g} its Lie algebra. For compact
groups globally and for noncompact locally every group element $g$ can be
represented through exponential, e.g. $g(t)$=$\exp (tX_{g})$ with $X_{g}\in $%
\textsf{g.} Accordingly, for the orbit we can write $O(X)\equiv X(t)=\exp
(tX_{g})X\exp (-tX_{g}).$ Since the Lie group is a manifold $\mathcal{M}$,
the Lie algebra forms the tangent bundle of vector fields at given point of
\ $\mathcal{M}$. In particular, the tangent vector to the orbit $X(t)$ is
determined, as usual, by $TO(X)=\frac{d}{dt}%
X(t)_{t=0}=[X_{g},X]=ad_{X_{g}}X. $ Now we have to take into account that,
actually, our orbit is made for vector $X$ \ which comes from the torus,
i.e. $T=\exp (tX).$ This means that when we consider the commutator $%
[X_{g},X]$ it will be zero for $X_{g_{i}}=h_{i}$ and nonzero otherwise.
Consider now the Killing form \ $\kappa (x,y)$\ for two elements $x$ and $y$
of the Lie algebra: 
\begin{equation}
\kappa (x,y)=tr(adx\text{ }ady).  \tag{3.3}
\end{equation}
From this definition it follows that 
\begin{equation}
\kappa ([x,y],z)=\kappa (x,[y,z]).  \tag{3.4}
\end{equation}
The roots of the Weyl group can be rewritten in terms of the Killing form
[11]. It effectively serves as scalar multiplication between vectors
belonging to the Lie algebra and, as such, allows to determine the notion of
orthogonality between these vectors. In particular if we choose $%
x\rightarrow X$ and $y,z$ $\in h_{i},$then, it is clear that the vector
tangential to the orbit $O(X)$ is going to be orthogonal to the subspace
spanned by the Cartan subalgebra. This result can be reinterpreted from the
point of view of symplectic geometry due to work of Kirillov [30].To this
purpose we would like to rewrite Eq.(3.4) in the equivalent form, i.e. 
\begin{equation}
\kappa (x,[y,z])=\kappa (x,\text{ad}_{y}z)=\kappa (\text{ad}_{x}^{\ast }y,z)
\tag{3.5}
\end{equation}
where in the case of compact Lie group ad$_{x}^{\ast }y$ actually coincides
with ad$_{x}y$ . The reason for introducing the asterisk * lies in the
following chain of arguments. Already in Eqs(A.1) and (2.9) we had
introduced vectors from the dual space. Such construction is possible as
soon as the scalar multiplication is defined. Hence, for the orbit Ad$_{g}X$
there must be a vector $\xi $ in the dual space \textsf{g}$^{\ast }$ such
that equation 
\begin{equation}
\kappa (\xi ,\text{Ad}_{g}X)=\kappa (\text{Ad}_{g}^{\ast }\xi ,X)  \tag{3.6}
\end{equation}
defines the \textit{coadjoint orbit} $O^{\ast }$($\xi $)=Ad$_{g}^{\ast }\xi
. $ Accordingly, for such an orbit there is also the tangent vector $%
TO^{\ast } $($\xi $)=$ad_{\mathsf{g}}^{\ast }\xi $ to the orbit and,
clearly, we have $\kappa (\xi ,ad_{X_{g}}X)=\kappa (ad_{\mathsf{g}}^{\ast
}\xi ,X).$ In the case if we are dealing with the flag space, the family of
coadjoint orbits also will represent the flag space structure. Next, let $%
x\in $\textsf{g}$^{\ast }$ and $\xi _{1},\xi _{2}\in TO^{\ast }$($x$), then
consider the properties of the (symplectic) form $\omega _{x}(\xi _{1},\xi
_{2})$ to be determined explicitly momentarily. For this one needs to
introduce notations, e.g. $ad_{\mathsf{g}}^{\ast }x=f(x,\mathsf{g})$ so that
for \textsf{g}$_{1}$ and \textsf{g}$_{2}\in $\textsf{g} one has $\xi
_{i}=f(x,\mathsf{g}_{i}),i=1,2.$ Then, one can claim that for compact Lie
group and the associated with it Lie algebra $\omega _{x}(\xi _{1},\xi
_{2})=\kappa (x,[\mathsf{g}_{1},\mathsf{g}_{2}]).$ Indeed, using Eq.(3.5)
one obtains: $\kappa (x,[\mathsf{g}_{1},\mathsf{g}_{2}])=\kappa (\xi _{1},%
\mathsf{g}_{2})=-\kappa (x,[\mathsf{g}_{2},\mathsf{g}_{1}])=-\kappa (\xi
_{2},\mathsf{g}_{1}).$ Thus constructed form defines the symplectic
structure on the coadjoint orbit $O^{\ast }$($x$) since it is closed, skew
-symmetric, nondegenerate and is effectively independent of the choice of 
\textsf{g}$_{1} $ and \textsf{g}$_{2}.$The proofs can be found in the
literature [39].Thus obtained symplectic manifold $\mathcal{M}_{x}$ is the
quotient \textsf{g}$/\mathsf{g}_{h}$ with \textsf{g}$_{h}$ being made of
vectors of Cartan subalgebra. Clearly, for such vectors, by construction, $%
\omega _{x}(\xi _{1},\xi _{2})=0$. From the point of view of symplectic
geometry, such points correspond to critical points for the velocity vector
field on the manifold $\mathcal{M}_{x}$ at which the velocity vanishes. They
are in one-to one correspondence with the fixed points of the orbit $O(X).$
This fact allows us to use the Poincare-Hopf index theorem (earlier used in
our works on dynamics of 2+1 gravity [40,41]) in order to obtain the Euler
characteristic $\chi $ \ for such manifold as sum of indices of vector
fields which can exist on $\mathcal{M}_{x}$. From the discussion presented
in Section 2.4. it should be clear that $\chi $ is proportional to $%
(n-1)^{d} $. It can be proven, Ref. [42], page 259, that it is equal exactly
to this number. From Section 1 it should be clear, however, that different
Coxeter-Weyl reflection groups may have different numbers of this type. And,
indeed, in Section 5 we shall discuss related \ important example of this
sort.

To complete the above discussion, following work by Atiyah [43], we notice
that every nonsigular algebraic variety in projective space is symplectic.
The symplectic (K\"{a}hler) structure is being inherited from that of
projective space. The complex K\"{a}hler structure for symplectic (Kirillov)
manifold is actually of Hodge type. This comes from the following arguments.
First, since we have used the Killing form to determine Kirillov's $\omega
_{x}$ symplectic form and since the same Killing form is effectively used in
Weyl reflection groups [39], e.g. see Eq.(A.1), the induced unitary\ one
dimensional representation of the torus subgroup of GL$_{n}(\mathbf{C})$ is
obtained \ according to Kirillov [30] by simply replacing $t$ by the root of
unity in Eq.(2.20). This is permissible if and only if the integral of
two-form $\int\nolimits_{\gamma }\omega _{x}$ taken over any two dimensional
cycle $\gamma $ on coadjoint orbit $O^{\ast }$($x$) is integer\footnote{%
An important example of such quantization is discussed in the next
subsection.}. But this is exactly the condition which makes the K\"{a}hler
complex structure that of the Hodge type [10].

\subsection{Construction of the moment map using methods of linear
programming}

In this subsection we are not going to employ the definition of moment
mapping used in symplectic geometry [5]\footnote{%
Although, essentailly, we are going to use the same thing.}. Instead, we
shall rely heavily on works by Atiyah [43,44] with only slightest
improvement coming from \ noticed connections with linear programming not
mentioned in his papers and in literature on symplectic geometry. In our
opinion, such connection is helpful for better physical understanding of
mathematical methods presented in this paper.

Using results and terminology of Section 2.1. and Appendix A (part c)) we
call a subset of $\mathbf{R}^{n}$ a polyhedron \ $\mathcal{P}$ if there
exist $m\times n$ matrix $\ A$ $($with $m<n)$ and a vector $b\in \mathbf{R}%
^{m}$ such that 
\begin{equation}
\mathcal{P}=\{\mathbf{x}\in \mathbf{R}^{n}\mid Ax\leq b\}.  \tag{3.7}
\end{equation}
Since each component of the inequalities $Ax\leq b$ determines the half
space while the equality $Ax=b$ -the underlying hyperplane, the polyhedron
is an intersection of finitely many halfspaces. The problem of linear
programming can be formulated as follows [45] $:$ for linear functional $%
\mathcal{\tilde{H}[}\mathbf{x}]$=$\mathbf{c}^{\text{T}}\cdot \mathbf{x}$
find $\max $ $\mathcal{\tilde{H}[}\mathbf{x}]$ on $\mathcal{P}$ provided
that the vector \textbf{c} is assigned. It should be noted that this problem
is just one of many related. It was selected only because of its immediate
relevance. Its relevance comes from the fact that the extremum of $\mathcal{%
\tilde{H}[}\mathbf{x}]$ is achieved at least at one of the vertices of $%
\mathcal{P}$ .The proof of this we omit since it can be found in any
standard textbook on linear programming, e.g. see [46] and references
therein. This result does not require the polyhedron to be centrally
symmetric\footnote{%
This fact is important for possible potential extension of the results of
this work which do require central symmetry.}. Only convexity of polyhedron
is of importance.

To connect the above optimization problem with the results of this paper we
constrain $\mathbf{x}$ variables to integers, i.e. to \textbf{Z}$^{n}.$ Such
restriction is known in literature as \textit{integer linear programming}.
In our case, it is equivalent to considering symplectic manifolds of
Hodge-type (e.g. read page 11 of Atiyah's paper, Ref.[43] ). As a warm up
exercise which, in part, we shall need in the next section as well,
following Fulton [26], consider a deformation retract of complex projective
space \textbf{CP}$^{n}$ which is the simplest possible toric variety [24].
Such retraction is achieved by using the map : 
\begin{equation*}
\text{ }\tau :\text{\ }\mathbf{CP}^{n}\rightarrow \mathbf{P}_{\geq }^{n}=%
\mathbf{R}_{\geq }^{n+1}\setminus \{0\}/\mathbf{R}^{+}
\end{equation*}
given explicitly by 
\begin{equation}
\tau :\text{ }(z_{0},...,z_{n})\mapsto \frac{1}{\sum\nolimits_{i}\left|
z_{i}\right| }(\left| z_{0}\right| ,...,\left| z_{n}\right|
)=(t_{0},...,t_{n})\text{ , }t_{i}\geq 0  \tag{3.8}
\end{equation}
so that, naturally, the mapping $\tau $ is onto the standard $n$-simplex : $%
t_{i}\geq 0$, $t_{0}+...+t_{n}=1.$\ To bring physics to this example,
consider the Hamiltonian for harmonic oscillator. In the appropriate system
of units we can write it as $\mathcal{H=}m(p^{2}+q^{2})$. More generally,
for the set of oscillators, i.e. for the ''truncated'' bosonic string, we
have:$\ $\ $\mathcal{H}[\mathbf{z}]=\sum\nolimits_{i}m_{i}\left|
z_{i}\right| ^{2},$ where, following Atiyah [43], we have introduced complex 
$z_{j}$ variables via $z_{j}=p_{j}+iq_{j}.$ \ Let now such Hamiltonian
system (string) possess finite fixed energy $\mathcal{E}$ . Then, we obtain: 
\begin{equation}
\mathcal{H}[\mathbf{z}]=\sum\nolimits_{i=0}^{n}m_{i}\left| z_{i}\right| ^{2}=%
\mathcal{E}\text{.}  \tag{3.9}
\end{equation}
It is not difficult to realize that the above equation actually represents
some surface in \textbf{CP}$^{n}$ since the points $z_{j}$ can be identified
with points $e^{i\theta }z_{j}$ in Eq.(3.9) while keeping the above
expression form- invariant. This observation is helpful in the rest of this
paper. As before, we can map such model living in \textbf{CP}$^{n}$ back
into simplex in an obvious way. We can do better this time, however, since 
\textbf{CP}$^{n}$ is the simplest case of toric variety [23]. If we let $%
z_{j}$ to ''live'' in such variety it will be affected by the torus action
as it was discussed in Section 2.4.This means that the masses in Eq.(3.9)
will change and accordingly the energy. Only if we constrain the torus
action to the roots of unity $\xi ^{j}$ will the energy be conserved. This
provides another justification for their use (e.g. see earlier discussion
next to Eq.(2.17)). Hence, \ from now on we are going to work with this case
only. If this is the case, then $\left| z_{i}\right| ^{2}$ is just some
positive number and \textit{the essence of the moment map lies exactly in
such} \textit{identification}. Hence, we obtain 
\begin{equation}
\mathcal{\tilde{H}}[\mathbf{x}]=\sum\nolimits_{i=0}^{n}m_{i}x_{i}, 
\tag{3.10}
\end{equation}
where we had removed the energy constraint for a moment thus making $%
\mathcal{\tilde{H}}[\mathbf{x}]$ to coincide with \ earlier defined linear
functional to be optimized. Now we have to find the \ convex polyhedron on
which such functional is going to be optimized. Thanks to works by Atiyah
[43,44] and Guillemin and Sternberg [47], this task is completed. Naturally,
the polyhedron emerges as intersection of images of the critical points
(i.e. those for which the two-form $\omega _{x}=0)$ of the moment map [48].
Then, the theorem of linear programing stated earlier guarantees that $%
\mathcal{\tilde{H}}[\mathbf{x}]$ achieves its maximum at \ least at some of
its vertices. Delzant [49] had demonstrated that this is the case without
use of linear programming language.

It is helpful to demonstrate the essence of above arguments on the simplest
but important example discussed originally by Frankel [50] and \ later
elaborated in [51] . Consider a two sphere $S^{2}$ of unit radius, i.e. $%
x^{2}+y^{2}+z^{2}=1$ and parametrize this sphere using coordinates $x=\sqrt{%
1-z^{2}}\cos \phi $, $y=\sqrt{1-z^{2}}\sin \phi $ , $z=z$.The Hamiltonian \
for the free particle ''living'' on such sphere is given by $\mathcal{H}[z]%
\mathcal{=}m\left( 1-\left| z\right| ^{2}\right) $ so that equations of
motion produce circles of latitude. These circles become (critical) points
of equilibria at the north and south pole of the sphere, i.e. for $z=\pm 1.$
Under such circumstances our polyhedron is the segment [-1,1] and its
vertices are located at $\pm 1$ (to be compared with Section1.1.). The
moment map $\mathcal{H}[x]\mathcal{=}m\left( 1-x\right) $ acquires its
maximum at $x=1$ and the value $x=1$ corresponds to two \ polyhedral
vertices located at 1 and -1 respectively. \ This doubling feature was
noticed and discussed in detail by Delzant [49] whose work contains all
needed proofs. These can be considered as elaborations on much earlier
results by Frankel [50]\footnote{%
E.g. read especially page 4 of Frankel's paper. His results, in turn, were
largely influenced by still much earlier seminal paper by Hopf and Samelson
[52].}. The circles on the sphere represent the torus action (e.g. read the
discussion following Eq.(3.9)) so that dimension of the circle is half of
that of the sphere. This happens to be a general trend : the dimension of
the Cartan subalgebra (more accurately, the normalizer of maximal torus) is
half of the dimension of the symplectic manifold $\mathcal{M}_{x}$ [48,50].
Incidentally, the integral of \ the symplectic two- form $\omega _{x}$ over $%
S^{2}$ is found to be 2 [51] so that the complex structure on the sphere is
that of Hodge type. Generalization of this example to the multiparticle case
can be found also in the same reference and will be used in Section 5.

The results discussed thus far although establish connection between the
singularities of symplectic manifolds and polyhedra do not allow to discuss
fine details which \ allow to distinguish \ between different polyhedra.
Fortunately, this has been to a large degree accomplished [39,53]. Such task
is equivalent to classification of all finite dimensional exactly integrable
systems. In the next sections we shall argue that such information is also
useful for classification of infinite dimensional exactly integrable systems
of conformal field theories. In this case the situation is analogous to that
encountered in solid state physics where free electrons with well defined
mass end energy \ in vacuum acquire effective mass and multitude of energies
when they live in periodic solids [36].

\section{\protect\bigskip New Veneziano-like amplitudes from old Fermat
(hyper) surfaces}

\subsection{\protect\bigskip Brief review of the Veneziano amplitudes}

\bigskip In 1968 Veneziano [54] postulated 4-particle scattering amplitude $%
A(s,t,u)$ given (up to a common constant factor) by 
\begin{equation}
A(s,t,u)=V(s,t)+V(s,u)+V(t,u),  \tag{4.1}
\end{equation}
where 
\begin{equation}
V(s,t)=\int\limits_{0}^{1}x^{-\alpha (s)-1}(1-x)^{-\alpha (t)-1}dx\equiv
B(-\alpha (s),-\alpha (t))  \tag{4.2}
\end{equation}
is Euler's beta function and $\alpha (x)$ is the Regge trajectory usually
written as $\alpha (x)=\alpha (0)+\alpha ^{\prime }x$ with $\alpha (0)$ and $%
\alpha ^{\prime }$ being the Regge slope and the intercept, respectively. In
the case of space-time metric with signature $\{-,+,+,+\}$ the Mandelstam
variables $s$, $t$ and $u$ entering the Regge trajectory are defined by [55] 
\begin{eqnarray}
s &=&-(p_{1}+p_{2})^{2},  \TCItag{4.3} \\
t &=&-(p_{2}+p_{3})^{2},  \notag \\
u &=&-(p_{3}+p_{1})^{2}.  \notag
\end{eqnarray}
The 4-momenta $p_{i}$\ are constrained by the energy-momentum conservation
law leading to relation between the Mandelstam variables: 
\begin{equation}
s+t+u=\sum\limits_{i=1}^{4}m_{i}^{2}.  \tag{4.4}
\end{equation}
Veneziano [54] noticed\footnote{%
To get our Eq.(4.5) from Eq.7 of Veneziano paper, it is sufficient to notice
that his $1-\alpha (s)$ corresponds to ours -$\alpha (s).$} that to fit
experimental data the Regge trajectories should obey the constraint 
\begin{equation}
\alpha (s)+\alpha (t)+\alpha (u)=-1  \tag{4.5}
\end{equation}
consistent with Eq.(4.4) in view of definition of $\alpha (s)\footnote{%
The Veneziano condition $\alpha (s)+\alpha (t)+\alpha (u)=-1$ can be
rewritten in the form consistent with that considered earlier in the text.
Indeed, let $m$, $n$, $l$ be some integers such that $\alpha (s)m+\alpha
(t)n+\alpha (u)l=0$, then by adding this equation to that written above we
obtain $\alpha (s)\tilde{m}+\alpha (t)\tilde{n}+\alpha (u)\tilde{l}=-1,$or,
even more generally, $\alpha (s)\tilde{m}+\alpha (t)\tilde{n}+\alpha (u)%
\tilde{l}+\tilde{k}\cdot 1=0.$Both types of the equations had been
extensively studied in the book by Stanley [56].}.$ He also noticed that the
amplitude $A(s,t,u)$ can be equivalently rewritten with help of this
constraint as follows 
\begin{equation}
A(s,t,u)=\Gamma (-\alpha (s))\Gamma (-\alpha (t))\Gamma (-\alpha (u))[\sin
\pi (-\alpha (s))+\sin \pi (-\alpha (t))+\sin \pi (-\alpha (u))].  \tag{4.6}
\end{equation}
The Veneziano amplitude looks strikingly similar to that suggested a bit
later by Virasoro [57]. The latter (up to a constant) is given by 
\begin{equation}
\bar{A}(s,t,u)=\frac{\Gamma (a)\Gamma (b)\Gamma (c)}{\Gamma (a+b)\Gamma
(b+c)\Gamma (c+a)}  \tag{4.7}
\end{equation}
with parameters $a=-\frac{1}{2}\alpha (s),$etc$.$ also subjected to the
constraint: 
\begin{equation}
\frac{1}{2}\left( \alpha (s)+\alpha (t)+\alpha (u)\right) =-1.  \tag{4.8}
\end{equation}
Use of the formulas 
\begin{equation}
\Gamma (x)\Gamma (1-x)=\frac{\pi }{\sin \pi x}  \tag{4.9a}
\end{equation}
and 
\begin{equation}
4\sin x\sin y\sin z=\sin (x+y-z)+\sin (y+z-x)+\sin (z+x-y)-\sin (x+y+z) 
\tag{4.9b}
\end{equation}
permits us to rewrite Eq.(4.7) in the alternative form (up to unimportant
constant): 
\begin{eqnarray}
\bar{A}(s,t,u) &=&[\Gamma (-\frac{1}{2}\alpha (s))\Gamma (-\frac{1}{2}\alpha
(t))\Gamma (-\frac{1}{2}\alpha (u))]^{2}\times  \notag \\
&&[\sin \pi (-\frac{1}{2}\alpha (s))+\sin \pi (-\frac{1}{2}\alpha (t))+\sin
\pi (-\frac{1}{2}\alpha (u))].  \TCItag{4.10}
\end{eqnarray}
Although these two amplitudes look deceptively similar, mathematically, they
are markedly different. Indeed, by using Eq.(4.6) conveniently rewritten as 
\begin{equation}
A(a,b,c)=\Gamma (a)\Gamma (b)\Gamma (c)[\sin \pi a+\sin \pi b+\sin \pi c] 
\tag{4.11}
\end{equation}
and exploiting the identity 
\begin{equation*}
\cos \frac{\pi z}{2}=\frac{\pi ^{z}}{2^{1-z}}\frac{1}{\Gamma (z)}\frac{\zeta
(1-z)}{\zeta (z)}
\end{equation*}
after some trigonometric calculations the following result is obtained: 
\begin{equation}
A(a,b,c)=\frac{\zeta (1-a)}{\zeta (a)}\frac{\zeta (1-b)}{\zeta (b)}\frac{%
\zeta (1-c)}{\zeta (c)},  \tag{4.12}
\end{equation}
provided that 
\begin{equation}
a+b+c=1.  \tag{4.13}
\end{equation}
For the Virasoro amplitude, apparently, no result like Eq.(4.12) can be
obtained. As the rest of this section demonstrates, the differences between
the Veneziano and the Virasoro amplitudes are much more profound. The
result, Eq.(4.12), is \ also remarkable in the sense that it allows us to
interpret the Veneziano amplitude from the point of view of number theory,
the theory of dynamical systems, etc. Ref. [2] contains additional details
of such \ interpretation. \ To our knowledge, no such interpretation is
possible for the Virasoro amplitudes. For this and other reasons described
below and in Appendix B we shall treat only Veneziano and Veneziano-like
amplitudes in this paper.

\subsection{From Veneziano amplitudes to hypergeometric functions and back}

The fact that the hypergeometric functions are the simplest solutions of the
Knizhnik-Zamolodchikov equations of CFT is well documented [58]. The
connection between these functions and the toric varieties had been also
developed in papers by Gelfand, Kapranov and Zelevinsky [59] (GKZ). Hence,
we see no need in duplication of their results in this work. Instead, we
would like to discuss other aspects of hypergeometric functions and their
connections with the Veneziano amplitudes emphasizing similarities and
differences between strings and CFT.

For reader's convenience, we begin by introducing some standard notations.
In particular, let 
\begin{equation*}
(a,n)=a(a+1)(a+2)\cdot \cdot \cdot (a+n-1)
\end{equation*}
and, more generally, $(a)=(a_{1},...,a_{p})$ and $(c)=(c_{1},...,c_{q}).$
Then with help of these notations the $p,q$-type hypergeometric function is
written as 
\begin{equation}
_{p}F_{q}[(a);(c);x]=\sum\limits_{n=0}^{\infty }\frac{(a_{1,}n)\cdot \cdot
\cdot (a_{p},n)}{(c_{1,}n)\cdot \cdot \cdot (c_{q},n)}\frac{x^{n}}{n!}. 
\tag{4.14}
\end{equation}
In particular,the hypergeometric function in the form known to Gauss is just 
$_{2}F_{1\text{ }}=$ $F[a,b;c;x]$. Practically all elementary functions and
almost all special functions can be obtained as special cases of the
hypergeometric function just defined [60].

We are interested \ in connections between the hypergeometric functions and
the Schwarz-Christoffel \ (S-C) mapping problem. The essence of this problem
lies in finding a function $\varphi (\zeta )=z$ which maps the upper half
plane $\func{Im}\zeta >0$ (or, which is equivalent, the unit circle) into
the exterior of the $n-$sided polygon located on the Riemann sphere
considered as one dimensional complex projective space $\mathbf{CP}^{1}$
(i.e. $z\in \mathbf{CP}^{1}).$ Traditionally, the pre images $%
a_{1},...,a_{n} $ of the\ polygon vertices \ located at points $%
b_{1},...,b_{n}$ in $\mathbf{CP}^{1}$ are placed onto $x$ axis of $\zeta $%
-plane so that $\varphi (a_{i})=b_{i}$ , $i=1-n$. Let the interior angles of
the polygon be $\pi \alpha _{1},...,\pi \alpha _{n\text{ }}$respectively.
Then the exterior angles $\mu _{i}$\ are defined trough relations $\pi
\alpha _{i}+\pi \mu _{i}=\pi $ , $i=1-n.$ The exterior angles are subject to
the constraint: $\sum\nolimits_{i=1}^{n}\mu _{i}=2.$ The above data \ allow
us to write for the S-C mapping function the following \ known expression: 
\begin{equation}
\varphi (\zeta )=C\int\limits_{0}^{\zeta }(t-a_{1})^{-\mu _{1}}\cdot \cdot
\cdot (t-a_{n})^{-\mu _{n}}+C^{\prime }.  \tag{4.15}
\end{equation}
If one of the points, say $a_{n},$ is located at infinity, it can be shown
that in the resulting formula for mapping the last term under integral can
be deleted.

Consider now the simplest but \ relevant example of mapping of the upper
half plane into triangle with angles $\alpha ,\beta $ and $\gamma $ subject
to Euclidean constraint: $\alpha +\beta +\gamma =1.$ Let, furthermore, $%
a_{1}=0,a_{2}=1$ and $a_{3}=\infty $ . Using Eq.(4.15) (with $C=1$) we
obtain for the length $c$ of the side of the triangle: 
\begin{equation}
c=\int\limits_{0}^{1}\left| \frac{d\varphi (\zeta )}{d\zeta }d\zeta \right|
=\int\limits_{0}^{1}z^{\alpha -1}(1-z)^{\beta -1}=B(\alpha ,\beta )=\frac{%
\Gamma (\alpha )\Gamma (\beta )}{\Gamma (1-\gamma )}.  \tag{4.16}
\end{equation}
Naturally, two other sides can be determined the same way. Much more
efficient \ is to use the \ familiar elementary trigonometry relation 
\begin{equation*}
\frac{c}{\sin \pi \gamma }=\frac{b}{\sin \pi \beta }=\frac{a}{\sin \pi
\alpha }.
\end{equation*}
Then, using Eq.(4.9a) we obtain for the sides the following results : $c=%
\frac{1}{\pi }\sin \pi \gamma \Gamma (\alpha )\Gamma (\beta )\Gamma (\gamma
);b=\frac{1}{\pi }\sin \pi \beta \Gamma (\alpha )\Gamma (\beta )\Gamma
(\gamma );a=\frac{1}{\pi }\sin \pi \alpha \Gamma (\alpha )\Gamma (\beta
)\Gamma (\gamma ).$ The perimeter length $\mathcal{L}=a+b+c$ \ of the
triangle is just the full Veneziano amplitude, Eq.(4.11). As is well known
[61], the conformal mapping with \ Euclidean constraint $\alpha +\beta
+\gamma =1$ can be performed only for $3$ sets of \textit{fixed} angles.
Another 4 sets of angles belong to the spherical case: $\alpha +\beta
+\gamma >1,$ while the countable infinity of angle sets exist for the
hyperbolic case: $\alpha +\beta +\gamma <1.$ Hence, the associated with such
mappings Fuchsian-type equations used in some string theory formulations
will not be helpful in deriving the Veneziano amplitudes. \ These equations
are useful however in the CFT as is well known [58].

Fortunately, there is alternative formalism. It allows us to treat both
string and CFT from the same mathematical standpoint. \ Although this
formalism is discussed in some detail in previous sections there is still
need to fill out some gaps before we can actually use it.

It is well documented that the bosonic string theory had emerged as an
attempt at multidimensional generalization of Euler's beta function [62].
Analogous development also took place in the theory of hypergeometric
functions where it was performed along two related lines. To illustrate the
key ideas, following Mostow and \ Deligne [63] \ consider the \ standard
hypergeometric function which, up to a constant\footnote{%
To indicate this we use symbol $\dot{=}$.}, is given by 
\begin{equation}
F[a,b;c;x]\dot{=}\int\limits_{1}^{\infty }u^{a-c}(u-1)^{c-b-1}(u-x)^{-a}du. 
\tag{4.17}
\end{equation}
The multidimensional (multivariable) analogue of the above function \
according to Picard (in notations of Mostow and Deligne) is given by 
\begin{equation}
F[x_{2},...,x_{n+1}]=\int\limits_{1}^{\infty }u^{-\mu _{0}}(u-1)^{-\mu
_{1}}\prod\limits_{i=2}^{n+1}(u-x_{i})^{-\mu _{i}}du  \tag{4.18}
\end{equation}
provided that $x_{0}=0,x_{1}=1$ and, as before, $\sum\nolimits_{i=0}^{n}\mu
_{i}=2.$ At the same time, using alternative representation of $F[a,b;c;x]$
given by 
\begin{equation}
F[a,b;c;x]\dot{=}\int\limits_{0}^{1}z^{b-1}(1-z)^{c-b-1}(1-zx)^{-a}dz 
\tag{4.19}
\end{equation}
one obtains as well the following multidimensional generalization: 
\begin{equation}
F[\alpha ,\beta ,\beta ^{\prime },\gamma ;x,y]\dot{=}\int \int\limits 
_{\substack{ u\geq 0,v\geq 0  \\ u+v\leq 1}}u^{\beta -1}v^{\beta ^{\prime
}-1}(1-u-v)^{\gamma -\beta -\beta ^{\prime }-1}(1-ux)^{-\alpha }(1-v\unit{y}%
)^{-\alpha ^{\prime }}dudv  \tag{4.20}
\end{equation}
This result was obtained by Horn already at the end of 19th century and \
was subsequently reanalyzed and extended by GKZ. Looking at the last
expression one can design \ by analogy the multidimensional extension of the
Euler's beta function. In view of Eq.(4.2), it is given by the following
integral attributed to Dirichlet: 
\begin{equation}
\mathcal{D}(x_{1},...,x_{k+1})=\int \int\limits_{\substack{ u_{1}\geq
0,...,u_{k}\geq 0  \\ u_{1}\text{ }+\cdot \cdot \cdot +u_{k}\leq 1}}\unit{u}%
_{1}^{x_{1}-1}\unit{u}_{2}^{x_{2}-1}...\unit{u}%
_{k}^{x_{k}-1}(1-u_{1}-...-u_{k})^{x_{k+1}-1}du_{1}...du_{k}.  \tag{4.21}
\end{equation}
This result is going to be reobtained and explicitly calculated below in
this section (e.g. see Eq.s($4.57$) and (5.36) below) thus leading to
multiparticle generalization of the Veneziano and Veneziano-like amplitudes.
Before doing so we need to reproduce some results from Bourbaki [7]. They
will be helpful in the next section as well.

\subsection{Selected exercises from Bourbaki(begining)}

To accomplish of our tasks we need \ to work out some problems listed at the
end of Chapter 5 paragraph 5 (problem set \# 3) of Bourbaki [7].
Fortunately, answers to these problems to a large extent (but not
completely!) can be extracted from the paper by Solomon [8]. In view of
their crucial mathematical and physical importance we reproduce major
relevant results in this subsection. Section 5 (Subsection 5.1.2.) contains
the rest of relevant results from this exercise.

Let $K$ be the field of characteristic zero (e.g.\textbf{C}) and $V$ be the
vector space of finite dimension $l$\textit{\ }over it. Let $G$ be a
subgroup of $GL_{l}(V)$ made of pseudo-reflections acting on $V$. Let $q$ be
the cardinality of $G$. Introduce \ now the symmetric $S(V)$ and exterior $%
E(V)$ algebra of $V$ and look for invariants of pseudo-reflection groups
made of $S(V)$ and $E(V)$. This task requires several steps. First,
multiplication of polynomials leads to the notion of graded ring $R$%
\footnote{%
Surely, once the definition of such ring is given, there is no need to use
polynomilals. But in the present case this analogy is useful.}. For example,
if the polynomial\ $P_{i}(x)$ of degree $i$ belongs to the polynomial ring $%
\mathbf{F}[x]$ then the product $P_{i}(x)P_{j}(x)\in P_{i+j}(x)\in \mathbf{F}%
[x]$. A graded ring $R$ is a ring with decomposition $R=\oplus _{j=\mathbf{Z}%
}R_{j}$ compatible with addition and multiplication. Next, for the vector
space $V$ if $x=x_{1}\otimes \cdot \cdot \cdot \otimes $ $x_{s}\in
V^{\otimes _{s}}$and $y=y_{1}\otimes \cdot \cdot \cdot \otimes $ $y_{t}\in
V^{\otimes _{t}}$, then the product $x\otimes y\in V^{\otimes _{s+t}}.$The
multitude of such type of tensor products forms noncommutative associative
algebra $T(V)$. Finally, the symmetric algebra $S(V)$ is defined by $%
S(V)=T(V)/I$ , where the ideal $I$ is made of $x\otimes y-y\otimes x$ (with
both $x$ and $y$ $\in V).$ In practical terms $S(V)$ is made of \ symmetric
polynomials \textbf{F}[$t_{1},...,t_{l}]$ with $t_{1},...,t_{l}$ being in
one-to-one correspondence with basis elements of $V$ (that is each of $%
t_{i}^{\prime }s$ is entering into $S(V)$ with power one). The exterior
algebra $E(V)$ can be defined analogously now. For this we need to map the
vector space $V$ into the Grassmann algebra of $V$ $.$This is done just like
in physics when one goes from bosonic functions (belonging to $S(V)$) to
fermionic functions (belonging to $E(V)$). In particular, if $x\in V$ then,
its image in the Grassmann algebra $\tilde{x}$ possess a familiar (to
physicists) property $:$ $\tilde{x}^{2}=0.$ The graded two- sided ideal $I$
can be defined now as 
\begin{equation}
I=\{\tilde{x}^{2}=0\mid \text{ }x\rightarrow \tilde{x};x\in V\text{ }\} 
\tag{4.23}
\end{equation}
so that $E(V)=T(V)/I$. \ To complicate matters a little bit consider a map $%
d:$ $x\rightarrow dx$ for $x\in V$ and $dx$ belonging to the Grassmann
algebra. If $t_{1},...,t_{l}$ is the basis of $V$ , then $dt_{i_{1}}\wedge
\cdot \cdot \cdot \wedge dt_{i_{k}}$ is the basis of $E_{k}(V)$ with $0\leq
k\leq l$ and , accordingly, the graded algebra $E(V)$ admits the following
decomposition: $E(V)$ $=\oplus _{k=0}^{l}E_{k}(V).$ Next, we need to
construct the invariants of pseudo-reflection group $G$ made out of $S(V)$
and $E(V)$ and, most importantly, out of tensor product $S(V)\otimes E(V)$.
Toward this goal we need to look if the action of the map $d:V\rightarrow
E(V)$ extends to a differential map 
\begin{equation}
d:S(V)\otimes E(V)\rightarrow S(V)\otimes E(V).  \tag{4.24}
\end{equation}
Clearly, $\forall x\in E(V)$ \ we have $d(x)=0.$Therefore, $\forall x,y\in
S(V)\otimes E(V)$ we can write $d(x,y)=d(x)y+xd(y)$. \ By combining these
two results together we obtain 
\begin{equation}
d:S_{i}(V)\otimes E_{j}(V)\rightarrow S_{i-1}(V)\otimes E_{j+1}(V) 
\tag{4.25}
\end{equation}
, i.e. the differentiation is compatible with grading. Now we are ready to
formulate the theorem by Solomon [8] which is of central importance for this
work. It is formulated in the form stated in Bourbaki [7].

\textbf{Theorem} \textbf{2} (Solomon [8] ) . \textit{Let }$P_{1},...,P_{k%
\text{ }}$\textit{\ be algebraically independent polynomial forms (made of
symmetric combinations of }$t_{1},...,t_{k}$\textit{\ ) generating the ring }%
$S(V)^{G}$\textit{\ of invariants of }$G$\textit{. Then, every invariant
differential p-form }$\omega ^{(p)}$\textit{\ may be written uniquely as a
sum} 
\begin{equation}
\omega ^{(p)}=\sum\limits_{i_{1}<...<i_{p}}c_{i_{1}...i_{p}}dP_{i_{1}}\cdot
\cdot \cdot dP_{i_{p}}\text{ \ ; }1\leq p\leq k  \tag{4.26}
\end{equation}
\textit{with }$c_{i_{1}...i_{p}}\in S(V)^{G}.$\textit{Moreover, actually,
the differential forms }$\Omega ^{(p)}=dP_{i_{1}}\wedge \cdot \cdot \cdot
\wedge dP_{i_{p}}$\textit{\ with }$1\leq p\leq k$\textit{\ generate the
entire algebra of }$G-$\textit{\ invariants of }$S(V)\otimes E(V).$

\textbf{Corollary\textit{\ }1.} Let \textit{\ }$t_{1},...,t_{k}$ be the
basis of $V$ . Furthermore, let $S(V)=$ \textbf{F}[$t_{1},...,t_{k}]$ be its
algebra of symmetric polynomials and $S(V)^{G}=$\textbf{F}[$P_{1},...,P_{l}]$
its finite algebra of $G-$invariants\footnote{%
The fact that the number of polynomial forms $P_{i}$ is equal to the rank 
\emph{l }of $G$ is not a trivial fact. The proof can be found in [64] p.128.
Incidentally, this proof implies immediately the result, Eq.(4.28), given
below.}. Then, since $dP_{i}=\sum\limits_{j}\frac{\partial P_{i}}{\partial
t_{j}}dt_{j}$ we have 
\begin{equation}
dP_{1}\wedge \cdot \cdot \cdot \wedge dP_{k}=J(dt_{1}\wedge \cdot \cdot
\cdot \wedge dt_{k})  \tag{4.27}
\end{equation}
where, up to a constant factor $c\in K$, the Jacobian $J$ \ is given by $%
J=c\Omega $ with 
\begin{equation}
\Omega =\prod\limits_{i=1}^{\nu }L_{i}^{c_{i}-1}  \tag{4.28}
\end{equation}
In this equation $L_{i}$ is $linear$ form defining $i-th$ reflecting
hyperplane $H_{i}$ (it is assumed that the set of $H_{1},...,H_{\nu }$
reflecting hyperplanes is associated with $G$) , i.e. $H_{i}=\{\alpha \in
V\mid L_{i}(\alpha )=0\}$ as defined in the Appendix A. The set of all
elements of $G$ fixing $H_{i}$ pointwise forms a cyclic subgroup of order $%
c_{i}$ generated by pseudoreflections\footnote{%
E.g. read Appedix A part c).}.

\textit{\ Remark} 4. \ The result given by Eq.(4.28) as well as the
proportionality : $J=c\Omega ,$ can be found in the paper by Stanley [65].
It can be also found in much earlier paper by Solomon [8] where it is
attributed to Steinberg and Shephard and Todd. Stanley's paper contains some
details missing in earlier papers however.

\bigskip

\textit{Remark} 5. Using Theorem 2 by Solomon, Ginzburg [9] proved the
following

\bigskip

\textbf{Theorem 3}.(Ginzburg [9], page 358) \textit{Let }$\omega _{x}(\xi
_{1},\xi _{2})$\textit{\ be a symplectic (Kirillov-Kostant) two- form
defined in Section 3.1, let }$\Omega ^{N}=\omega _{x}^{N}$\textit{\ be its }$%
N$\textit{-th exterior power -the volume form, with }$N$\textit{\ being the
number of positive roots of the associated Weyl-Coxeter reflection group,
then} 
\begin{equation*}
\ast \left( \Omega ^{N}\right) =const\cdot dP_{1}\wedge \cdot \cdot \cdot
\wedge dP_{k}
\end{equation*}
\textit{where the star }$\ast $\textit{\ denotes the standard }$Hodge-type$ 
\textit{star operator .}

\textbf{Corollary 2. }As it was stated in Section 3.1., every nonsingular
algebraic variety in projective space is symplectic. The symplectic
structure gives raise to the complex K\"{a}hler structure which, in turn, is
of Hodge-type for Kirillov -type symplectic manifolds.

\textit{Remark}\textbf{\ }6. In seminal work, Ref. [66], Atiyah and Bott had
argued that $\omega ^{(p)}$ \ can be used as basis of the equivariant
cohomology ring. This result will be used in Section 5. We refer our readers
to the monograph [67] by Guillemin and Sternberg where all concepts of
equivariant cohomology are \ pedagogically explained.

Obtained results are sufficient for reobtaining the Dirichlet integral,
Eq.(4.21), and, more generally, the \ multiparticle Veneziano-like
amplitudes. They also provide needed mathematical background for adequate
physical interpretation. The next subsection makes all these statements
explicit.

\subsection{Veneziano amplitudes from Fermat hypersurfaces}

\subsubsection{General considerations}

Let us begin with observation that all integrals of Section 4.2. can be
reobtained with help of theorem by Solomon (Theorem 2). Indeed, let $V$ be
the complex affine space of dimension $\mathit{l}$\emph{\ }and let $%
L_{i}(v),v\in V$ be the linear form defining $i-$th hyperplane H$_{i},$ i.e. 
\begin{equation}
\text{H}_{i}=\{v\in V\mid L_{i}(v)=0\}\text{ , }i=1,\text{..., }\mathit{l}%
\emph{.}  \tag{4.29}
\end{equation}
As in the theory of linear programming discussed in Section 3.2., the set of
linear equations given above defines a polyhedron $\mathcal{P}$ . In view of
Eq.s(4.26)-(4.29), consider an integral $I$ of the type 
\begin{equation}
I=\int\limits_{\mathcal{P}}c_{i_{1}...i_{p}}dP_{i_{1}}\wedge \cdot \cdot
\cdot \wedge dP_{i_{p}}\text{ \ ; }1\leq p\leq l.  \tag{4.30}
\end{equation}
This is a typical integral of general hypergeometric type. All integrals of
Section 4.2. are of this type [60]. They can be obtained as solutions of the
associated with them systems of differential equations of the Picard-Fuchs
(P-F) type [60] also used in the mirror symmetry calculations [68].

In this work, we shall discuss another, more direct, option which avoid use
of the P-F type equations. This enable us to find \ physical applications of
mathematical results not discussed to our knowledge in mathematical physics
literature.

We begin with the simplest example borrowed from the paper by Griffiths
[69]. His paper begins with calculation of the following integral for the
period $\pi $ 
\begin{equation}
\pi (\lambda )=\oint_{\Gamma }\frac{dz}{z(z-\lambda )}  \tag{4.31}
\end{equation}
along the closed contour $\Gamma $ in the complex $z$-plane. Since this
integral depends upon parameter $\lambda $ the period $\pi (\lambda )$ is
some function of $\lambda .$ It can be determined by straightforward
differentiation of $\pi (\lambda )$ with respect to $\lambda $ thus leading
to the desired differential equation 
\begin{equation}
\lambda \pi ^{\prime }(\lambda )+\pi (\lambda )=0  \tag{4.32}
\end{equation}
enabling us to calculate $\pi (\lambda ).$ This simple result can be vastly
generalized to cover the case of period integrals of the type 
\begin{equation}
\Pi (\lambda )=\oint\limits_{\Gamma }\frac{P\left( z_{1},...,z_{n}\right) }{%
Q\left( z_{1},...,z_{n}\right) }dz_{1}\wedge dz_{2}\cdot \cdot \cdot \wedge
dz_{n}.  \tag{4.33}
\end{equation}
The equation $Q\left( z_{1},...,z_{n}\right) =0$ determines algebraic
variety. It may conatain a parameter (or parameters) $\lambda $ so that the
polar locus of values of $z^{\prime }s$ satisfying equation $Q=0$ depends
upon this parameter(s). By analogy with Eq.(4.32) it is possible to obtain a
set of differential equations of P-F type. This was demonstrated originally
by Manin [70]. In this work we are not going to develop this line of
research however. Instead, following Griffiths [69], we want to analyze in
some detail the nature of the expression under the sign of integral in
Eq.(4.33).

If $\mathbf{x}=(x_{0,}...,x_{n})$ are homogenous coordinates of a point in 
\textit{projective space}\textbf{\ }and $z=(z_{1},...,z_{n})$ are the
associated coordinates of the point in the \textit{affine space} where $%
z_{i}=x_{i}/x_{0}$ , then, the rational $n-$form $\omega $ is given in the 
\textit{affine} space by 
\begin{equation}
\omega =\frac{P\left( z_{1},...,z_{n}\right) }{Q\left(
z_{1},...,z_{n}\right) }dz_{1}\wedge dz_{2}\cdot \cdot \cdot \wedge dz_{n} 
\tag{4.34}
\end{equation}
with rational function $P/Q$ being a quotient of two homogenous polynomials
of the \textit{same} degree. Upon substitution $z_{i}=x_{i}/x_{0}$ the form $%
dz_{1}\wedge dz_{2}\cdot \cdot \cdot \wedge dz_{n}$ changes to 
\begin{equation*}
dz_{1}\wedge dz_{2}\cdot \cdot \cdot \wedge dz_{n}=\left( x_{0}\right)
^{-(n+1)}\sum\limits_{i=0}^{n}(-1)^{i}x_{i}dx_{0}\wedge ...\wedge d\hat{x}%
_{i}\wedge ...\wedge dx_{n}
\end{equation*}
where the hat on the top of $x_{i}$ means that it is excluded from the
product. It is convenient now to define the form $\omega _{0}$ via 
\begin{equation*}
\omega _{0}:=\sum\limits_{i=0}^{n}(-1)^{i}x_{i}dx_{0}\wedge ...\wedge d\hat{x%
}_{i}\wedge ...\wedge dx_{n}
\end{equation*}
so that in terms of \textit{projective space} coordinates the form $\omega $
can be rewritten as 
\begin{equation}
\omega =\frac{p(\mathbf{x})}{q(\mathbf{x})}\omega _{0}  \tag{4.35}
\end{equation}
where $p(\mathbf{x})=P(\mathbf{x})$ and $q(\mathbf{x})=Q(\mathbf{x}%
)x_{0}^{n+1}$ or, in more symmetric form, $q(\mathbf{x})=Q(\mathbf{x}%
)x_{0}\cdot \cdot \cdot x_{n}.$ In this case the degree of denominator of
the rational function $p/q$ is that of numerator $+(n+1)$.This is the result
of Corollary 2.11 of Griffith's paper [69].Conversely, each homogenous
differential form $\omega $ in projective space can be written in affine
space upon substitution : $x_{0}=1$ and $x_{i}=z_{i},i\neq 0.$

\ We would like to take advantage of this fact now. To this purpose, as an
example, we would like to study the period integrals associated with
equation describing Fermat hypersurface in complex projective space 
\begin{equation}
\mathcal{F(}N):x_{0}^{N}+\cdot \cdot \cdot +x_{n}^{N}+x_{n+1}^{N}=0. 
\tag{4.36}
\end{equation}
We would like now to combine Griffith's Corollary 2.11. with Solomon's
Theorem 2. That is to say we would like to consider the set of independent
linear forms $x_{i}^{<c_{i}>}$ , $i=0-(n+1),$where $<c_{i}>$ denotes
representative of $c_{i}$ in \textbf{Z} such that $1\leq <c_{i}>\leq N-1%
\footnote{%
These limits for $<c_{i}>$ are in accord with Gross [71], page 198.
Subsequently, they will be changed to $1\leq <c_{i}>\leq N$, e.g. see
Eq.(4.60) and the discussion around it.}$ \ and $x_{i}$ =H$_{i}$ ,
Eq.(4.29). They belong to\ the set of \ hyperplanes in $\mathbf{C}^{n+1}$
whose complement is complex algebraic torus $T$ (Definition 9, Section 2.4).
We want to consider the form $\omega $ living at the intersection of $T$
with $\mathcal{F}$. To this purpose introduce $<c>$ as 
\begin{equation}
<c>=\frac{1}{N}\sum\limits_{i}<c_{i}>  \tag{4.37}
\end{equation}
where the numbers $c_{i}$ have been defined earlier, after Eq.(4.28). They
belong to the set $X(S^{1})$: 
\begin{equation}
X(S^{1})=\{\bar{c}\in (\mathbf{Z}/N\mathbf{Z})^{n+2}\equiv (\mathbf{Z}/N%
\mathbf{Z})\times \cdot \cdot \cdot \times (\mathbf{Z}/N\mathbf{Z})\mid \bar{%
c}=(c_{0},...,c_{n+1}),\tsum\limits_{i}c_{i}=0\func{mod}N\}  \tag{4.38}
\end{equation}
fixing hyperlanes H$_{i}$ pointwise. The condition $\tsum\limits_{i}c_{i}=0%
\func{mod}N$ is in accord with Eq.(2.22) as required. Its role and true
meaning is illustrated \ by using the Fermat hypersurface $\mathcal{F}$ as
an example. By combining Solomon's Theorem 2 with Griffiths Corollary 2.11.
the form $\omega ,$Eq$.(4.35),$ is given by 
\begin{equation}
\omega =\frac{x_{0}^{<c_{0}>-1}\cdot \cdot \cdot x_{n+1}^{<c_{n+1}>-1}}{%
\left( x_{0}^{N}+\cdot \cdot \cdot +x_{n}^{N}+x_{n+1}^{N}\right) ^{<c>}}%
\text{ }\omega _{o}.  \tag{4.39}
\end{equation}
By design, it satisfies all the requirements just described.

\subsubsection{The 4-particle Veneziano-like amplitude}

Using Eq.(4.39), let us consider the simplest \ but important case : $n=1.$
It is relevant for calculation of 4-particle Veneziano-like amplitude.
Converting $\omega $ into affine form \ according to Griffiths prescription
and taking into account Solomon's Theorem 2 we obtain the following result
for the period integral: 
\begin{equation}
I_{aff}=\oint\limits_{\Gamma }\frac{1}{\left( x_{1}^{N}+x_{2}^{N}\mp
1\right) }dx_{1}^{<c_{1}>}\wedge dx_{2}^{<c_{2}>}.  \tag{4.40}
\end{equation}
The $\pm $ sign in the denominator requires some explanation. Indeed, let us
for a moment restore the projective form of this integral. We obtain the
following integral: 
\begin{equation}
I_{proj}=\oint\limits_{\Gamma }\frac{%
z_{1}^{<c_{1}>}z_{2}^{<c_{2}>}z_{0}^{<c_{0}>}}{\left( z_{1}^{N}+z_{2}^{N}\pm
z_{0}^{N}\right) }(\frac{dz_{1}^{{}}}{z_{1}}\wedge \frac{dz_{2}^{{}}}{z_{2}}-%
\frac{dz_{0}^{{}}}{z_{0}}\wedge \frac{dz_{2}^{{}}}{z_{2}}+\frac{dz_{0}^{{}}}{%
z_{0}}\wedge \frac{dz_{1}^{{}}}{z_{1}}).  \tag{4.40a}
\end{equation}
By construction, it is manifestly symmetric with respect to permutation of
its arguments. In addition, the integrand is manifestly torus action
invariant in the sense of Eq.s (2.16),(2.22). If we use $\xi ^{j}=\exp (\pm i%
\frac{2\pi j}{N})$ with $1\leq j\leq N-1$ the numerator of the integrand
above acquires the factor $\exp \{i\frac{2\pi }{N}%
(<c_{1}>j+<c_{2}>k+<c_{0}>l)\}$. Eq.(2.22) \ when combined with Griffiths
Corollary 2.11 imposes the constraint 
\begin{equation}
<c_{1}>j+<c_{2}>k+<c_{0}>l=N.  \tag{4.41a}
\end{equation}
\ We shall call it the ''Veneziano condition'' while Eq.(4.41b) (below) we
shall call the ''Shapiro-Virasoro'' condition.\footnote{%
These names are given by analogy with the existing terminilogy for open
(Veneziano) and closed (Shapiro-Virasoro) bosonic strings. Clearly, in the
present context they emerge for reasons different from that used in
conventional formulations.} It is needed to make the entire integrand torus
action invariant. Transition from projective to affine space brakes the
permutational symmetry firstly because of selecting, say, $z_{0}($and
requiring it to be one) and, secondly, by possibly switching sign in front
of $z_{0}.$ The permutational symmetry can be restored in the style of
Veneziano, e.g. see Eq.(4.1). The problem of switching the sign in front of $%
z_{0}$ can be treated similarly but requires extra care. This is so because
instead of the factor $\xi ^{j}$ used above we could use $\varepsilon ^{j}$
as well. Here $\varepsilon =\exp (\pm i\dfrac{\pi }{N})$. Use of such factor
makes the integral $I_{proj}$ also torus action invariant. For this case the
condition, Eq.(4.41a), has to be changed to 
\begin{equation}
<c_{1}>j+<c_{2}>k+<c_{0}>l=2N  \tag{4.41b}
\end{equation}
in accord with Lemma 1 of Gross [71]. By such change we are in apparent
disagreement with Corollary 2.11. We write ''apparent'' because,
fortunately, there is a way to reconcile Corollary 2.11. by Griffiths with
Lemma 1 by Gross. It will be discussed below. Already assuming that this is
the case, we notice that there are at least two different classes of
transformations leaving $I_{proj}$ unchanged. When switching to the affine
form these two classes are not equivalent: the first leads to differential
forms of the first kind while the second-to the second kind. Both are living
on the Jacobian variety $J(N)$ associated with the Fermat surface $\mathcal{%
F(}N)$ : $z_{1}^{N}+z_{2}^{N}\pm 1=0.$ It happens, that physically more
relevant are the forms of the second kind. We would like to describe them
now.

We begin by noticing that in switching from projective to affine space the
following set of $3N$ points (at infinity) should be deleted from the Fermat
curve $z_{1}^{N}+z_{2}^{N}+z_{3}^{N}$ $=0$.These are: $(\varepsilon \xi
^{j},0,1),$ $(0,\varepsilon \xi ^{j},1),(\varepsilon ^{2}\xi
^{j},\varepsilon \xi ^{j},0)$ respectively. By assuming that this is the
case and paramerizing: $z_{1}=$ $\varepsilon t_{1}^{\frac{1}{N}}$ and $%
z_{2}=\varepsilon t_{2}^{\frac{1}{N}}$( by analogy with Eq.(3.8)), we obtain
the simplex equation $t_{1}+t_{2}=1$ as deformation retract for $\mathcal{F(}%
N).$ After this, Eq.(4.40) acquires the following form : 
\begin{equation}
I_{aff}=\xi ^{j<c_{1}>+k<c_{2}>}\frac{1}{N^{2}}\oint\limits_{\Gamma }\frac{%
\varepsilon ^{<c_{1}>}t_{1}^{\frac{<c_{1}>}{N}}\varepsilon ^{<c_{2}>}t_{2}^{%
\frac{<c_{2}>}{N}}}{\left( t_{1}+t_{2}-1\right) }\frac{dt_{1}}{t_{1}}\wedge 
\frac{dt_{2}}{t_{2}}.  \tag{4.42}
\end{equation}
The overall phase factor guarantees the linear independence of the above
period integrals in view of the well-known result: $1+\xi ^{r}+\xi
^{2r}+...+\xi ^{(N-1)r}=0$ . It will be omitted for brevity in the rest of
our discussion.

To calculate $I_{aff}$ we need to use generalization of method of residues
for multidimensional complex integrals as developed by Leray [72] and
discussed in physical context e.g. by Hwa and Teplitz [73] and others. From
this reference we find that taking the residue can be achieved either by
dividing the differential form in Eq.(4.42) by $ds=t_{1}dt_{1}+t_{2}dt_{2}$
or, what is equivalent, by writing instead of Eq.(4.42) the following
physically suggestive result 
\begin{equation}
I_{aff}=\frac{1}{N^{2}}\oint\limits_{\Gamma }\varepsilon ^{<c_{1}>}t_{1}^{%
\frac{<c_{1}>}{N}}\varepsilon ^{<c_{2}>}t_{2}^{\frac{<c_{2}>}{N}}\frac{dt_{1}%
}{t_{1}}\wedge \frac{dt_{2}}{t_{2}}\delta (t_{1}+t_{2}-1)  \tag{4.43}
\end{equation}
to be discussed further in Section 5\footnote{%
Keeping in mind that $\delta (x)=\delta (-x)$ gives additional support in
favour of $\varepsilon -$factors.}. For the time being, taking into account
that $t_{2}=1-t_{1}$,\ after calculating the Leray residue we obtain, 
\begin{equation}
I_{aff}=\frac{1}{N^{2}}\int\limits_{0}^{1}\unit{u}^{\frac{<c_{1}>}{N}-1}(1-%
\text{u})^{\frac{<c_{2}>}{N}-1}d\text{u}=\frac{1}{N^{2}}B(a,b)  \tag{4.44}
\end{equation}
where $B(a,b)$ is Euler's beta function (as in Eq.(4.2))\ with $a=\frac{%
<c_{1}>}{N}$ and $b=\frac{<c_{2}>}{N}.$ The phase factors had been \
temporarily suppressed for the sake of comparison with the results of
Rohrlich \ [71] (published as an Appendix to paper by Gross and also
discussed in the book by\ Lang [74]). To make such comparison, we need to
take into account the multivaluedness of the integrand above if it is
considered in the standard complex plane. Referring our readers to Ch-r 5 of
Lang's book [74] allows us to avoid rather long discussion about the
available choices of integration contours. Proceeding in complete analogy
with the case considered by Lang, we obtain the period $\Omega (a,b)$ of the
differential form $\omega _{a,b}$ of the \textit{second} kind living on $%
J(N):$ 
\begin{equation}
\frac{\Omega (a,b)}{N}=\frac{1}{N}\int\limits_{\kappa }\omega _{a,b}=\frac{1%
}{N^{2}}(1-\varepsilon ^{<c_{1}>})(1-\varepsilon ^{<c_{2}>})B(a,b). 
\tag{4.45}
\end{equation}
The Jacobian $J(N)$ is related to the Fermat curve $\mathcal{F(}N)$
considered as Riemann surface of genus $g=\frac{1}{2}(N-1)(N-2).$\ Obtained
result differs from that by Rohrlich only by phase factors : $\varepsilon
^{\prime }s$ instead of $\xi ^{\prime }s.$ The number of such periods is
determined by the inequalities of the type $1\leq <c_{i}>\leq N-1.$ In
addition to the differential forms of the second kind there are also the
differential forms of the \textit{third} kind living on $\mathcal{F(}N).$%
They can be easily obtained from that of the second kind by relaxing the
condition $1\leq <c_{i}>\leq N-1$ to $1\leq <c_{i}>\leq N$ , Lang [74], page
39. The differential forms of the second kind are associated with the de
Rham cohomology classes H$_{DR}^{1}($ $\mathcal{F(}N),\mathbf{C})$ (Gross
[71], Lemma 1)$.$ The differentials forms of the first kind, discussed in
the book by Lang [74], by design do not have any poles while the
differentials of the second kind by design do not have residues. Only
differentials of the third kind have poles of order $\leq $ 1 with
nonvanishing residues and hence, are physically interesting. We shall be
dealing mostly with differentials of the 2nd kind \ converting them
eventually into that of the third kind. The differentials of the third kind
are linearly independent from that of the first kind according to Lang [74].

Symmetrizing our result, Eq.(4.45 ), following Veneziano we obtain the
4-particle Veneziano-like amplitude 
\begin{equation}
A(s,t,u)=V(s,t)+V(s,u)+V(t,u)  \tag{4.46}
\end{equation}
where, for example, upon analytical continuation $V(s,t)$ is given by 
\begin{equation}
V(s,t)=(1-\exp (i\frac{\pi }{N}(-\alpha (s))(1-\exp (i\frac{\pi }{N}(-\alpha
(t))B(\frac{-\alpha (s)}{N},\frac{-\alpha (t)}{N}),  \tag{4.47}
\end{equation}
provided that we had identified $<c_{i}>$ with $\alpha (i),etc.$\ \
Naturally, in arriving at Eq.(4.47) we had extended the differential forms
from that of the second kind to that of the third. The analytical properties
of such designed Veneziano-like amplitudes are discussed in detail the
subsection on multiparticle amplitudes (below) and in Appendix B.

\subsubsection{Connection with CFT trough hypergeometric functions and the
Kac--Moody-Bloch-Bragg condition}

\ To complete our treatment of the 4-particle Veneziano amplitude several
items still need to be discussed. In particular, we would like to compare
the hypergeometric function, Eq.(4.19), with the beta function in the light
of just obtained results. Taking into account that [60] 
\begin{equation*}
(1-zx)^{-a}=\sum\limits_{n=0}^{\infty }\frac{(a,n)}{n!}\left( zx\right) ^{n}
\end{equation*}
Eq.(4.19) can be rewritten as follows: 
\begin{eqnarray}
F[a,b;c;x] &\dot{=}&\sum\limits_{n=0}^{\infty }\frac{(a,n)}{n!}%
x^{n}\int\limits_{0}^{1}z^{b+n-1}(1-z)^{c-b-1}dz  \notag \\
&&\sum\limits_{n=0}^{\infty }\frac{(a,n)}{n!}x^{n}B(b+n,c-b).  \TCItag{4.48}
\end{eqnarray}
This result is to be compared with Eq.(4.44).To this purpose it is
convenient to rewrite Eq.(4.44) in the following more general form (up to a
constant factor): 
\begin{equation*}
I(m,l)\dot{=}\int\limits_{0}^{1}\unit{u}^{\frac{<c_{1}>-N+mN}{N}}(1-\text{u}%
)^{\frac{<c_{2}>-N+lN}{N}}d\text{u}=B(a+m,b+l),
\end{equation*}
where $m,l=0,\pm 1,\pm 2,..$ \ It is clear, that the phase factors entering
into Eq.(4.45) will either remain unchanged or will change sign upon such
replacements. At the same time the Veneziano condition, Eq.(4.41), will
change into 
\begin{equation}
<c_{0}>+<c_{1}>+<c_{2}>=N+mN+lN+kN.  \tag{4.49}
\end{equation}
This result can be explained physically with help of some known facts from
solid state physics, e.g. read Ref.[36]. To this purpose, using Sections 2.2
and 2.4 let us consider the result of torus action on the form $\omega $,
Eq.(4.39). If we demand this action to be invariant in accord with both
Eq.(2.16) and Theorem 2 by Solomon, then we obtain, 
\begin{equation}
\sum\limits_{i}<c_{i}>m_{i}=0\func{mod}N  \tag{4.50}
\end{equation}
with integers $m_{i}$ having the same meaning as numbers $\left( \mathit{l}%
_{j}\right) _{i}$ in Eq.(2.9). Consider Eq.(4.50) for a special case of
4-particle Veneziano amplitude. Then, according to the footnote after
Eq.(4.5) the Veneziano condition can be rewritten as 
\begin{equation}
<c_{0}>m_{0}+<c_{1}>m_{1}+<c_{2}>m_{2}=0\func{mod}N  \tag{4.51}
\end{equation}
to be compared with Eq.(2.22). But, in view of the Griffiths Corollary 2.11
the condition $\func{mod}N$ \ (or \ $\func{mod}2N)$\ \ \ for the Veneziano
amplitudes should actually be replaced by $N$ (or $2N$). At the same time
for hypergeometric functions in view of Eq.s(4.49), (4.51), we should write
instead

\begin{equation}
<c_{0}>m_{0}+<c_{1}>m_{1}+<c_{2}>m_{2}=mN+lN+kN.  \tag{4.52}
\end{equation}
Such condition is known in solid state physics as Bragg equation [36] . This
equation plays the central role in determining crystal structure by X-ray
diffraction. Lattice periodicity reflected in this equation affects
kinematics of scattering processes for phonons and electrons in crystals.
Under such conditions the concepts of particle energy and momentum loose
their usual meaning and should be amended to account for lattice
periodicity. The same type of amendments should be made when comparing
elementary scattering processes in CFT against those in high energy physics.
In view of Eq.(2.22) and results of Appendix A we shall call Eq (4.52) the 
\textit{Kac-Moody-Bloch-Bragg (K--M-B-B) equation}.

\subsubsection{The multiparticle Veneziano amplitudes and their analytic
properties}

By analogy with the 4-particle case, the Fermat variety $\mathcal{F}_{aff}$ $%
(N)$ in the affine form in the multiparticle case is given by the following
equation 
\begin{equation}
\text{ }\mathcal{F}_{aff}(N):\text{\ \ }Y_{1}^{N}+\cdot \cdot \cdot
+Y_{n+1}^{N}=1,\text{ }Y_{i}=x_{i}/x_{0}\equiv z_{i}.  \tag{4.53}
\end{equation}
As before, use of parametrization $f:$ $z_{i}=t_{i}^{\frac{1}{N}}\exp (\pm 
\frac{\pi i}{N})$ such that $\sum\nolimits_{i}t_{i}=1$ allows us to reduce
the Fermat variety $\mathcal{F}_{aff}(N)$ to its deformation retract which
is $n+1$ simplex $\Delta .$ I.e. $f($ $\mathcal{F}_{aff}(N))=\Delta $, where 
$\Delta :\sum\nolimits_{i}t_{i}=1$. The period integrals of type given by
Eq.(4.33) with $\omega $ form given by Eq.(4.39) after taking the Leray-type
residue are reduced to the following standard form (up to a constant): 
\begin{equation}
I\dot{=}\int\limits_{\Delta }t_{1}^{\frac{<c_{1}>}{N}-1}\cdot \cdot \cdot
t_{n+1}^{\frac{<c_{n+1}>}{N}-1}dt_{1}\wedge \cdot \cdot \cdot \wedge dt_{n},
\tag{4.54}
\end{equation}
where, again, all phase factors had been suppressed temporarily. For $n=1$
this integral coincides with that given in Eq.(4.44) (up to a constant) as
required. As part of preparations for calculation of this integral for $n>1$
let us have another look at the case $n=1$ first where we have integrals of
the type 
\begin{equation*}
I=\int\limits_{0}^{1}dxx^{a-1}(1-x)^{b-1}=B(a,b)=\frac{\Gamma (a)\Gamma (b)}{%
\Gamma (a+b)}.
\end{equation*}
Alternatively, we can look at 
\begin{equation}
\Gamma (a+b)I=\int\limits_{0}^{\infty }\int\limits_{0}^{\infty
}dx_{1}dx_{2}x_{1}^{a-1}x_{2}^{b-1}\exp (-x_{1}-x_{2}).  \tag{4.55}
\end{equation}
In the double integral on the r.h.s. let us consider change of variables: $%
x_{1}=\hat{x}_{1}t$ , $x_{2}=\hat{x}_{2}t$ so that $x_{1}+x_{2}=t$ provided
that $\hat{x}_{1}+\hat{x}_{2}=1.$ Taking $t$ and $\hat{x}_{1}$ as new
variables and taking into account that the Jacobian of such transformation
is one we obtain the following result, 
\begin{equation*}
\Gamma (a+b)I=\int\limits_{0}^{\infty }dtt^{a+b-1}\exp
(-t)\int\limits_{0}^{1}d\hat{x}_{1}\hat{x}_{1}^{a-1}(1-\hat{x}_{1})^{b-1},
\end{equation*}
as expected. Going back to our original integral, Eq.(4.54) and introducing
notations $a_{i}=\frac{<c_{i}>}{N}$ we obtain 
\begin{equation}
\Gamma (\sum\nolimits_{n=1}^{n+1}a_{i})I\dot{=}\int\limits_{0}^{\infty }%
\frac{dt}{t}t^{\sum\nolimits_{i=1}^{n+1}a_{i}}\exp (-t)\int\limits_{\Delta
}t_{1}^{a_{1}-1}\cdot \cdot \cdot t_{n+1}^{a_{n+1}-1}dt_{1}\wedge \cdot
\cdot \cdot \wedge dt_{n}.  \tag{4.56}
\end{equation}
By analogy with the case $n=1$ we introduce new variables :$s_{i}=tt_{i}$.
Naturally, we expect $\sum\nolimits_{n=1}^{n+1}s_{i}=t$ since $t_{i}$
variables are subject to the simplex constraint $\sum%
\nolimits_{i=1}^{n+1}t_{i}=1$.With such replacements we obtain 
\begin{eqnarray*}
\Gamma (\sum\nolimits_{n=1}^{n+1}a_{i})I &=&\int\limits_{0}^{\infty }\cdot
\cdot \cdot \int\limits_{0}^{\infty }\exp
(-\sum\nolimits_{n=1}^{n+1}s_{i})s_{1}^{a_{1}}\cdot \cdot \cdot
s_{n+1}^{a_{n+1}}\frac{ds_{1}}{s_{1}}\wedge \cdot \cdot \cdot \wedge \frac{%
ds_{n+1}}{s_{n+1}} \\
&=&\Gamma (a_{1})\cdot \cdot \cdot \Gamma (a_{n+1}).
\end{eqnarray*}
Using this result, finally, the $n-$particle contribution to the Veneziano
amplitude is given by the following expression: 
\begin{equation}
I\dot{=}\frac{\prod\limits_{i=1}^{n+1}\Gamma (a_{i})}{\Gamma
(\sum\nolimits_{n=1}^{n+1}a_{i})}.  \tag{4.57}
\end{equation}

\textit{Remark}\textbf{\ }7. \ Eq.(4.57) can be found in paper by Gross
[71], page 206, where it is suggested (postulated) without derivation.
Eq.(4.57) provides complete explicit calculation of the Dirichlet integral,
Eq.(4.21), and, as such, can be found, for example, in the book by Edwards
[75] published in 1922. Calculations similar to ours also can be found in
lecture notes by Deligne [76]. We shall use some additional results from his
notes below.

Our calculations are far from being complete however. To complete our
calculations we need to introduce the appropriate phase factors. In
addition, we need to discuss carefully the analytic continuation \ of just
obtained expression for amplitude to negative values of parameters $a_{i}$\
. Fortunately, the phase factors can be reinstalled in complete analogy with
the 4-particle case in view of the following straightforwardly verifiable
identity 
\begin{equation}
B(x,y)B(x+y,z)B(x+y+z,u)\cdot \cdot \cdot B(x+y+...+t,l)=\frac{\Gamma
(x)\Gamma (y)\cdot \cdot \cdot \Gamma (l)}{\Gamma (x+y+...+l)}  \tag{4.58}
\end{equation}
Because of this identity, the multiphase problem is reduced to that we had
considered already for the 4-particle case and, hence, can be considered as
solved. The analytic continuation problem connected with multiphase problem
is much more delicate and requires longer explanations.

The first difficulty we encounter is related to the constraints imposed on $%
<c_{i}>$ factors discussed in connection with the 4-particle case, e.g.
restriction :$1\leq <c_{i}>\leq N-1$ $($or $1\leq <c_{i}>\leq N).$ To
resolve this difficulty we shall follow Deligne's lecture notes [76].We
begin with Eq.(4.39). The Veneziano condition, Eq.(4.41a), extended to the
multivariable case is written as 
\begin{equation}
1=<c>=\frac{1}{N}\sum\limits_{i}<c_{i}>  \tag{4.59}
\end{equation}
whereas Griffiths Corollary 2.11. \textit{does not} require this constraint
to be imposed. To satisfy this corollary, it is sufficient only to require $%
m=<c>$ for some integer $m$ to be specified below. Clearly, such requirement
accordingly will change the total sum of exponents in \ the numerator of
Eq.(4.39). In particular, for $m=2$ we would reobtain back Eq.(4.41b). It
should be noted at this point that Lemma 1 by Gross [71] although imposes
such constraint but was proven \textit{not} in connection with the period
differential form, Eq.(4.39). This lemma implicitly assumes that the Leray
residue \textit{was taken already} and deals with the differential forms
occurring as result of such operation. To avoid guessing in the present case
we need to initiate our analysis again from Eq.(4.39) taking into account
Corollary 2.11.

Following Deligne [76], let us discuss what happens if we replace $<c>,$
Eq.(4.37), by $<-c>$. In view of definition of the bracket sign $<...>$ we
obtain, 
\begin{equation}
<-c>=\frac{1}{N}\sum\limits_{i}<-c_{i}>=\frac{1}{N}\sum%
\limits_{i}<-c_{i}+N>=n+2-<c>,  \tag{4.60}
\end{equation}
where the factor $n+2$ comes from the sum $\sum\nolimits_{i}1$ and $<c>$ is
the same as in Eq.(4.37) provided that $1\leq <c_{i}>\leq N.$ This result
implies that the number $m$ defined above can be only in the range 
\begin{equation}
\frac{n+2}{N}\leq m\leq n+2.  \tag{4.61}
\end{equation}

\textit{Remark} 8. Fermat variety $\mathcal{F}(N)$, Eq.(4.36), is of \textit{%
Calabi-Yau} type if and only if $n+2=N$ [77], page 531. Clearly, this
requirement is equivalent to the Veneziano condition, Eq.(4.59), i.e. $m=1$.

\textit{Remark} 9. By not imposing this condition we can still get many
interesting physically relevant results using Deligne's notes [76].We had
encountered this already while using Eq.(4.41b). Clearly, this equation is
reducible to Eq.(4.41a) anyway but earlier we obtained crucially physically
important phase factor $\varepsilon $ (instead of $\xi )$ by working with
Eq.(4.41b). It should be obvious by now that $m$ is responsible for change
in phase factors: from $\xi $ ( for $m=1$)-to $\varepsilon $ $($ for $m=2$%
)-to $\hat{\varepsilon}_{m}=\exp (i\frac{2\pi }{mN})$ ( for $m>2$).$\ $%
Physical significance of these phase factors is discussed in the Appendix B.

To extend these results we need to introduce several new notations now. Let $%
V_{\mathbf{C}}$ be finite dimensional vector space over \textbf{C}. A 
\textbf{C}-rational Hodge structure of weight $n$ on $V$ is a decomposition $%
V_{\mathbf{C}}=\bigoplus\limits_{p+q=n}V^{p,q}$ such that $\bar{V}%
^{p,q}=V^{q,p}$ \ We extend the definition of the torus action given by
Eq.s(2.14),(2.15) in order to accommodate the complex conjugation : $%
(t,V^{p,q})$=$t^{-p}\bar{t}^{-q}V^{p,q}.$ Next, we define the filtration
(the analog of the flag decomposition, Eq.(2.19)) via $F^{p}V=\bigoplus%
\limits_{p^{\prime }>p}V^{p^{\prime },q^{\prime }}$ so that $\cdot \cdot
\cdot \supset F^{p}V\supset F^{p+1}V\supset \cdot \cdot \cdot $ is a
decreasing filtration on $V$. The differential form, Eq.(4.39), belongs to
space $\Omega _{m}^{n+1}(\mathcal{F})$ of differential forms such that $%
\omega =\dfrac{p(\mathbf{z})}{q(\mathbf{z})^{m}}\omega _{0}$ , where $p(%
\mathbf{z})$ is homogenous polynomial of degree $m\deg (q)-(n+2).$ Such
differential forms have a pole of order $\leq m.$ As in the standard \
complex analysis one can define the multidimensional analogue of residue via
map $R(\omega )$: $\Omega _{m}^{n+1}(\mathcal{F})\rightarrow $H$^{n}(%
\mathcal{F},\mathbf{C})$ given by 
\begin{equation}
<\sigma ,R(\omega )>=\frac{1}{2\pi i}\int\nolimits_{\sigma }\omega \text{ , }%
\sigma \in \text{H}_{n}(\mathcal{F},\mathbf{C}).  \tag{4.62}
\end{equation}
Deligne proves that:

a) H$^{n}(\mathcal{F},\mathbf{C})=\bigoplus\limits_{\bar{c}\neq 0}$H$^{n}(%
\mathcal{F},\mathbf{C})_{\bar{c}}$ \ where

b)H$^{n}(\mathcal{F},\mathbf{C})_{\bar{c}}\subset F^{<c>-1}$H$^{n}(\mathcal{F%
},\mathbf{C})$ \ while the complex conjugate of H$^{n}(\mathcal{F},\mathbf{C}%
)_{\bar{c}}$

\ \ \ is given by H$^{n}(\mathcal{F},\mathbf{C})_{-\bar{c}}\subset
F^{n-<c>+1}$H$^{n}(\mathcal{F},\mathbf{C}).$

Thus, by construction, H$^{n}(\mathcal{F},\mathbf{C})_{\bar{c}}$ is of
bidegree \ $(p,q)$ with $p=<c>-1,q=n-p,$ while its complex conjugate H$^{n}(%
\mathcal{F},\mathbf{C})_{-\bar{c}}$ is of bidegree $(q,p).$ Obtained
cohomologies are nontrivial and of Hodge-type only when $<c>\neq 1.$Finally,
the procedure of extracting the residue from integral in Eq.(4.62) with $%
\omega $ containing pole of order $m$ is described in the book by Hwa and
Teplitz [73] and in spirit is essentially the same as in standard one
variable complex analysis. Therefore, we end up again with the differential
form $\omega $, Eq.(4.39), with $<c>=1.$However, this form will be used with
phase factors $\hat{\varepsilon}_{m}$ instead of $\xi .$Physical
consequences of this replacement are considered in the Appendix B.

The obtained results provide needed support for use of method of coadjoint
orbits, Section 3.1., and are in accord with the Theorem 3 by Ginzburg. They
provide justification for existence of physical models ( discussed in the
next section) associated with periods of $\mathcal{F}(N)$.

\section{Designing physical model for the Veneziano-like amplitudes}

\bigskip

\subsection{Some auxiliary results}

In this section we would like to demonstrate that results obtained in
previous sections contain \ sufficient information for restoration of the
underlying physical model producing the Veneziano-like amplitudes.

Let us begin with calculation of the volume of $k$-dimensional simplex $%
\Delta _{k}$. To do so, it is sufficient to consider calculation of the
integral of the type\footnote{%
In view of Eq.s(4.30) and (4.43) we actually need to consider integrals of
the type $\int dx_{1}^{<c_{1}>}\wedge \cdot \cdot \cdot \wedge
dx_{k+1}^{<c_{k+1>}}\delta (1-x_{1}^{N}-\cdot \cdot \cdot -x_{k+1}^{N})$ .
They indeed will be discussed later in this section. Because their
computation and interpretation is almost the same as that given by Eq.(5.1)
we prefer to discuss Eq.(5.1) first.} 
\begin{equation}
\text{vol}(\Delta _{k})=\int_{x_{i}\geq 0}dx_{1}\cdot \cdot \cdot
dx_{k+1}\delta (1-x_{1}-\cdot \cdot \cdot -x_{k+1}).  \tag{5.1}
\end{equation}
Using results from symplectic geometry [20,48] it is straightforward to show
that the above integral (up to unimportant constant) is just the
microcanonical partition function for the system of $k+1$ harmonic
oscillators with the total energy equal to $1$. To calculate such partition
function it is sufficient to take into account the integral representation
of the delta function. Then, the standard manipulations with integrals
produce the following anticipated result: 
\begin{equation}
\text{vol}(\Delta _{k})=\frac{1}{2\pi }\oint \frac{dy\exp (iy)}{(iy)^{k+1}}=%
\frac{1}{k!}.  \tag{5.2a}
\end{equation}
Clearly, for the dilated volume we would obtain instead vol$(n\Delta _{k})=%
\dfrac{n^{k}}{k!}$ where $n$ is the dilatation coefficient. From this
calculation we can obtain as well the volume of $k$-dimensional hypercube
(or, more generally the convex polytope)as 
\begin{equation}
n^{k}=k!vol(n\Delta _{k}).  \tag{5.2b}
\end{equation}
This result was obtained by Atiyah [43] \ who was inspired by earlier result
by Koushnirenko [78]. It has the following interpretation. If $n$ denotes
the number of segments, say in x direction, in \textbf{Z}$^{k}$ then $n+1$
is the number of lattice points associated with such segments. As in
Eq.(1.4), Eq.(5.2b) produces the total number of points inside the hypercube
and at its faces. This number comes as solution of the combinatorial problem
of finding the total number of points with integral coordinates inside of
the dilated simplex $n\Delta _{k}.$ In Appendix B we had introduced 
\begin{equation}
p(k,n)=\left| n\Delta _{k}\cap \mathbf{Z}^{k}\right| =\frac{(n+1)(n+2)\cdot
\cdot \cdot (n+k)}{k!}  \tag{5.3a}
\end{equation}
as the total number of points with integral coordinates inside of the
dilated simplex $n\Delta _{k}$. According to Stanley [79], this result
provides also the number of solutions in non negative integers of the
equation $x_{1}+...+x_{k}\leq n.$ When n$\rightarrow \infty $ Gelfand et al
[19] had proved (for \textit{any} convex integral polytope $\Delta _{k})$
that 
\begin{equation}
p(k,n)=\left| n\Delta _{k}\cap \mathbf{Z}^{k}\right| =\frac{\text{vol}\Delta
_{k}}{k!}n^{k}+O(n^{k-1}).  \tag{5.3.b}
\end{equation}
If we, as Gelfand et al, put the volume of the simplex $\Delta _{k}$ equal
to one then, $p(k,n)\simeq \dfrac{n^{k}}{k!},$ as before. This result is not
limited to hypercubes, of course. In Section 3.2. we had discussed general
case appropriate for arbitrary convex polytope. We shall return to this
topic later in this section.

The number $p(k,n)$ can be obtained with help of the generating function $%
P(k,t)$ given by 
\begin{equation}
P(k,t)\equiv \frac{1}{\left( 1-t\right) ^{k+1}}=\sum\limits_{n=0}^{\infty
}p(k,n)t^{n}.  \tag{5.4}
\end{equation}
This generating function can be interpreted as partition function of some
statistical mechanical system. Indeed, in the standard bosonic string theory
similar generating function 
\begin{equation}
P(t)=\prod\limits_{n=1}^{\infty }\frac{1}{1-t^{n}}=\sum\limits_{n=0}^{\infty
}p(n)t^{n}  \tag{5.5}
\end{equation}
is used as partition function for the bosonic string\footnote{%
The fact that $P(t)$ should be raised to 24th power [55] does not change
much in this regard.} .

According to Apostol [80] the number $p(n)$ can be understood as follows.
Every non negative number $n$ can be represented as 
\begin{equation*}
n=(1+...+1)+(2+...+2)+...+(m+...+m)+..=k_{1}+2k_{2}+...+mk_{m}+...
\end{equation*}
where $k_{1}$ gives number of ones in first parenthesis, $k_{2}$ gives
number of twos, and so on. Hence, $p(n)$ is a partition of $n$ into positive
summands whose number is $\leq n$. Clearly, $p(k,n)$ and $p(n)$ are related
to each other. The nature of their relation will be discussed below in
Section 5.1.2. In the meantime, let us suppose that we can approximate the
''exact'' partition function $P(t)$ by $P(k,t)$ at least for large $k$.
Then, using Eq.s (5.4) and results of the Appendix B we replace $k$ by $%
\alpha (s)$ and use this expression in beta function representing Veneziano
amplitude $V(s,t).$ Performing term- by- term integration we obtain
Eq.(B.14) which is familiar form of the Veneziano amplitude [55,62].
Incidentally, it is straightforward to demonstrate that such integration is
equivalent to taking the Laplace transform of $P(k,t)$ with respect to $t-$%
variable in Eq.(5.4). From general results of Regge theory it is known that
such procedure (when applied to the partition function) leads to poles
revealing the spectrum.

Based on the results just presented several conclusions can be drawn: a) the
partition function $P(k,t)$ \ can be adequately used in connection with the
Veneziano amplitude; b) such partition function has geometric and
group-theoretic meaning. In particular, $p(k,n)$ is exactly equal to the
Kostant multiplicity formula, Eq.(1.45), as it is proven in Guillemin et all
[47], Chr7., and Guillemin [81], Chapter 4.\footnote{%
Actually, these authors had proven the following . Let $N(\alpha )$ denote
the number of integer solutions of the linear equation $\alpha =n_{1}\alpha
_{1}+...+n_{k}\alpha _{k},$ for some prescribed $n_{i}\geq 0,$ $\alpha
_{i}\in \mathbf{Z}^{k}$ $(i=1,...,k),$ then $N(\alpha )=k!$vol$(\alpha
\Delta _{k})$, e.g. see Eq.(5.2b) where, $\Delta _{k}$ is convex polytope
associated with solution of the above linear equation (as explained in
Section 3.2). Such obtained function $N(\alpha )$ happen to coincide with
Kostant multiplicity formula, Eq.(1.45).}; c) although the Kostant
multiplicity formula was extended by Kac to the infinite dimensional case (
Ref.[12] and Appendix A) in the present case we are dealing with $finite$%
\textbf{\ }$dimensional$ version of the Kostant multiplicity formula because
it is adequate.

Just presented arguments will be considerably extended and reinforced below.

\subsubsection{Additional facts from the theory of pseudo-reflection groups}

In Appendix A, part c), we had listed some basic facts related to
pseudo-reflection groups. At this point we would like to extend this
information in connection with results obtained thus far. To this purpose we
would like to use some facts from the classical paper by Shepard and Todd
[82] (S-T). We shall use it along with the monograph by McMullen [83]
containing the up to date developments related to S-T work.

Adopting S-T notations, the unitary group $G(N,p,n)$ is defined as follows.
Let $N\geq 1,n\geq 2$, and let $p$ be a divisor of $N$, i.e. $N=pq$. Let $%
\xi $ be a primitive $N$-th root of unity. Then $G(N,p,n)$ is the group of
all monomial transformations in \textbf{C}$^{n}$ of the form (e.g. see
Section 2.4) 
\begin{equation}
x_{i}^{\prime }=\xi ^{\nu _{i}}x_{\sigma (i)},\text{ }i=1,...,n,  \tag{5.6}
\end{equation}
where $\sigma (1),...,\sigma (n)$ is permutation $\sigma $ of $(1,,...,n)$ ,
i.e. $\sigma \in S_{n},$and 
\begin{equation}
\sum\nolimits_{i}\nu _{i}=0(\func{mod}N).  \tag{5.7a}
\end{equation}
(the Veneziano condition). In the case if $N=pq$ the above condition should
be changed to 
\begin{equation}
\sum\nolimits_{i}\nu _{i}=0(\func{mod}p)  \tag{5.7b}
\end{equation}
(the closed string condition). The group $G(N,p,n)$ has order(or
cardinality) $\left| G\right| =$ $qN^{n}n!$. This follows easily from the
same arguments as were discussed in Section 2.4. In addition, the factor $n!$
is coming from permutational symmetry $S_{n}$ while the factor $q$ is
counting extra possibilities coming from the decomposition of $N$.
Alternatively, the order of the group $G(N,p,n)$ can be reobtained by
considering the set of 2-fold reflections given by 
\begin{equation}
x_{i}^{\prime }=\xi ^{\nu }x_{j},\text{ \ }x_{j}^{\prime }=\xi ^{-\nu }x_{i},%
\text{ \ }x_{k}^{\prime }=x_{k}\text{ },k\neq i,j.  \tag{5.8}
\end{equation}
Such set generates a normal subgroup of order $N^{n}n!$ . The other
reflections, if any, are of the form 
\begin{equation}
x_{j}^{\prime }=\xi ^{\tfrac{\nu N}{r}}x_{j},\text{ }x_{i}^{\prime }=x_{i}%
\text{ for }i\neq j,  \tag{5.9}
\end{equation}
where $j=1,...,n$, ($\nu ,N/r)=1$and $r\mid q$ if $q>1$. The following
theorem can be found in McMullen's book [83], page 292.

\textbf{Theorem} \textbf{4}. \textit{If }$n\geq 2$\textit{\ , then up to
conjugacy within the group of all unitary transformations, the only \textbf{%
finite} irreducible unitary reflection groups in }\textbf{C}$^{n}$\textit{\
which are imprimitive are the groups }$G(m,p,n)$\textit{\ with }$\mathit{m}%
\geq 2,p\mid m$\textit{\ and }$(m,p,n)\neq (2,2,2).$

\textbf{Definition 16.} A group G of unitary transformations of \textbf{C}$%
^{n}$ is called imprimitive if \textbf{C}$^{n}$ is the direct sum \textbf{C}$%
^{n}=E_{1}\oplus \cdot \cdot \cdot \oplus E_{k}$ of non-trivial proper
linear subspaces $E_{1},...,E_{k}$ such that the family$\{E_{1},...,E_{k}\}$
is invariant under $G$.

\textit{Remark }10. Clearly, the flag decomposition discussed in Section
2.4. fits this criteria.

\textit{Remark}\textbf{\ }11$.$ The above Theorem 4 provides justification
for developments presented in this paper thus far and to those which follow.
Clearly, it is just a restatement of the results presented earlier in
Section 2.4.

For our purposes, it is sufficient to consider only the case $p=1$.The group 
$G(N,1,n)$, traditionally denoted as $\gamma _{n}^{N}$ , is the group of
symmetries of the complex $n-$ cube. Actually, it is the same as that for
real $n$-cube [82,84] which is inflated by the factor of $N.$ Thus, we have
reobtained the result, Eq.(1.4), of Section 1.1. (given that the length $N$
contains $N+1$ points (including the origin)) The cubic symmetry is also
discussed in the Appendix A. \ In spite of these facts, in order to utilize
them efficiently we need to reobtain the same results from yet another
perspective. To this purpose, following S-T we introduce an auxiliary
function $g_{r}(m,p,n)$ describing the number of operations of $G(m,p,n)$
which leave fixed every point of subspace of dimensionality $%
n-r,r=0,1,...,n. $ By definition, $g_{0}(m,p,n)=1$ and, moreover, let 
\begin{equation}
G(m,p,n;t)\equiv \sum\limits_{r=0}^{n}g_{r}(m,p,n)t^{r}.  \tag{5.10}
\end{equation}
On one hand, the above equation serves as a definition of the generating
function $G(m,p,n;t)$, on another- in view of the results of Section 1, we
can interpret the r.h.s. as the Weyl character formula. Shepard and Todd
calculate $G(m,p,n;t)$ explicitly. Their derivation is less physically
adaptable however than that of Solomon [8]. Hence, we would like to discuss
Solomon's results now.

\subsubsection{Selected exercises from Burbaki (end)}

Section 4.3. contains some solutions to the problem set \#3 given at the end
of Chapter 5 (paragraph 5) of Bourbaki [7]. In view of their physical
significance,we would like to present the rest of the solutions now. The
problem set \# 3 deals with results obtained by Solomon [8]. In this
subsection we would like to reobtain his results in physically more
illuminating way. In doing so we shall freely use notations and results of
Section 4.3. In particular, we need to use the ring $S(V)^{G}$ of symmetric
invariants composed of algebraically independent polynomial forms $%
P_{1},...,P_{l}$ made of symmetric combinations of \ $t_{1},...t_{l}$ raised
to some powers $d_{i}$ , $i=1,...,l$ , different for different reflection
groups [84]. The ring of invariants is graded and it admits decomposition
(which is actually always finite) :$S(V)^{G}$ =$\bigoplus\nolimits_{j=0}^{%
\infty }S_{j}(V)^{G}$. Let $\dim _{K}V_{j}^{G}$ be the dimension of the
graded invariant subspace $S_{j}(V)^{G}$ defined over the field $K$. Then,
the Poincar$e^{\prime }$ polynomial $P(S(V)^{G},t)$ is defined by 
\begin{equation}
P(S(V)^{G},t)=\sum\limits_{i=0}^{\infty }(\dim _{K}V_{j}^{G})t^{i}. 
\tag{5.11}
\end{equation}
The Poincar$e^{\prime }$ polynomial possesses the splitting property \
[84](the most useful in K-theory) of the following nature. If the total
vector space $M$ is made as a product $V\otimes _{K}V^{\prime }$ of vector
spaces $V$ and $V^{\prime }$ then, the Poincar$e^{\prime }$ polynomial is
given by 
\begin{equation}
P(V\otimes _{K}V^{\prime },t)=P(V,t)P(V^{\prime },t).  \tag{5.12}
\end{equation}
This splitting property is of topological nature and is extremely useful in
actual calculations. In particular, consider the polynomial ring F$[x]$ made
of monomials of degree $d$ which can be identified with the graded vector
space $V$. Then, the Poincare polynomial for such space is given by 
\begin{equation}
P(V,t)=1+t^{d}+t^{2d}+....=\frac{1}{1-t^{d}}.  \tag{5.13}
\end{equation}
Consider now the multivariable polynomial ring F[$x_{1},...,x_{n}]$ made of
monomials of respective degrees $d_{i}$ . Then, using the splitting property
we obtain at once 
\begin{equation*}
V^{T}=\text{F[}x_{1}]\otimes _{K}\text{F[}x_{2}]\otimes _{K}\text{F[}%
x_{3}]\otimes _{K}\cdot \cdot \cdot \otimes _{K}\text{F[}x_{n}]
\end{equation*}
and, of course, 
\begin{equation}
P(V^{T},t)=\frac{1}{1-t^{d_{1}}}\cdot \cdot \cdot \frac{1}{1-t^{d_{n}}}. 
\tag{5.14}
\end{equation}
In particular, if all $d_{i}$ in Eq.(5.14) are equal to one, which is
adequate for $S(V),$e.g. read Ref.[84] , page 171, then we reobtain back
Eq.(5.4) (with $n$ in Eq.(5.14) being replaced by $k+1$). At the same time,
in the case of cubic symmetry $B_{n+1}($useful for the purposes of this
work), S-T find for the group $G(N,1,n)$ the exponents $d_{i}^{{}}=Ni$ with $%
i=1,...,n$ [82-84]. In the typical case of real space and cubic symmetry we
have $N=2$ (e.g. see Eq.(A.6) of Appendix A) so that these exponents $%
d_{i}^{{}}$coincide with those listed in the book by Humphreys[85], page 59,
as required. If we replace $t$ by $t^{2}=x$ (or, more generally, $t^{N}=x$)\
\ \ in the associated Poincar$e^{\prime }$ polynomial, then we obtain, 
\begin{equation}
P(V^{T},x)=\prod\limits_{i=1}^{k+1}\frac{1}{1-x^{i}}=\sum\limits_{n=0}^{%
\infty }\hat{p}(k,n)x^{n}.  \tag{5.15}
\end{equation}
In the limit $k\rightarrow \infty $ this result coincides with Eq.(5.5).
According to the Theorem 4 cited above the arbitrary large $n$ still leads
to \textbf{finite} unitary reflection group. It is tempting to choose the
Poincare' polynomial, Eq.(5.15), as partition function for ''new'' bosonic
string. This is not the case however. The new string partition function is
not given by Eq.(5.15). It does not correspond to the truncated bosonic
string model and, in fact, it is not even bosonic! This is so because of the
results obtained by Solomon [8] in 1963 to be discussed now.

To make our presentation self contained and focused on \ physics at the same
time we need to discuss few auxiliary facts from the theory of invariants of
the Weyl-Coxeter reflection groups. Let $G\subset GL(V)$ be one of such
groups. Denote $\left| G\right| =\prod\limits_{i}d_{i}$ and introduce the
averaging operator $Av:V\rightarrow V$ via 
\begin{equation}
Av(x)=\frac{1}{\left| G\right| }\sum\limits_{\varphi \in G}\varphi \circ x. 
\tag{5.16}
\end{equation}
By definition, $x$ is the group invariant, $x$ $\in V^{G}$ , if $Av(x)=x.$
In particular, 
\begin{equation}
\dim _{K}V_{j}^{G}=\frac{1}{\left| G\right| }\sum\limits_{\varphi \in
G}tr(\varphi _{j}).  \tag{5.17}
\end{equation}
By combining this result with Eq.(5.11) we obtain: 
\begin{eqnarray}
P(S(V)^{G},t) &=&\sum\limits_{i=0}^{\infty }(\dim
_{K}V_{j}^{G})t^{i}=\sum\limits_{i=0}^{\infty }\frac{1}{\left| G\right| }%
\sum\limits_{\varphi \in G}tr(\varphi _{i})t^{i}  \notag \\
&=&\frac{1}{\left| G\right| }\sum\limits_{\varphi \in
G}[\sum\limits_{i=0}^{\infty }tr(\varphi _{i})t^{i}]=\frac{1}{\left|
G\right| }\sum\limits_{\varphi \in G}\frac{1}{\det (1-\varphi t)}. 
\TCItag{5.18}
\end{eqnarray}
The obtained result is known as the Molien theorem [84]. It is based on the
following nontrivial identity 
\begin{equation}
\sum\limits_{i=0}^{\infty }tr(\varphi _{i})t^{i}=\frac{1}{\det (1-\varphi t)}
\tag{5.19}
\end{equation}
valid for the upper triangular matrices, i.e. for matrices which belong to
the Borel subgroup $B$ (Section 2). \ For such matrices 
\begin{equation}
tr(\varphi _{i})=\sum\limits_{j_{1}+j_{2}+...+j_{n}=i}\lambda
_{1}^{j_{1}}\cdot \cdot \cdot \lambda _{n}^{j_{n}}  \tag{5.20}
\end{equation}
where the Borel-type matrix $\varphi $ of dimension $n$ has $\lambda
_{1},...,\lambda _{n}$ on its diagonal. In view of Eq.s(2.14),(2.18), \ and
following Humphreys [85], it is useful to re interpret Eq.(5.20) as follows.
We consider action of the averaging operator, Eq.(5.16), on the monomials 
\begin{equation*}
x=z_{1}^{j_{1}}\cdot \cdot \cdot z_{n}^{j_{n}},\text{ where \ }%
j_{1}+j_{2}+...+j_{n}=i.
\end{equation*}
These are the eigenvectors for $\varphi _{i}$ with corresponding eigenvalues 
$\lambda _{1}^{j_{1}}\cdot \cdot \cdot \lambda _{n}^{j_{n}}.$The sum of
these eigenvalues is the trace of the linear operator $Av(x)$ on $%
S_{i}(V)^{G}$. But, according to Eq.(5.17), this is just the dimension of
space $S_{i}(V)^{G}.$ This dimension has the following meaning. If, as we
had argued after Eq.(2.22), the eigenvalues in Eq.(5.20) are made of $i-th$
roots of unity then, by combining Eq.s (2.16), (2.20),(2.22), Appendix A,
part c), and Eq.(5.16) we arrive at the Veneziano condition 
\begin{equation}
\sum\limits_{k}m_{k}j_{k}=i  \tag{5.21a}
\end{equation}
again. Since $m_{i}=d_{i}-1\func{mod}i,$ the above equation is equivalent to 
\begin{equation}
\sum\limits_{k}d_{k}j_{k}=0\func{mod}i  \tag{5.21b}
\end{equation}
with the exponents $d_{i}$ defined earlier. In particular, if $%
\sum\limits_{k}d_{k}j_{k}=i$ then, using this equation along with
Eq.s(5.19),(5.20), we arrive at the following physically suggestive result: 
\begin{eqnarray}
\sum\limits_{i=0}^{\infty }tr(\varphi _{i})t^{i}
&=&\sum\limits_{i=0}^{\infty
}[\sum\limits_{j_{1}d_{1}+j_{2}d_{2}+...+j_{n}d_{n}=i}\lambda
_{1}^{j_{1}d_{1}}\cdot \cdot \cdot \lambda _{n}^{j_{n}d_{n}}]t^{i}  \notag \\
&=&\sum\limits_{j_{1}=0}^{\infty
}t^{j_{1}d_{1}}\sum\limits_{j_{2}=0}^{\infty }t^{j_{2}d_{2}}\cdot \cdot
\cdot \sum\limits_{j_{n}=0}^{\infty }t^{j_{n}d_{n}}  \notag \\
&=&\prod\limits_{i=1}^{n}\frac{1}{1-t^{d_{i}}}  \TCItag{5.22a}
\end{eqnarray}
to be compared with earlier obtained Eq.(5.14). Alternatively, 
\begin{equation}
\sum\limits_{i=0}^{\infty }tr(\varphi
_{i})t^{i}=\sum\limits_{j_{1}=0}^{\infty }\lambda _{1}^{j_{1}}t^{j_{1}}\cdot
\cdot \cdot \sum\limits_{j_{n}=0}^{\infty }\lambda _{n}^{j_{n}}t^{j_{n}}=%
\frac{1}{\det (1-\varphi t)}.  \tag{5.23}
\end{equation}
We have gone through all details in order to demonstrate the bosonic nature
of the obtained result: by replacing $t$ with exp($-\varepsilon )$ with $%
0\leq \varepsilon \leq \infty $ and associating numbers $j_{i}$ with the \
Bose statistic occupation numbers we have obtained the partition function
for the set of $n$ independent harmonic oscillators (up to zero point
energy). By combining these results with Eq.(5.18) we obtain yet another
physically suggestive (to those who prefer to use the bosonic path
integrals) result: 
\begin{equation}
\frac{1}{\left| G\right| }\sum\limits_{\varphi \in G}\frac{1}{\det
(1-\varphi t)}=\prod\limits_{i=1}^{n}\frac{1}{1-t^{d_{i}}}.  \tag{5.24}
\end{equation}
The results just obtained are auxiliary however. Our main objective is to
obtain the explicit form of $G(m,p,n;t)$ defined in Eq.(5.10) and to explain
its physical meaning. To this purpose, let us recall that according to the
Theorem 2 by Solomon the differential form $\omega ^{(p)}$, Eq.(4.26), \
belongs to the set of $G$-invariants of the product $S(V)\otimes E(V)$. The
splitting property, Eq.(5.12), of the Poincar$e^{\prime }$ polynomials
requires some minor changes for the present case. In particular, if by
analogy with $S(V)^{G\text{ }}$decomposition we would write ($S(V)\otimes
E(V))^{G}=\bigoplus\limits_{i,j}S_{i}(V)^{G}\otimes E_{j}(V)^{G}$ , then,
the associated Poincar$e^{\prime }$ polynomial is given by 
\begin{equation}
P((S(V)\otimes E(V))^{G};x,y)=\sum\limits_{i,j\geq 0}(\dim
_{K}S_{i}^{G}\otimes E_{j}^{G})x^{i}y^{j}.  \tag{5.25}
\end{equation}
By analogy with Eq.(5.17), following Solomon [8], we introduce 
\begin{equation}
\dim _{K}S_{i}^{G}\otimes E_{j}^{G}=\frac{1}{\left| G\right| }%
\sum\limits_{\varphi \in G}tr(\varphi _{i})tr(\varphi _{j}).  \tag{5.26}
\end{equation}
In order to use this result we need to take into account Eq.(1.18). That is
we need to take into account that 
\begin{equation}
\sum\limits_{j=0}^{n}tr(\varphi _{j})y^{j}=\det (1+\varphi y)  \tag{5.27}
\end{equation}
to be contrasted with Eq.(5.19). To prove that this is the case it is
sufficient to recall that for fermions the occupation numbers $j_{i}$ are
just $0$ and $1.$ Hence, 
\begin{eqnarray}
\sum\limits_{j=0}^{n}tr(\varphi _{j})y^{j}
&=&\sum\limits_{j=0}^{n}[\sum\limits_{j_{1}+j_{2}+...+j_{n}=j}\lambda
_{1}^{j_{1}}\cdot \cdot \cdot \lambda _{n}^{j_{n}}]y^{j}  \notag \\
&=&\prod\limits_{i=1}^{n}\sum\limits_{j_{i}=0}^{1}\lambda
_{i}^{j_{i}}y^{j_{i}}=\prod\limits_{i=1}^{n}(1+\lambda _{i}y),  \TCItag{5.28}
\end{eqnarray}
to be compared with Eq.(5.22b).Using Eq.(5.26) in (5.25) and taking into
account the rest of the results, \ in accord with Bourbaki [7] the following
expression for the Poincar$e^{\prime }$ polynomial \ is obtained : 
\begin{equation}
P((S(V)\otimes E(V))^{G};x,y)=\frac{1}{\left| G\right| }\sum\limits_{\varphi
\in G}\frac{\det (1+\varphi y)}{\det (1-\varphi x)}=\prod\limits_{i=1}^{n}%
\frac{(1+yx^{d_{i}-1})}{(1-x^{d_{i}})}.  \tag{5.29a}
\end{equation}
To check its correctness we can: a) put $y=0$ thus obtaining back Eq.(5.23)
or b) put $y=-x$ thus obtaining identity $1=1$.

\textit{Remark }$12$. Comparison of Eq.(5.29a) with Ruelle zeta function,
Eq.(1.24), suggests the interpretation of this result from the point of view
of evolution of dynamical systems. Additional details concerning such
interpretation can be found in our earlier work, Ref. [2].

The result, Eq.(5.29a), can now be used for several tasks. First, for
completeness of presentation we would like to recover the major S-T result: 
\begin{equation}
G(m,p,n;t)=\prod\limits_{i=1}^{n}(m_{i}t+1)  \tag{5.30}
\end{equation}
used heavily in theory of hyperplane arrangements [60,86]. For $t=1$ above
equation produces: $G(m,p,n;t=1)=\left| G\right|
=\prod\limits_{i=1}^{n}d_{i} $ as required. Moreover, in view of \
Eq.(5.10), it allows us to recover $g_{r}(m,p,n)$. Second, after this is
done, we need to discuss its physical meaning.

To recover the S-T results let us rewrite Eq.(5.29) as follows 
\begin{equation}
\sum\limits_{\varphi \in G}\frac{\det (1+\varphi y)}{\det (1-\varphi x)}%
=\left| G\right| \prod\limits_{i=1}^{n}\frac{(1+yx^{d_{i}-1})}{(1-x^{d_{i}})}
\tag{5.29b}
\end{equation}
and let us treat the right (R) and the left (L) hand sides separately.
Following Bourbaki [7] we put $y=-1+t(1-x)$. Substitution of this result to
R produces at once 
\begin{equation}
R\mid
_{x=1}=\prod\limits_{i=1}^{n}(d_{i}-1+t)=\prod\limits_{i=1}^{n}(m_{i}+t) 
\tag{5.31}
\end{equation}
To do the same for L requires us to keep in mind that $\det AB=\det A\det B$
and, hence, $\det AA^{-1}=1$ leads to $\det A^{-1}=1/\det A$ .Therefore,
after few steps we arrive at 
\begin{equation}
L\mid _{x=1}=\sum\limits_{r=1}^{n}h_{r}t^{r}.  \tag{5.32}
\end{equation}
Equating L with R, replacing $t$ by $1/T$ and relabeling $1/T$ again by $t$
\ and $h_{l}$ by $\tilde{h}_{l}$ = $g_{r}(m,p,n)$ we obtain the S-T result,
Eq.(5.10). To obtain physically useful result we have to take into account
that for the cubic symmetry we had already : $d_{i}=iN.$ Let therefore $%
y=-x^{Nq+1}$ in Eq.(5.29a) so that we obtain 
\begin{equation}
P((S(V)\otimes E(V))^{G};z)=\prod\limits_{i=1}^{n}\frac{1-z^{q+i}}{1-z^{i}} 
\tag{5.33}
\end{equation}
\textit{Remark}\textbf{\ }$13$. The result almost identical to our Eq.(5.33)
was obtained some time ago in the paper by Lerche et al.[87], page 444. To
obtain their result, it is sufficient to replace $z^{q+i}$ by \ $z^{q-i}.$
Clearly, such substitution is not permissible in our case because for cubic
symmetry in the limit $z=x^{N}=1$ we obtain 
\begin{equation}
P((S(V)\otimes E(V))^{G};z=1)=\frac{(q+1)(q+2)\cdot \cdot \cdot (q+n)}{n!} 
\tag{5.34}
\end{equation}
in accord with Eq.(5.3a)\footnote{%
The correctness of this result can be checked straightforwardly. Indeed, if
taking into account the Theorem 2 by Solomon, we consider products of the
type $dx_{1}^{p_{1}}\cdot \cdot \cdot dx_{n}^{p_{n}}$ with $%
p_{1}+...+p_{n}=q $ , then Eq.(5.34) gives the number of such products.}.

Obtained result suggest us to choose the Poincar$e^{\prime }$ polynomial,
Eq.(5.29a), as partition function associated with the Veneziano amplitude.
From our detailed derivation of Eq.(5.29a) it follows that: a) the
underlying physical model should be supersymmetric; b) it should be finite
dimensional; c) in view of the results of Section 1 the Poincar$e^{\prime }$
polynomial, Eq.(5.33), could be identified with the Weyl character formula;
d) accordingly, following Guillemin et al [20,48,81], the result Eq.(5.34),
can be identified with the Kostant multiplicity formula.

Being armed with such results, many\ additional details are provided in the
next subsection

\subsection{The multiparticle Veneziano-like amplitudes in the light of
Shepard-Todd results}

Earlier, when we discussed the volume integral, Eq.(5.1), we noticed that
its calculation closely resembles that of the Dirichlet-Veneziano integral
given below 
\begin{equation}
I=\int_{x_{i}\geq 0}dx_{1}^{<c_{1}>}\wedge \cdot \cdot \cdot \wedge
dx_{k+1}^{<c_{k+1>}}\delta (1-x_{1}^{N}-\cdot \cdot \cdot -x_{k+1}^{N}). 
\tag{5.35}
\end{equation}
We would like to demonstrate now that this is indeed the case. In view of
the results of Appendix B let $<c_{i}>=p_{i}$ and let $N$ divide all $%
p_{i}^{\prime }s$ so that $n_{i}=\frac{N}{p_{i}}$ . Let $x^{p_{i}}=z_{i}$
then, using the integral representation of delta function we obtain (up to a
constant as before) 
\begin{equation}
I=\frac{1}{2\pi }\oint dy[\prod\limits_{i=1}^{k+1}\left( \frac{1}{iyn_{i}}%
\right) ^{\dfrac{1}{n_{i}}}\Gamma (\frac{1}{n_{i}})]\exp (iy)\dot{=}\frac{%
\Gamma (\frac{p_{1}}{N})\cdot \cdot \cdot \Gamma (\frac{p_{k+1}}{N})}{\Gamma
(\sum\limits_{i}\frac{p_{i}}{N})}  \tag{5.36}
\end{equation}
in agreement with Eq.(4.57). In order to make connection with the results
earlier obtained we need to reconsider the calculation we just made. To this
purpose, using again Appendix B we remove the bracket sign \ $<...>$ from $%
c_{i}^{\prime }s$ and assume that $p_{i}/N$ can have any rational value
which we denote by $x_{i}$ . With such notations our integral $I$ acquires
the form of the Dirichlet integral, Eq.(4.21). In this integral let $%
t=u_{1}+...+u_{k}$. This allows us to use already familiar expansion,
Eq.(5.4), where now $t^{n}$ on the r.h.s. of Eq.(5.4) we have to replace by
the following identity 
\begin{equation}
t^{n}=(u_{1}+...+u_{k})^{n}=\sum\limits_{n=(n_{1},...,n_{k})}\frac{n!}{%
n_{1}!n_{2}!...n_{k}!}u_{1}^{n_{1}}\cdot \cdot \cdot u_{k}^{n_{k}} 
\tag{5.37}
\end{equation}
with restriction $n=n_{1}+...+n_{k}.$This type of identity was used earlier
in our work on Kontsevich-Witten model [32]. Moreover, from the same paper
we obtain the alternative and very useful form of the above expansion 
\begin{equation}
(u_{1}+...+u_{k})^{n}=\sum\limits_{\lambda \vdash k}f^{\lambda }S_{\lambda
}(u_{1},...,u_{k})  \tag{5.38}
\end{equation}
where the Schur polynomial $S_{\lambda }$ is defined by 
\begin{equation}
S_{\lambda }(u_{1},...,u_{k})=\sum\limits_{n=(n_{1},...,n_{k})}K_{\lambda
,n}u_{1}^{n_{1}}\cdot \cdot \cdot u_{k}^{n_{k}}  \tag{5.39}
\end{equation}
with coefficients $K_{\lambda ,n}$ known as Kostka numbers [89], $\
f^{\lambda }$ being the number of standard Young tableaux of shape $\lambda $
and the notation $\lambda \vdash k$ meaning that $\lambda $ is partition of $%
k$. Through such connection with Schur polynomials one can develop
connections with Kadomtsev-Petviashvili (KP) hierarchy of nonlinear exactly
integrable systems on one hand\ and with the theory of Schubert varieties on
another [32]. Due to the length of this paper, detailed analysis of such
connections is left for future work. Nevertheless, we shall encounter below
additional examples of striking similarities between the results of this
paper and those of the Kontsevich-Witten model.

Using of either Eq.(5.37) or (5.38) in Eq.(5.4) and substitution of the
r.h.s, of Eq.(5.4) into the Dirichlet integral, Eq.(4.21), produces (after
effectively performing the multiple Laplace transform) the following part of
the multiparticle Veneziano amplitude 
\begin{equation}
A(1,...k)=\frac{\Gamma _{n_{1}...n_{k}}(\alpha (s_{k+1}))}{(\alpha
(s_{1})-n_{1})\cdot \cdot \cdot (\alpha (s_{k})-n_{k})}.  \tag{5.40}
\end{equation}
This result for the multiparticle Veneziano amplitude clearly should be
symmetrized in order to obtain the full multiparticle Veneziano amplitude.
This is accord with results by Veneziano for the 4-particle case. It also
should be corrected by the appropriate inclusions of the phase factors as it
was discussed already in connection with Eq.(4.58) and in Appendix B. Since
the result, Eq.(5.40), was obtained with help of Eq.(5.4) all earlier
arguments developed for the four-particle amplitude remain unchanged in the
multiparticle case. However, this time we can do much better job with help
of the result by Solomon, Eq.(5.29a).

To this purpose \ we would like to demonstrate that using Eq.(5.29a) we can
obtain the generating (or partition) function $\mathbf{\Theta }(t,y,k)$ for
the Veneziano-like amplitudes. Following Hirzebruch and Zagier [90], page
190, we replace Eq.(5.29a) by the related expression, $\mathbf{\tilde{\Theta}%
}(t,y,k)$ 
\begin{equation}
\mathbf{\tilde{\Theta}}(t,y,k)=\prod\limits_{i=1}^{k}\frac{1+yt^{d_{i}}d%
\text{u}_{i}^{d_{i}}}{1-t^{d_{i}}\text{u}_{i}^{d_{i}}}  \tag{5.41}
\end{equation}
containing the differential form $d$u$_{i}^{d_{i}}\dot{=}$u$_{i}^{d_{i}-1}d$u%
$_{i}$ . If we assign the weight $d_{i}$ to the power u$_{i}^{d_{i}}$ then
the auxiliary parameter $t$ will keep track of such weight. In particular,
should we use u$_{i}^{d_{i}-1}$ instead of $d$u$_{i}^{d_{i}}$ we would need
to replace $t^{d_{i}}$ by $t^{d_{i}-1_{{}}}$ in the numerator. For u$_{i}=1$ 
$\forall i$ , we would reobtain back Eq.(5.29a). It is straightforward to
demonstrate that the partition function $\mathbf{\tilde{\Theta}}(t,y,k)$
reproduces the entire Solomon's algebra of differential forms $\omega ^{(p)}$%
, Eq.(4.26). Indeed, by expanding both the numerator and the denominator and
taking into account Eq.(5.25) let us consider the representative term of
such expansion proportional to $y^{p}t^{l}$ where $0\leq p\leq k$ .
Evidently, the numerator of Eq.(5.41) will supply $t^{d_{i_{1}}}\cdot \cdot
\cdot t^{d_{ip}}d$u$_{i_{1}}^{d_{i_{1}}}\cdot \cdot \cdot d$u$%
_{i_{p}}^{d_{i_{p}}}$ contribution while the denominator will supply $%
t^{e_{1}d_{i_{1}}}\cdot \cdot \cdot t^{e_{p}d_{i_{p}}}$u$%
_{i_{1}}^{e_{1}d_{i_{1}}}\cdot \cdot \cdot $u$_{i_{p}}^{e_{p}d_{i_{p}}}$
contribution to such term with numbers $e_{i}$ being nonnegative integers. \
Multiplication of these two factors should be done under the condition that
such term (as well as any other term in the expansion of $\mathbf{\tilde{%
\Theta}}(t,y,k)$ ) is to be $G-$invariant and, in our case,of weight $l.$
This implies that the sum $%
e_{1}d_{1}+...+e_{p}d_{p}=l-d_{i_{1}}-...-d_{i_{p}}$. Finally,
antisymmetrization of the product $d$u$_{i_{1}}^{d_{i_{1}}}\cdot \cdot \cdot
d$u$_{i_{p}}^{d_{i_{p}}}$ would lead to $\frac{1}{p!}d$u$%
_{i_{1}}^{d_{i_{1}}}\wedge \cdot \cdot \cdot \wedge d$u$_{i_{p}}^{d_{i_{p}}}$%
. Fortunately, in Eq.(4.26) this is done already so that the product $d$u$%
_{i_{1}}^{d_{i_{1}}}\cdot \cdot \cdot d$u$_{i_{p}}^{d_{i_{p}}}$ is
sufficient.

To make physical sense out of the results just presented we need to change
the rules slightly. These are summarized in the following partition function 
\begin{equation}
\mathbf{\Theta }(t,y,k)=\frac{1}{\alpha _{1}\cdot \cdot \cdot \alpha _{k}}%
\prod\limits_{i=1}^{k}\pi (k;j)\frac{1+y(k)d\text{u}_{i}^{\alpha _{i}}}{%
1-t^{d_{i}}\text{u}_{j}^{d_{i}}}  \tag{5.42}
\end{equation}
where the notation $y(k)$ symbolizes the fact that we are interested only in
terms proportional to $k-th$ power of $y$. Also,we have replaced the group
exponents $d_{i}$ in numerator by arbitrary nonzero integers $\alpha _{i}$
(\ in view of Eq.(4.21) and Appendix B) and introduced the symmetrization
operator $\pi (k;j)$ which \ is essentially of same nature as symmetrization
in Eq.(5.37). By expanding both numerator and denominator and collecting
terms proportional to $y^{k}t^{n}$ we obtain the following expression for
these terms 
\begin{equation*}
t^{e_{1}d_{i_{1}}}\cdot \cdot \cdot t^{e_{k}d_{i_{k}}}\text{u}%
_{i_{1}}^{e_{1}d_{i_{1}}}\cdot \cdot \cdot \text{u}_{i_{k}}^{e_{k}d_{i_{k}}}%
\{\text{u}_{1}^{\alpha _{1}}\cdot \cdot \cdot \text{u}_{k}^{\alpha _{k}}\}d%
\text{u}_{1}\cdot \cdot \cdot d\text{u}_{k}.
\end{equation*}
As before, we have to require $e_{1}d_{1}+...+e_{k}d_{k}=n.$ We had
encountered and discussed this expression earlier (after Eq.(5.5)) since the
exponents $d_{i}$ are given by $i$ with $i=1,...,k.$Once we have a
combination of nonnegative integers $n_{1}+...+n_{k}=n$ it can be
represented as $d_{1}e_{1}+...+d_{k}e_{k}=n$ so that the number of ways to
compose $n$ out of $n_{i}$'s is given by Eq.(5.3a).This factor should be
superimposed with the symmetrizing factor of the type given in Eq.(5.37)
before the above expression is ready for integration. Integrating each $%
u_{i} $ variable within limits [0,1] we reobtain back the portion of the
Veneziano amplitude given by Eq.(5.40).

\textit{Remark}\textbf{\ }14. a) The factor $\frac{1}{\alpha _{1}\cdot \cdot
\cdot \alpha _{k}}$ in front of Eq.(5.42) could be avoided should we replace 
$d$u$_{i}^{\alpha _{i}}$ in the numerator of Eq.(5.42) by u$_{i}^{\alpha
_{i}-1}$ and use the factors $d$u$_{i}$ together with integration operation.
b) To represent just obtained canonical formalism results in the grand
canonical form causes no additional problems and, hence, for the sake of
space is left for the readers. c) Obtained canonical expression does not
contain the essential phase factors. These should be reinstated in order the
above expression can be actually used.

Based on these results, the Veneziano-Shepard-Todd-Solomon canonical
partition function is given by

\begin{equation}
\mathbf{\Theta }(t,y,k)=\prod\limits_{i=1}^{k}\pi (k;j)\frac{1+y(k)\text{u}%
_{i}^{\alpha _{i}-1}}{1-t^{d_{i}}\text{u}_{j}^{d_{i}}}.  \tag{5.43}
\end{equation}

\subsection{The moment map, the Duistermaat-Heckman formula and the
Khovanskii-Pukhlikov correspondence}

In Section 3.2 we had introduced and discussed the moment map. Based on the
results obtained thus far we are ready to use this map now. To this purpose
it is sufficient to replace the integral $I$ in Eq.(5.35) by much simpler
integral 
\begin{equation}
I(E)=\int_{x_{i}\geq 0}dx_{1}^{c_{1}}\cdot \cdot \cdot
dx_{k+1}^{c_{k+1}}\delta (E-x_{1}-\cdot \cdot \cdot -x_{k+1})  \tag{5.44}
\end{equation}
where the parameter $E$ is introduced for further convenience. The trivial
case, when $1=c_{1}=...=c_{k+1},$ was discussed already, e.g. see Eq.s(5.1)
and (5.2), so that for this case: $I(E)=E^{k}/k!$ .The general case can be
treated similarly and it will be discussed in the next subsection. In the
meantime, for this ''trivial'' case the partition function $\Xi $,
Eq.(1.42), can be easily obtained as 
\begin{equation}
\Xi (\beta )=\int\limits_{0}^{\infty }dEI(E)\exp (-\beta E)=\beta ^{-(k+1)}.
\tag{5.45}
\end{equation}
By combining Eq.s(5.44) and (5.45) this result can be rewritten in the
alternative form 
\begin{equation}
\Xi (\beta )=\int dx_{1}\cdot \cdot \cdot dx_{k+1}\exp (-\beta
(x_{1}+...+x_{k+1})).  \tag{5.46}
\end{equation}
Taking into account that $x_{i}\geq 0$ $\forall i$ we reobtain $\Xi (\beta
)=\beta ^{-(k+1)}$ as required. Such seemingly trivial results can be
rewritten in the form compatible with the Duistermaat-Heckman \ (D-H)
formula [20,48,81]. In order to use and to understand this formula several
definitions need to be introduced.

In particular, as it was mentioned already in Ref. [2], the Fermat variety $%
\mathcal{F}(N)$ defined by Eq.(4.36) is a special case of the Brieskorn-Pham
(B-P) variety $V_{B-P}$ defined by 
\begin{equation}
V_{B-P}(f)=\{\mathbf{z}\in \mathbf{CP}^{n+1}\mid f(\mathbf{z})=0\} 
\tag{5.47}
\end{equation}
where $f(\mathbf{z})=z_{0}^{a_{0}}+...+z_{n+1}^{a_{n+1}}$ with $%
a_{0},...,a_{n+1}$ being a $n+2$ tuple of positive integers greater or equal
than 2.Clearly, $\mathcal{F}(N)$ is a special case for which all $%
a_{i}^{\prime }$s are the same and equal to $N$. By analogy with Eq.(2.14)
consider now the torus action (the monodromy) map $h:V_{B-P}(f)\rightarrow
V_{B-P}(f)$ defined by 
\begin{equation}
h_{t}(\mathbf{z})\equiv h_{t}(z_{0},...,z_{k+1})=\{\exp
(it/a_{0})z_{0},...,\exp (it/a_{k+1})z_{k+1}\}.  \tag{5.48}
\end{equation}
Let now $I:$ $0\leq t\leq 2\pi $ so that the mapping torus $T_{h}$ fiber
bundle can be constructed according to the following recipe: 
\begin{equation}
T_{h}=\frac{V_{B-P}(f)\times I}{i},  \tag{5.49}
\end{equation}
where the identification map $i$ is defined by the rule 
\begin{equation}
i:(\mathbf{z},0)=(h_{2\pi }(\mathbf{z}),2\pi ).  \tag{5.50}
\end{equation}
Thus constructed fiber bundle has topological and dynamical meaning (by the
way, this also provides yet another reason for inclusion of the Ruelle
transfer operator Eq.(1.13) in Section 1) discussed in earlier work [2]. For
the sake of space we refer our readers to this work for additional details.

Since such constructed fiber bundle ''lives'' in the complex \ projective
space it also has a meaning of a symplectic manifold $M$. Using the same
methods as in Sections 3.2. and 4.4.2. we construct a moment map $\mathcal{H}%
[\mathbf{z}]$ : $M\rightarrow P$ which reduces $M$ to a polytope $P$ (in our
case to a simplex $\Delta $ which is also a polytope). This can be easily
understood if in the period integral $I=\oint \omega $ with $\omega $ given
by Eq.(4.39) we (for the sake of argument) choose $<c>=1$ in the denominator
then, use the identity $1/a=\int\nolimits_{0}^{\infty }dt\exp (-at)$ to
bring the denominator into exponent, and then, perform the same deformation
retract operations which resulted in Eq.(4.54). On such deformation retract
the indentification map $i$ still works. In general, for each fiber $f_{i}(%
\mathbf{z})=e_{i}\in P$ (so that $\sum\nolimits_{i}m_{i}e_{i}=\mathcal{E}$ ,
e.g. see Eq.(3.10)) we\ introduce the reduced phase space known as the
Mardsen-Weinstein symplectic quotient $M_{red}(\mathcal{E})=\{f_{i}(\mathbf{z%
})=e_{i}\mid \mathbf{z}\rightarrow h_{2\pi }(\mathbf{z})\}$ [51]. Clearly,
since the vertices of $P$ are fixed points of $M$, $M_{red}(\mathcal{E})$
contains singularities associated with these fixed points. Take now the
moment map, Eq.(3.10), and consider the integral 
\begin{equation}
I[\mathbf{m}]=\int\limits_{M_{red}}\exp (-<\mathbf{m}\cdot \mathbf{f(x)}>)d%
\mathbf{x}  \tag{5.51}
\end{equation}
where $d\mathbf{x}$ is the symplectic measure ( in case of Eq.(5.44) it can
be reduced to $dx_{1}...dx_{k+1},$ e.g. see the next subsection ) and $<%
\mathbf{m}\cdot \mathbf{f(x)}>=\sum\limits_{i=1}^{k+1}m_{i}f_{i}(\mathbf{z})$
(which in the case of Eq.(5.46) is just $\sum\limits_{i=1}^{k+1}x_{i}).$ The
related integral, known as the D-H formula, can be written as follows
[20,51]: 
\begin{equation}
I[\mathbf{\beta ;\xi }]\mathbf{=}\int\limits_{P}\exp (-i\beta <\mathbf{\xi }%
\cdot \mathcal{E}>)h(\mathcal{E})d\mathcal{E}\text{ .}  \tag{5.52}
\end{equation}
It can be thought as the Fourier transform of the D-H function $h(\mathcal{E}%
)$ which is just the characteristic function of $P($the push forward of the
symplectic measure$)$, i.e. it is $1$ for points $\mathcal{E}$ $\in P$ and $%
0 $ otherwise. In agreement with the result by Atiyah [43], 
\begin{equation}
I[\mathbf{\beta }\rightarrow 0]=volM_{red}(\mathit{P})\equiv volP, 
\tag{5.53}
\end{equation}
where in our case $volP$=$p(k,n)$ (with $n=\mathcal{E}$) was defined in
Eq.(5.3a)\footnote{%
Comparing this ''quantum''result against ''classical'', Eq.(5.2), we notice
that the ''semiclassical approximation'': n$\rightarrow \infty $ of quantum
result produces the classical result as in standard quantum mechanics.}.
This is so because $M_{red}(\mathcal{E})$ is made of collection of single
points as discussed in Ref. [48], page 71, and Section 2.4. In particular,
in Subsection 2.4. it was argued that if $M_{red}(\mathcal{E})$ can be
thought of as the projective toric variety, then such variety is made of
points. For non negative integer $\mathcal{E}$ each such point represents
solution in nonegative integers of the equation $x_{1}+...+x_{k}=\mathcal{E}$
as discussed earlier in this section. Moreover, in Section 3.1.it was
demonstrated that $volM_{red}(\mathcal{E})=\chi $ where $\chi $ is the Euler
characteristic of such type of manifolds\footnote{%
Alternative derivation of this fact involving rather sophisticated methods
of algebraic geometry can be found in Ref.[23].}. This information is to be
used in the next subsection. In the meantime, D-H had also demonstrated that 
\begin{equation}
I[\beta ;\mathbf{\xi }]\mathcal{=}\sum\limits_{p}\frac{\exp (-i\beta <%
\mathbf{\xi }\cdot \mathbf{f(}p\mathbf{)}>)}{(i\beta
)^{k}\prod\limits_{i}^{k}<\mathbf{m}_{p}\cdot \mathbf{\xi }>}%
=\int\limits_{M_{red}}\exp (-i\beta <\mathbf{\xi }\cdot \mathbf{f(x)}>)d%
\mathbf{x}  \tag{5.54}
\end{equation}
with vector $\mathbf{m}_{p}$ representing the set of weights associated with 
$p$-th solution of Eq.(3.9) (with $\left| z_{i}\right| ^{2}$ being replaced
by $x_{i}$ ) while $\xi $ is any vector consistent with definition of the
rational polyhedral cone (Section 2.1). By expanding both sides of the above
identity in power series in auxiliary parameter $\beta $ and equating $\beta
-$independent terms we obtain 
\begin{equation}
volP=(-1)^{k}\sum\limits_{p}\frac{<\mathbf{\xi }\cdot \mathbf{f(}p\mathbf{)}%
>^{k}}{k!\prod\limits_{i}^{k}<\mathbf{m}_{p}\cdot \mathbf{\xi }>}  \tag{5.55}
\end{equation}
in accord with result by Vergne[51] given without derivation. Because this
equality should be independent of the choice for $\xi $ we must require 
\begin{equation}
k!\prod\limits_{i}^{k}<\mathbf{m}_{p}\cdot \mathbf{\xi }>=(-1)^{k}<\mathbf{%
\xi }\cdot \mathbf{f(}p\mathbf{)}>^{k}  \tag{5.56}
\end{equation}
in order to be in accord with Eq.(5.53). Instead of analyzing this equation
we choose another (much shorter) route aimed at proof of Eq.(5.55) by
utilizing some results from the work of Khovanskii and Pukhlikov [91].
Following these authors, we introduce auxiliary functions 
\begin{equation}
i(x_{1},...,x_{k};\xi _{1},...,\xi _{k})=\frac{1}{\xi _{1}...\xi _{k}}\exp
(\sum\limits_{i=1}^{k}x_{i}\xi _{i}),  \tag{5.57a}
\end{equation}
\begin{equation}
s(x_{1},...,x_{k};\xi _{1},...,\xi _{k})=\frac{1}{\prod\limits_{i=1}^{k}(1-%
\exp (-\xi _{i}))}\exp (\sum\limits_{i=1}^{k}x_{i}\xi _{i}).  \tag{5.57.b}
\end{equation}
The authors demonstrate that these functions are connected to each other
with help of the Todd transform, i.e. 
\begin{equation}
Tdi(y_{1},...,y_{k};\mathbf{\xi })_{y_{i}=x_{i}}=s(x_{1},...,x_{k};\mathbf{%
\xi }),  \tag{5.58}
\end{equation}
where $\mathbf{\xi }=(\xi _{1},...,\xi _{k})$ and the Todd operator $Td(z)$
is defined by 
\begin{equation}
Td(z)=\prod\limits_{i=1}^{k}\frac{z_{i}}{1-\exp (-z_{i})}\mid
_{z_{i}\rightarrow \frac{\partial }{\partial z_{i}}}.  \tag{5.59}
\end{equation}
The significance of this result comes from the following observations.
First, the expression $\sum\limits_{i=1}^{k}x_{i}\xi _{i}$ in Eq.(5.57)
represents a convex polyhedral cone, e.g. see Eq.(2.2). As Fig.2 suggests,
summation over cones forming complete fan is equivalent to summation over
the vertices of the polytope. If such summation is made we arrive at the
r.h.s. of the identity, Eq.(1.7), and, hence, we can use the l.h.s. of
Eq.(1.7) as well. Second, since we have connected the identity, Eq.(1.7),
with the Weyl character formula in Section 1, the l.h.s. of Eq.(1.7) is just
the character of the Weyl reflection group, e.g. see Eq.s(1.38) and (1.39).
Such character can be associated with quantum mechanical partition function,
Eq.(1.40).Third, looking at the D-H result, Eq.(5.54), replacing factor $%
i\beta $ by $-1,$readjusting \textbf{m}$_{p}$ and comparing such D-H sum
with Eq.(5.57a) (summed over vertices) we arrive at complete equivalence
between these two expressions. Moreover, by expanding (summed over the
vertices) the denominator of the r.h.s. of Eq.(5.57b) in powers of $\xi $ in
the limit of small $\xi ^{\prime }s$ we reobtain back the D-H result,
Eq.(5.54). Based on these facts, we arrive at important

\textbf{Corollary 3.}The\textbf{\ }Todd transform defined by Eq.(5.58) (with
summation over the vertices) provides direct link between classical and
quantum mechanical dynamical systems in accord with classical-quantum
mechanical correspondence which follows from the method of coadjoint orbits
discussed in Section 3.1.

Clearly, such correspondence holds for dynamical systems described by
semisimple Lie groups (algebras) and, most likely, it can be extended to all
(pseudo) reflection Weyl-Coxeter groups. In fact, for pseudo-reflection
groups \ this fact can be considered as proven in view of Eq.s
(5.10),(5.29),(5.30),(5.33) and (5.34) and results of Section 1. Moreover,
this claim is supported by results of Vergne [51] who had independently
obtained the central result: $volP$=$p(k,n))$ using in part the D-H
formalism. No connections with Solomon-Shepard-Todd results had been
mentioned in her work. For justification of the results of this paper such
connections are essential.

All this can be brought to more elegant mathematical form with help of work
by Atiyah and Bott [66] inspired by earlier work of Witten [92]. This is
discussed in the next subsection.

\subsection{From Riemann-Roch-Hirzebruch to Witten and Lefschetz via Atiyah
and Bott}

\bigskip

Let $E$ be a vector bundle over variety $X$ so that $ch(E)$ is the Chern
character of $E$ and $Td(X)$ is the Todd class of $X$ , then the
Hirzebruch--Grotendieck- Riemann- Roch formula the Euler characteristic $%
\chi (E)$ is given by [93] 
\begin{equation}
\chi (E)=\int\limits_{X}ch(E)\wedge Td(X)  \tag{5.60}
\end{equation}
This formula is too formal for immediate applications. To connect this
result with what was obtained earlier we need to use some results by
Guillemin [94,95] inspired by earlier work by Pukhlikov and Khovanskii [91].
To understand the mathematical significance of the results of both
references reading of earlier paper by Atiyah and Bott [66] is the most
helpful.\footnote{%
In the Remark 6. we had mentioned that all details of Atiyah and Bott paper
[66] are pedagogically explained in the monograph by Guillemin and Sternberg
[67].} Hence, we would like to make few observations related to this
historic paper first. Since the rest of our paper is not written in the
language of symplectic geometry needed for the current discussion, we
briefly introduce few relevant notations first. In particular, for the
Hamiltonian of planar harmonic oscillator discussed in Section 3.2 the
standard symplectic two-form $\omega $ can be written in several equivalent
ways 
\begin{equation*}
\omega =dx\wedge dy=rdr\wedge d\theta =\frac{1}{2}dr^{2}\wedge d\theta =%
\frac{i}{2}dz\wedge d\bar{z}
\end{equation*}
Since we are interested only in rotationally invariant observables this
means that $\theta $ dependence can be dropped, i.e.[20,48,81] 
\begin{equation*}
\frac{1}{2\pi }\int \int f(\alpha r^{2})rdr\wedge d\theta =\frac{1}{2}\int
f(\alpha x)dx.
\end{equation*}
This fact was used already starting with Eq.(5.1). For collection of $k$
oscillators the symplectic form $\Omega $ is given, as usual, by $\Omega $ =$%
\sum\nolimits_{i=1}^{k}dx_{i}\wedge dy_{i}=\frac{i}{2}\sum%
\nolimits_{i=1}^{k}dz_{i}\wedge d\bar{z}_{i}$ so that its $n$-th power is
given by $\Omega ^{n}=\Omega \wedge \Omega \wedge \cdot \cdot \cdot \wedge
\Omega $ =$dx_{1}\wedge dy_{1}\wedge \cdot \cdot \cdot dx_{n}\wedge dy_{n}$
. In polar coordinates the symplectic volume form is given by $(2\pi
)^{-n}\Omega ^{n}/n!$ . Clearly, at this level of our presentation we can
drop the factor of $\left( 2\pi \right) ^{-n}.$ With help of these results,
it is convenient to introduce the differential form 
\begin{equation}
\exp \Omega =1+\Omega +\frac{1}{2!}\Omega \wedge \Omega +\frac{1}{3!}\Omega
\wedge \Omega \wedge \Omega +\cdot \cdot \cdot  \tag{5.61}
\end{equation}
with the convention that $\int\limits_{M}\hat{\omega}=0$ if the degree of $%
\hat{\omega}$ differs from the dimension of $M.$ If the manifold $M$ does
not have singularities then, according to Liouville theorem, the form $%
\Omega $ is closed, i.e. $d\Omega =0.$ But if $M$ contains singularities
(e.g. fixed points) this is no longer so. To fix the problem Atiyah and Bott
\ had suggested to use the amended symplectic form 
\begin{equation}
\Omega ^{\ast }=\Omega -f\cdot u  \tag{5.62}
\end{equation}
with $f$ being some function(s) (to be determined momentarily) and $u$ being
some indeterminate(s) also to be determined momentarily. The above form is
(equivariantly) closed with respect to the following action of the operator $%
d_{X}$ : 
\begin{equation}
d_{X}\Omega ^{\ast }=\left( i(X)\Omega -df\right) \cdot u  \tag{5.63}
\end{equation}
if and only if $i(X)\Omega =df$ . Here $i(X)$ means the standard contraction
(i.e. operation inverse to exterior differentiation). As is well known [50],
the condition $i(X)\Omega =df$ is equivalent to the Hamiltonian equations
and, hence, the Hamiltonian $f$ is the moment map [5,20,48]. Consider now
again the D-H integral, Eq.(5.51). In view of the results just obtained it
can be rewritten as 
\begin{equation}
I[\mathbf{m}]=\int\limits_{M_{red}}\exp \Omega ^{\ast
}=\int\limits_{M_{red}}\exp (\Omega -<\mathbf{m}\cdot \mathbf{f(x)}>). 
\tag{5.64}
\end{equation}
Since, when written in terms of complex variables, the form $\Omega $
represents the first Chern class, it is only natural to associate exp($-<%
\mathbf{m}\cdot \mathbf{f(x)}>)$ with Chern character $ch(E)$ (with some
caution). Details can be found in cited references. Clearly, the form $%
\mathbf{m}\cdot \mathbf{f(x)}$ corresponds to Atiyah and Bott's combination$%
\ \ f\cdot u.$ It is convenient now to make formal redefinitions: $%
f_{i}\rightleftharpoons c_{i},$ with $c_{i}$ denoting Chern class of the
i-th complex line bundle\footnote{%
The splitting principle mentioned earlier allows to ''disect'' manifold into
product of simpler manifolds, e.g. \textbf{C}$^{n}=\mathbf{C}\times \mathbf{C%
}\cdot \cdot \cdot \mathbf{C,}$etc. each having its own first Chern
class.This will be explained more carefully below.}. Next, we temporarily
replace the vector \textbf{m} by indeterminate vector -\textbf{h. }After
this we can use the D-H formula, Eq.(5.54), in order to rewrite it as
follows 
\begin{equation}
\int\limits_{M_{red}}\exp (\Omega +<\mathbf{h}\cdot \mathbf{c(x)}%
>)=\sum\limits_{p}\frac{\exp (<\mathbf{h}\cdot \mathbf{c(}p\mathbf{)}>)}{%
\prod\limits_{i}^{k}h_{i}^{p}}.  \tag{5.65}
\end{equation}
Now we can apply the Khovanskii-Pukhlikov-Todd operator, Eq.(5.58), to both
sides of Eq.(5.65). Clearly, the l.h.s. acquires the form 
\begin{equation}
I[\mathbf{h}]=\sum\limits_{p}\int\limits_{M_{red}}\exp (\Omega +<\mathbf{h}%
\cdot \mathbf{c(}p\mathbf{)}>)\prod\limits_{i=1}^{k}\frac{c_{i}(p)}{1-\exp
(-c_{i}(p))}  \tag{5.66}
\end{equation}
which looks essentially the same as the r.h.s of \ Eq.(5.60) while at the
same time the r.h.s. will formally coincide with the r.h.s of Eq.(1.7) and,
therefore, as it is shown in Section 1, in principle can be brought to the
form coinciding with the r.h.s. of Eq.(5.33) (with $n$ replaced by $k$). The
Todd transform by Khovanskii and Pukhlikov allows us to obtain the Lie group
characters, i.e. quantum objects, using ''classical'' \ partition functions.
In particular, for $\mathbf{h}=0$ the l.h.s. should produce the Euler
characteristic $\chi $ because the r.h.s. is obviously reproducing $\chi .$
Moreover, according to Guilemin [81] , $\chi $ is equal to the dimension $%
Q=Q^{+}-Q^{-}$ of the \ quantum Hilbert space associated with classical
system described by the moment map Hamiltonian. Before discussing the
meaning of subspaces $Q^{+}(Q^{-})$ we would like for a moment to focus our
attention on the mathematical meaning of the Veneziano-like amplitudes in
the light of just obtained results\footnote{%
In Section 4 we had argued that these are periods associated with
differential forms living on Fermat hypersurfaces. We shall demonstrate that
the arguments to be used \ in the remainder of this paper are in support of
this fact .}. To this purpose, following Khovanskii and Pukhlikov again, we
notice that for \textit{any} polynomial $P(z)$ we can write the identity 
\begin{equation}
P(z_{1,...,}z_{N})\exp (\sum\limits_{i=1}^{N}p_{i}z_{i})=P(\frac{\partial }{%
\partial p_{1}},...,\frac{\partial }{\partial p_{N}})\exp
(\sum\limits_{i=1}^{N}p_{i}z_{i}).  \tag{5.67}
\end{equation}
By applying identity Eq.(5.67), to Eq.(5.65) and assuming \textbf{h}=0 at
the end of calculations we reobtain the result of Guillemin, Ref. [81], page
73,or Ref. [94], which is essentially the same thing as Eq.(7.17) by Atyiah
and Bott. Explicitly, 
\begin{equation}
\int\limits_{M_{red}}P(c_{1},...,c_{k})=\sum\limits_{p}\frac{%
P(e(p)_{1,...,}e(p)_{k})}{\prod\limits_{i}^{k}h_{i}^{p}},  \tag{5.68}
\end{equation}
where the numbers $e(p)_{i}$ had been defined in the previous subsection.
This relation had been used by Kontsevich [96] and others [97] for
enumeration of rational curves on algebraic varieties. For us it is
important that such relation can provide the intersection numbers which are
averages over $M_{red}$ of\ the products of the 1st Chern classes of the
tautological line bundles\footnote{%
For the sake of space we refere our readers to Ref.[6] where these conceps
are beautifully explained.}. This observation formally makes calculation of
the Veneziano-like amplitudes similar in spirit to that earlier encountered
in connection with the Witten-Kontsevich model [32]. From \ the discussion
we had so far it should be clear that the l.h.s. of Eq.(5.68) is analogous
to Eq.(4.54). However, below we suggest better computational alternative.

To describe this alternative, we need to make a connection with 1982 work by
Witten [92]. Such connection is essential.\ It also was emphasized in the
paper by Atiyah and Bott [66]. Fortunately, most of what we would like to
say about Witten's paper is already well \ developed and documented
[93].This allows us to squeeze our discussion to the absolute minimum
emphasizing mainly features absent in standard treatments but needed for
this work. In particular, we expect our readers to be familiar with the
basic facts about Hodge-de Rham theory as described in 2nd edition of book
by Wells [10].

We begin with the following observations. Let $X$ be a complex Hermitian
manifold and let $\mathcal{E}^{p+q}(X)$ denote the complex -valued
differential forms (sections) of type $(p,q)$ $,p+q=r,$ living on $X$. The
Hodge decomposition insures that $\mathcal{E}^{r}(X)$=$\sum\nolimits_{p+q=r}%
\mathcal{E}^{p+q}(X).$ The Dolbeault operators $\partial $ and $\bar{\partial%
}$ act on $\mathcal{E}^{p+q}(X)$ according to the rule \ $\partial :\mathcal{%
E}^{p+q}(X)\rightarrow \mathcal{E}^{p+1,q}(X)$ and $\bar{\partial}:\mathcal{E%
}^{p+q}(X)\rightarrow \mathcal{E}^{p,q+1}(X)$ , so that the exterior
derivative operator is defined as $d=\partial +\bar{\partial}$. Let now $%
\varphi _{p}$,$\psi _{p}\in \mathcal{E}^{p}$. By analogy with traditional
quantum mechanics we define (using Dirac's notations) the inner product 
\begin{equation}
<\varphi _{p}\mid \psi _{p}>=\int\limits_{M}\varphi _{p}\wedge \ast \bar{\psi%
}_{p}  \tag{5.69}
\end{equation}
where the bar means the complex conjugation and the star $\ast $ means the
usual Hodge conjugation. Use of such product is motivated by the fact that
the period integrals, e.g. those for the Veneziano-like amplitudes, are
expressible through such inner products [6,10]. Fortunately, such product
possesses properties typical for the finite dimensional quantum mechanical
Hilbert spaces. In particular, 
\begin{equation}
<\varphi _{p}\mid \psi _{q}>=C\delta _{p,q}\text{ and }<\varphi _{p}\mid
\varphi _{p}>>0,  \tag{5.70}
\end{equation}
where $C$ is some positive constant. With respect to such defined scalar
product it is possible to define all conjugate operators, e.g. $d^{\ast }$,
etc. and, most importantly, the Laplacians 
\begin{eqnarray}
\Delta &=&dd^{\ast }+d^{\ast }d,  \notag \\
\square &=&\partial \partial ^{\ast }+\partial ^{\ast }\partial , 
\TCItag{5.71} \\
\bar{\square} &=&\bar{\partial}\bar{\partial}^{\ast }+\bar{\partial}^{\ast }%
\bar{\partial}.  \notag
\end{eqnarray}
All this was known to mathematicians before Witten's work [92]. The
unexpected twist occurred when Witten suggested to extend the notion of the
exterior derivative $d$. Within the de Rham picture (valid for both real and
complex manifolds) let $M$ be a compact Riemannian manifold and $K$ be the
Killing vector field which is just one of the generators of the isometry of $%
M$ then, Witten had suggested to replace the exterior derivative operator $d$
by the extended operator 
\begin{equation}
d_{s}=d+si(K)  \tag{5.72}
\end{equation}
where the operator $i(K)$ has the same meaning as in Eq.(5.63) and $s$ is
real nonzero parameter. Witten argues that one can construct the Laplacian
(the Hamiltonian in his formulation) $\Delta $ by $\Delta
_{s}=d_{s}d_{s}^{\ast }+d_{s}^{\ast }d_{s}$ . This is possible if and only
if $d_{s}^{2}=d_{s}^{\ast 2}$ $=0$ or, since $d_{s}^{2}=s\mathcal{L}(K)$ ,
where $\mathcal{L}(K)$ is the Lie derivative along the field $K$, if the Lie
derivative acting on the corresponding differential form vanishes. The
details are beautifully explained in the much earlier paper by Frankel [50]
discussed already in Section 3.2. What is important for us is the
observation by Atiyah and Bott that if one treats the parameter $s$ as
indeterminate and identifies it with $f$ \ (in Eq.(5.62)), then the
derivative $d_{X}$ \ in Eq.(5.63) is exactly Witten's $d_{s}$ This
observation provides the link between the D-H formalism discussed earlier in
this subsection and Witten's supersymmetric quantum mechanics still to be
discussed further.

In Section 4 we provided arguments explaining why the Veneziano-like
amplitudes should be considered as period integrals associated with homology
cycles on Fermat (hyper) surfaces. According to Wells[10 ], and also
Ref.[6], the inner scalar product, Eq,(5.69), prior to normalization can be
associated with such period integrals. Once such correspondence is
established, we would like to return to our discussion of the quantum
Hilbert space $Q$ and the associated with it spaces $Q^{+}$ and $Q^{-}.$
Following Ref.[48,51.81] we consider the (Dirac) operator $\acute{\partial}=%
\bar{\partial}+\bar{\partial}^{\ast }$ and its adjoint with respect to
scalar product, Eq.(5.66), then 
\begin{equation}
Q=\ker \acute{\partial}-co\ker \acute{\partial}^{\ast }=Q^{+}-Q^{-}=\chi 
\tag{5.73}
\end{equation}
Such definition was used by Vergne[51] (e.g.see discussion after Corollary
3) to reproduce Eq.(5.3a). We would like to arrive at the same result using
different arguments. As a by product, we shall obtain the alternative
quantum mechanical interpretation of the Veneziano-like amplitudes.

To this purpose we notice first that according to Theorem 4.7. by Wells [10]
we have $\Delta =2\square =2\bar{\square}$ with respect to the K\"{a}hler
metric on $X$. Next, according to Corollary 4.11. of the same reference $%
\Delta $ commutes with $d,d^{\ast },\partial ,\partial ^{\ast },\bar{\partial%
}$ and $\bar{\partial}^{\ast }.$ From these facts it follows immediately
that if we, in accord with Witten, choose $\Delta $ as our Hamiltonian, then
the supercharges can be selected as Q$^{+}=d+d^{\ast }$ and $Q^{-}=i\left(
d-d^{\ast }\right) .$ Evidently, this is not the only choice as Witten
indicates. If the Hamiltonian \ H is acting in finite dimensional Hilbert
space we can require axiomatically that : a) there is a vacuum state (or
states) $\mid \alpha >$ such that H$\mid \alpha >=0$ (i.e. this state is
harmonic differential form) and Q$^{+}\mid \alpha >=$Q$^{-}\mid \alpha >=0$
. This requires, of course, that [H,Q$^{+}]=[$H,Q$^{-}]=0.$ Finally, \ once
again, following Witten we require that $\left( Q^{+}\right) ^{2}=\left(
Q^{-}\right) ^{2}=$H. Then, the equivariant extension, Eq.(5.72), leads to $%
\left( Q_{s}^{+}\right) ^{2}=$ H+$2is\mathcal{L}$($K$). Fortunately, the
above \ supersymmetry algebra can be extended. As it was mentioned in
section 3.1., there are operators acting on differential forms living on
K\"{a}hler (or Hodge) manifolds whose commutators are isomorphic to $sl_{2}(%
\mathbf{C})$ Lie algebra. It was mentioned in the same section that \textit{%
all} semisimple Lie algebras are made of copies of $sl_{2}(\mathbf{C})$. Now
we can exploit these observations further using the Lefschetz isomorphism
theorem whose exact formulation is given as Theorem 3.12 in Wells [10]. We
are only going to use some parts of it in this work.

In particular, using notations of Ref.[10] we introduce operator $L$
commuting with $\Delta $ and its adjoint $L^{\ast }\equiv \Lambda $ .\ It
can be shown [10], page 159, that $L^{\ast }=w\ast L\ast $ where, as before, 
$\ast $ denotes the Hodge star operator and the operator $w$ can be formally
defined through the relation $\ast \ast =w$, [10], page 156. From these
definitions it should be clear that $L^{\ast }$ also commutes with $\Delta $
on the space of harmonic differential forms (in accord with page 195 of
Wells). In Section 3.1. we have mentioned the Jacobson-Morozov theorem. This
theorem essentially guarantees that relations given by Eq.s(3.2a-c) can be
brought to form 
\begin{equation}
\lbrack h_{\alpha },e_{\alpha }]=2e_{\alpha }\text{ , }[h_{\alpha
},f_{\alpha }]=-2f_{\alpha }\text{ , \ }[e_{\alpha },f_{\alpha }]=h_{\alpha
}\   \tag{5.74}
\end{equation}
upon appropriate rescaling, e.g. see [11], page 37. Here $\alpha \in \Delta $
(Appendix A) with $\Delta $ being the root system (this notation should not
cause confusion ). As part of the preparation for proving of the Lefschetz
isomorphism theorem, it can be shown [10] that 
\begin{equation}
\lbrack \Lambda ,L]=B\text{ and }[B,\Lambda ]=2\Lambda \text{, }[B,L]=-2L. 
\tag{5.75}
\end{equation}
Comparison between the above two expressions leads to the Lie algebra
endomorphism, i.e. the operators $h_{\alpha },f_{\alpha }$ and $e_{\alpha }$
act on the vector space $\{v\}$ to be described below while the operators $%
\Lambda ,L$ and $B$ obeying the same commutation relations act on the space
of differential forms. It is possible to bring Eq.s(5.74) and (5.75) to even
closer correspondence. To this purpose, following Dixmier [98],Ch-r 8, we
introduce the operators $h=\sum\nolimits_{\alpha }a_{\alpha }h_{\alpha }$, $%
e=\sum\nolimits_{\alpha }b_{\alpha }e_{\alpha }$ , $f=\sum\nolimits_{\alpha
}c_{\alpha }f_{\alpha }.$ Then, provided that the constants are subject to
constraint: $b_{\alpha }c_{\alpha }=a_{\alpha }$ , the commutation relations
between the operators $h$, $e$ and $f$ are \textit{exactly the same} as for $%
B$, $\Lambda $ and $L$ respectively. To avoid unnecessary complications we
choose $a_{\alpha }=b_{\alpha }=c_{\alpha }=1$. Next, following Serre [38],
ch-r 4, we need to introduce the notion of the \textit{primitive} vector (or
element).This is the vector $v$ such that $hv$=$\lambda v$ but $ev=0.$ The
number $\lambda $ is the weight of the module $V^{\lambda }=\{v\in V\mid hv$=%
$\lambda v\}.$ If the vector space is \textit{finite dimensional}, then $%
V=\sum\nolimits_{\lambda }V^{\lambda }$ . Moreover, only if $V^{\lambda }$
is finite dimensional it is straightforward to prove that the primitive
element does exist. The proof is based on the observation that if $x$ is the
eigenvector of $h$ with weight $\lambda ,$then $ex$ is also the eigenvector
of $h$ with eigenvalue $\lambda -2,$ etc. Moreover, from the book by Kac
[12], Chr.3, it follows that if $\lambda $ is the weight of $V$ then $%
\lambda -<\lambda ,\alpha _{i}^{\vee }>\alpha _{i}$ is also the weight with
the same multiplicity. Since according to Eq.(A.2) $<\lambda ,\alpha
_{i}^{\vee }>\in \mathbf{Z}$\textbf{\ ,\ }Kac introduces another module: $%
U=\sum\nolimits_{k\in \mathbf{Z}}$ $V^{\lambda +k\alpha _{i}}$ \ . Such
module is finite for finite reflection groups and is infinite for the affine
reflection groups. We would like to argue that for our purposes it is
sufficient to use only the finite reflection (or pseudo-reflection) groups.
This is so in view of the Theorem 2 by Solomon valid for finite reflection
groups. It should be clear, however, from reading of book by Kac that the
infinite dimensional version of the module $U$ leads straightforwardly to
all known string-theoretic results. In the case of CFT this is essential, as
we had argued earlier in Sections 2.4. and 4.4.3., but for calculation of
the Veneziano-like amplitudes this is not essential since by accepting such
option we\ \ loose our connections with the Lefschetz isomorphism theorem (
relying heavily on the existence of primitive elements) and, therefore, with
the Hodge theory on which arguments of our presentation are based. Hence,
below we choose to work with finite dimensional option only and in Section 6
we briefly discuss possible modifications which could be caused by
variations of complex Hodge structure (i.e. motions in moduli space).

Let $v$ be a primitive element of weight $\lambda $ then, following Serre,
we let $v_{n}=\frac{1}{n!}e^{n}v$ for $n\geq 0$ and $v_{-1}=0,$ so that 
\begin{eqnarray}
hv_{n} &=&(\lambda -2n)v_{n}  \TCItag{5.76} \\
ev_{n} &=&(n+1)v_{n+1}  \notag \\
fv_{n} &=&(\lambda -n+1)v_{n-1}.  \notag
\end{eqnarray}
Clearly, the operators $e$ and $f$ are the creation and the annihilation
operators according to existing in physics terminology while the vector $v$
is being interpreted as the vacuum state vector. The question arises: how
this vector is related to earlier introduced vector $\mid \alpha >?$ Before
providing the answer to this question we need, following Serre, to settle
the related issue. In particular, we can either: a) assume that for all $%
n\geq 0$ the first of Eq.s(5.76) has solutions and all vectors $%
v,v_{1},v_{2} $ , ...., are linearly independent or b) beginning from some $%
m+1\geq 0,$ all vectors $v_{n\text{ }}$are zero, i.e. $v_{m}\neq 0$ but $%
v_{m+1}=0.$ The first option leads to the infinite dimensional
representations associated with Kac-Moody affine algebras just mentioned.
The second option leads to finite dimensional representations and to the
requirement $\lambda =m$ with $m$ being an integer. Following Serre, this
observation can be exploited further thus leading us to crucial physical
identifications. Serre observes that with respect to $n=0$ Eq.s(5.76)
possess a (''super'')symmetry . That is the linear mappings 
\begin{equation}
e^{m}:V^{m}\rightarrow V^{-m}\text{ and \ }f^{m}:V^{-m}\rightarrow V^{m} 
\tag{5.77}
\end{equation}
are isomorphisms and the dimensionality of $V^{m}$ and $V^{-m}$ are the
same. Serre provides an operator (the analog of Witten's $F$ operator) $%
\theta =\exp (f)\exp (e)\exp (-f)$ such that $\theta \cdot f=-e\cdot \theta $%
, $\theta \cdot e=-\theta \cdot f$ and $\theta \cdot h=-h\cdot \theta .$ In
view of such operator, it is convenient to redefine $h$ operator : $%
h\rightarrow \hat{h}=h-\lambda $. Then, for such redefined operator the
vacuum state is just $v$. Since both $L$ and $L^{\ast }=\Lambda $ commute
with the supersymmetric Hamiltonian H and because of group endomorphism we
conclude that the vacuum state $\mid \alpha >$ for H corresponds to the
primitive state vector $v$.

Now we are ready to apply yet another isomorphism following Ginzburg [9],
ch-r 4\footnote{%
Unfortunately, the original sourse contains minor mistakes. These are easily
correctable. The corrected results are given in the text.}. To this purpose
we make the following identification 
\begin{equation}
e_{m}\rightarrow t_{m+1}\frac{\partial }{\partial t_{m}}\text{ , }%
f_{m}\rightarrow t_{m}\frac{\partial }{\partial t_{m+1}}\text{ , }%
h_{m}\rightarrow t_{m+1}\frac{\partial }{\partial t_{m+1}}-t_{m}\frac{%
\partial }{\partial t_{m}}  \tag{5.78}
\end{equation}
Such operators are acting on the vector space made of monomials of the type 
\begin{equation}
v_{k}\rightarrow \mathcal{F}_{n}=\frac{n!}{n_{1}!n_{2}!\cdot \cdot \cdot
n_{k}!}t_{1}^{n_{1}}\cdot \cdot \cdot t_{k}^{n_{k}}  \tag{5.79}
\end{equation}
where $n_{1}+...+n_{k}=n$ . This result should be compared with Eq.(5.37).
Eq.s (5.76) have now their analogs 
\begin{eqnarray}
h_{m}\ast \mathcal{F}_{n}(m) &=&(n_{m+1}-n_{m})\mathcal{F}_{n}(m)  \notag \\
e_{m}\ast \mathcal{F}_{n}(m) &=&n_{m}\mathcal{F}_{n}(m+1)  \TCItag{5.80} \\
\text{ }f_{m}\ast \mathcal{F}_{n}(m) &=&n_{m+1}\mathcal{F}_{n}(m-1)  \notag
\end{eqnarray}
where, clearly, one should make the following consistent identifications: $%
m-2n=n_{m+1}-n_{m}$ , $n_{m}=n+1$ and $m-n+1=n_{m+1}.$ Next, we define the
total Hamiltonian: $h=$ $\sum\nolimits_{m=1}^{k}h_{m}$ and consider its
action on the total wave function, e.g. see Eq.(5.37), $\sum\limits_{n=%
\left( n_{1,}...,n_{k}\right) }\frac{n!}{n_{1}!n_{2}!...n_{k}!}%
t_{1}^{n_{1}}\cdot \cdot \cdot t_{k}^{n_{k}}$. Upon redefining the
Hamiltonian as before we obtain the ground state degeneracy equal to $p(k,n)$
in accord with Eq.(5.3.a). According to Witten [92], this degeneracy
determines the Euler characteristic $\chi $, e.g. see Eq.s(5.60),(5.73).
Vergne [51] also obtained $p(k,n)$ using \ the standard procedure based on
Atiyah-Singer-Hirzebruch index calculations. It involves calculation of
traces of the Dirac operator as it is given by Eq.(5.73). Obtained results
provide alternative explanation of the quantum mechanical nature of the
multiparticle Veneziano (and Veneziano-like) amplitudes.

\bigskip

\section{Instead of Discussion}

\bigskip

Although this paper came out as rather long, in reality, much more remains
to be done. In our previous work, Ref. [2], on Veneziano amplitudes we
noticed connections with singularity theory, number theory, knot theory,
dynamical systems theory. Present work adds to this list the theory of
exactly integrable systems, K-theory, algebraic geometry, combinatorics,
etc. These are mathematical aspects of the unfolding story. There are
however physical aspects no less fascinating. For instance, in his 1982
paper [92] Witten had noticed remarkable connections between the
sypersymmetry and the Lorentz invariance. This connection can be extended
now based on the results of this work. Indeed, pseudo-reflections used
frequently in this work are isometries of the complex hyperbolic space [99].
We have discussed properties of real hyperbolic space earlier in connection
with mathematical problems related to AdS-CFT correspondence [100 ] and
dynamics of 2+1 gravity [101 ]. The connection between the Lorentz
space-time and the real hyperbolic geometry is beautifully summarized in
Thurston's book [102]. As results of our earlier work [100 ] indicate, the
hyperbolic ball model of real hyperbolic space is quite adequate for
description of meaningful physical models and the boundary of hyperbolic
space plays a crucial role in such description. For instance, the
infinitesimal\ variations at the boundary of the Poincare$^{\prime }$ disc
model -the simplest model of hyperbolic space H$^{2}-$ naturally produce the
Virasoro algebra (e.g. see our work, Ref.[100 ], Section 7, for details).
Extension of the method producing this algebra to, say, H$^{3}$ is
complicated by the Mostow rigidity theorem also discussed in our paper [100
]. This theorem tells unequivocally that the Teichm\"{u}ller space for the
hyperbolic 3-manifolds without boundaries is just a point. That is all
hyperbolic surfaces in hyperbolic spaces H$^{n}$, $n>2$ are rigid
(nonbendable). This restriction can be lifted in certain cases discussed in
our earlier work. As it is demonstrated in Goldman's monograph [99], the
real hyperbolic space is just part of the complex hyperbolic space which,
not too surprisingly, can be modelled by the complex hyperbolic ball model.
What is surprising, however, is the fact that the group of isometries of the
boundary of such ball is the Heisenberg group. The method of coadjoint
orbits discussed in Section 3.1. and the Todd transform, Eq.(5.58), designed
by Khovanskii-Pukhlikov are indications of much tighter connections between
classical and quantum mechanics. If, at least locally, our space is complex
hyperbolic, then the isometries of this space generated by pseudo-reflection
groups are inseparable from quantum mechanical description of reality.
Moreover, the reader should not be left with impression that treatment of
CFT can avoid use of these pseudo-reflection groups. The monograph by Kane
[84] contains important references on earlier works by Kac and Kac and
Petersen indicating profound importance of these groups for CFT as well. The
intriguing problem still remains: if in the case of real hyperbolic space H$%
^{n}$, $n>2$, the Mostow rigidity theorem forbids deformations of almost all
real hyperbolic structures what could be said about the analog of such
theorem for complex hyperbolic spaces and the rigidity of complex structures
in such spaces? Although at this moment we are not yet aware of the detailed
answer to this question, still several remarks are appropriate at this
point. In particular, in the monograph [103] on quantum groups by Chari and
Pressley on page 435 the deformation of $sl_{2}(\mathbf{C})$ is discussed
and on page 439 it is stated that such deformation leads to failure of
Kirillov-Kostant method of coadjoint orbits. As results of this work
indicate, for the purposes of calculation of the Veneziano-like amplitudes,
there is no need for such drastic measures. Moreover, the projectivised
version of $SL_{2}(\mathbf{C})$ associated with isometries of H$^{3}$ is
connected part of the Lorentz group O(3,1) which, in turn, is isomorphic to $%
PSL_{2}(\mathbf{C})$ as stated in our earlier work, Ref. [100 ] on AdS-CFT
correspondence. Hence, it is impossible simultaneously to deform $SL_{2}(%
\mathbf{C})$, to keep the Lorentz invariance and to have quantum mechanics
(the Heisenberg group) in its traditional form.

\bigskip

\textbf{Acknowledgments}. The author would like to thank Professors Richard
Kane (U.of. Western Ontario ), Victor Ginzburg (U.of Chicago), Anatoly
Libgober (U.of Ill.at Chicago) and Predrag Cvitatovi\v{c} (Ga.Tech) for
helpful correspondence. This work was partially supported by NSF/DMS grant
\# 0306887.

\pagebreak

\bigskip

\textbf{Appendix A. Some results from the theory of Weyl-Coxeter reflection
and pseudo-reflection groups}

\bigskip

a) \textit{The Weyl group}

As in Section 1.2., let $V$ be a finite dimensional vector space endowed
with a scalar product, Eq.(1.6), which is positive-definite symmetric
bilinear form. For each nonzero $\alpha \in V$ \footnote{%
In Section 1.2 we have used symbol $u_{i}$ instead of $\alpha .$}let $%
r_{\alpha }$ denote the orthogonal reflection in the hyperplane H$_{\alpha }$
through the origin perpendicular to $\alpha $ \ (i.e. set of hyperplanes H$%
_{\alpha }$ is in one-to one correspondence with set of $\alpha ^{\prime }s)$
so that for $v\in V$ we obtain 
\begin{equation}
r_{\alpha }(v)=v-<v,\alpha ^{\vee }>\alpha ,  \tag{A.1.}
\end{equation}
where $\alpha ^{\vee }=2\alpha /<\alpha ,\alpha >$ is the vector \textit{dual%
} to $\alpha .$ Thus defined reflection is an orthogonal transformation in a
sense that \ $<r_{\alpha }(v),r_{\alpha }(\mu )>=$\ $<\nu ,\mu >.$ In
addition,\ $\left[ r_{\alpha }(v)\right] ^{2}=1$ $\forall \alpha ,\nu .$
Conversely, these two properties imply the transformation law, Eq.(A.1).

\ From these results it follows as well that for $v=\alpha $ we get $%
r_{\alpha }(\alpha )=-\alpha $ that is reflection in the hyperplane with
change of \ vector orientation. If the set of vectors which belong to $V$ is
mutually orthogonal, then $r_{\alpha }(v)=v$ for $v\neq \alpha $ but, in
general, the orthogonality is not required. Because of this, one introduces
the \textit{root system }$\Delta $ of vectors which span $V$. Such system is 
\textit{crystallographic} if for each pair $\alpha ,\beta \in \Delta $ one
has 
\begin{equation}
<\alpha ^{\vee },\beta >\in \mathbf{Z}\text{ and }r_{\alpha }(\beta )\in
\Delta .  \tag{A.2.}
\end{equation}
Thus, each reflection $r_{\alpha }$ ($\alpha \in \Delta )$ permutes $\Delta
. $ Finite collection of such reflections forms a group $W$ called \textit{%
Weyl group of }$\mathit{\Delta }$\textit{\ }. The vectors $\alpha ^{\vee }$
(for $\alpha \in \Delta )$ form a root system $\Delta ^{\vee }$ \textit{dual}
to $\Delta $. Let $v\in \Delta $ be such that $<v$,$\alpha >\neq 0$ for each 
$\alpha \in \Delta $. Then, the set $\Delta ^{+}$ of roots $\alpha \in
\Delta $ such that $<v$,$\alpha >>0$ is called a system of \textit{positive}
roots of $\Delta $. A root $\alpha \in \Delta ^{+}$\ is \textit{simple} if
it is not a sum of two elements from $\Delta ^{+}.$The number of simple
roots coincides with the dimension of the vector space $V$ and the root set $%
\Delta $ is made of disjoint union $\Delta =\Delta ^{+}\amalg \Delta ^{-}$.
The integral linear combinations of roots, i.e. $\sum\limits_{i}m_{i}\alpha
_{i}$ \ with $m_{i}^{\prime }s$ being integers, forms a root \textit{lattice}
$Q$($\mathcal{\Delta )}$ in $V$ (that is free abelian group of rank $n=\dim
V).$ Clearly, the simple roots form a basis $\Sigma $ of $Q$($\mathcal{%
\Delta )}$. Accordingly, $Q$($\mathcal{\Delta }^{+}\mathcal{)}$ is made of
combinations $\sum\limits_{i}m_{i}\alpha _{i}$ with $m_{i}^{\prime }s$ being
nonnegative integers.

In view of one-to -one correspondence between the set of hyperplanes $\cup
_{\alpha }$H$_{\alpha }$ and the set of roots $\Delta $ it is convenient
sometimes to introduce \textbf{chambers} as connected components of the
complement of $\cup _{\alpha }$H$_{\alpha }$ in $V$. In the literature, Ref.
[83], page 70, this complement is known also as the \textit{Tits cone}%
.Accordingly, for a given chamber $C_{i}$ its \textbf{walls} are made of
hyperplanes H$_{\alpha }.$The roots in $\Delta $ can therefore be
characterized as those roots which are orthogonal to some wall of $C_{i}$
and directed towards interior of this chamber. A \textbf{gallery }is a
sequence \ ( $C_{0}$, $C_{1}$,..., $C_{\mathit{l}}$\ )\ of chambers each of
which is adjacent to and distinct from the next. Let $%
w=r_{i_{1}}...r_{i_{l}} $ then, treating the Weyl group $W$ as a chamber
system, a gallery from 1 to $w$ can be formally written as ($1$, $%
r_{i_{1}},r_{i_{1}}r_{i_{2}},...,r_{i_{1}}...r_{i_{l}})$. If this gallery is
of the shortest possible $lenght$ $\mathit{l}$($w)$ then one is saying that $%
r_{i_{1}}...r_{i_{l}}$ is \textit{reduced decomposition} for the word $w$
made of ''letters'' $r_{i_{j}}$. Let $C_{x}$ and $\ C_{y}$ be some distinct
chambers which we shall call $x$ and $y$ for brevity. One can introduce the
distance function $d(x,y)$ so that, for example, if $w=r_{i_{1}}...r_{i_{l}}$
is the reduced decomposition, then $d(x,y)=w$ if and only if there is a
gallery of the type $r_{i_{1}}...r_{i_{l}}$ from $x$ to $y$. If, for
instance, $d(x,y)=r_{i}$ , this means simply that $x$ and $y$ are distinct
and $i-$adjacent. A \textbf{building} $\mathcal{B}$ is a chamber system
having distance function $d(x,y)$ taking values in Weyl-Coxeter group $W$.
Finally, an \textbf{apartment} in a building $\mathcal{B}$ is a subcomplex $%
\mathcal{\hat{B}}$ of $\mathcal{B}$ which is isomorphic to W. That is there
is a bijection $\varphi :$ $W\rightarrow \mathcal{\hat{B}}$ such that $%
\varphi (w)$ and $\varphi (w^{\prime })$ are $i-$adjacent in $\mathcal{\hat{B%
}}$ if and only if $w$ and $w^{\prime }$ are adjacent in $W$, e.g. see
Ref.[104].

\bigskip

b) \textit{The} \textit{Coxeter group}

The Coxeter group is related to the Weyl group through the obviously looking
type of relation between reflections 
\begin{equation}
\left( r_{\alpha }r_{\beta }\right) ^{m(\alpha ,\beta )}=1  \tag{A.3}
\end{equation}
where, evidently, $m(\alpha ,\alpha )=1$ and $m(\alpha ,\beta )\geq 2$ for $%
\alpha \neq \beta $. In particular, for \textit{finite} Weyl groups\ \ $%
m(\alpha ,\beta )\in \{2,3,4,6\},$Ref. [85], page39, while for the affine
Weyl groups (to be discussed below) $m(\alpha ,\beta )\in \{2,3,4,6,\infty
\} $, e.g. read Ref.[85], page 136, and Proposition A.1$.$below. Clearly,
different refection groups will have different matrix $m(\alpha ,\beta )$
and, clearly, the matrix $m(\alpha ,\beta )$ is connected with bilinear form
(Cartan matrix, see below) for the Weyl's group $W$ [105].

As an example of \ use of the concept of building in the Weil group,
consider the set of \textit{fundamental} \textit{weights} defined as
follows.\ For the root basis $\Sigma $ (or $\Sigma ^{\vee })$ the set of
fundamental weights $\mathcal{D}=$\{$\omega _{1},...,\omega _{n}\}$\textit{\
with respect to} $\Sigma $ is defined by the rule: 
\begin{equation}
<\alpha _{i}^{\vee },\omega _{j}>=\delta _{ij}.  \tag{A.4}
\end{equation}
The usefulness of such defined fundamental weights lies in the fact that
they allow to introduce the concept of the \textit{highest weight }$\lambda $
\textit{(}sometimes also known as dominant weight, [106] page 203). Thus
defined $\lambda $ can be presented as $\lambda
=\sum\nolimits_{i=1}^{d}a_{i}\omega _{j}$ with all $a_{i}\geq 0.$ Sometimes
it is convenient to relax the definition of fundamental weights to just
weights by comparing Eq.s(A.2) and (A.4). That is $\beta ^{\prime }s$ in
Eq.(A.2) are just weights. Thus, for instance, we have $\Delta $ as building
and subcomplex $\mathcal{D}$ of fundamental weights as an apartment complex.

To illustrate some of these concepts let us consider examples which are
intuitively appealing and immediately relevant to the discussion in the main
text. These are the root system $B_{d}$ and $C_{d}.$ They are made of vector
set $\{u_{1},...,u_{d}\}$ constituting an orthonormal basis of the $d-$%
dimensional cube. The vectors $u_{i}$ should not be necessarily of unit
length, Ref. [84], page 27. It is important only that they all have the same
length. For B$_{d}$ system one normally chooses, Ref. [84], page 30, 
\begin{equation}
\Delta =\{\pm u_{i}\pm u_{j}\mid i\neq j\}\amalg \{\pm u_{i}\}.  \tag{A.5}
\end{equation}
In this case, the reflections corresponding to elements of $\Delta $ can be
described by their effect on the set $\{u_{1},...,u_{d}\}.$ Specifically, 
\begin{eqnarray*}
r_{u_{i}-u_{j}} &=&\text{permutation which interchanges }u_{i}\text{ and u}%
_{j}; \\
r_{u_{i}} &=&\text{sign change of }u_{i}; \\
r_{u_{i}+u_{j}} &=&\text{permutation which interchanges }u_{i}\text{ and u}%
_{j}\text{ and changes their sign.}
\end{eqnarray*}
The action of the Weyl group on $\Delta $ can be summarized by the following
formula 
\begin{equation}
W(\Delta )=\left( \mathbf{Z}/2\mathbf{Z}\right) ^{d}\unlhd \Sigma _{d} 
\tag{A.6}
\end{equation}
with \ $\unlhd $ representing the semidirect product between the permutation
group $\Sigma _{d}$ and the dihedral group $\left( \mathbf{Z}/2\mathbf{Z}%
\right) ^{d}$ of sign changes both acting on $\{u_{1},...,u_{d}\}.$Thus
defined product constitutes the full symmetry group of the $d-$cube, Ref.
[84], page 31. The same symmetry information is contained in $C_{d}$ root
system defined by 
\begin{equation}
\Delta =\{\pm u_{i}\pm u_{j}\mid i\neq j\}\amalg \{\pm 2u_{i}\}  \tag{A.7}
\end{equation}
Both systems possess the same root decomposition: $\Delta =\Delta ^{+}\amalg
\Delta ^{-}$ \ [84], page 37 . In particular, considering a square as an
example we obtain the basis $\Sigma _{B_{2}}$ of $Q$($\mathcal{\Delta )}$ as 
\begin{equation}
\Sigma _{B_{2}}=\{u_{1}-u_{2},u_{2}\}.  \tag{A.8a}
\end{equation}
From here the dual basis is given by 
\begin{equation}
\Sigma _{B_{2}}^{\vee }=\{u_{1}-u_{2},2u_{2}\}.  \tag{A.8b}
\end{equation}
Using Eq.(A.4) we obtain the fundamental weights as $\omega _{1}=u_{1}$ \
and $\omega _{2}=\frac{1}{2}(u_{1}+u_{2})$ respectively which by design obey
the orthogonality condition , Eq.(A.4). The Dynkin diagram, Ref. [84], page
122, for $B_{2}$ provides us with coefficients \ $a_{1}=1$ and $a_{2}$=$2$
obtained for the case when expansion $\lambda
=\sum\nolimits_{i=1}^{d}a_{i}\omega _{j}$ is relaxed to $\lambda
=\sum\nolimits_{i=1}^{d}a_{i}\beta _{j}$ as discussed above. In view of
Eq.(A.8a) this produces at once: $\lambda _{B_{2}}=u_{1}+u_{2}.$
Analogously, for C$_{2}$ we obtain: 
\begin{equation}
\Sigma _{C_{2}}=\{u_{1}-u_{2},2u_{2}\}.  \tag{A.9}
\end{equation}
with coefficients $a_{1}=2$ and $a_{2}$=$2$ thus leading to $\lambda
_{C_{2}}=2\left( u_{1}+u_{2}\right) .$

For the square, these results are intuitively obvious. Evidently, the $d-$
dimensional case can be treated accordingly. The physical significance of
the highest weight should become obvious if one compares the Weyl-Coxeter
reflection\ group algebra with that for the angular momentum familiar to
physicists. In the last case, the highest weight means simply the largest
value of the projection of the angular momentum onto z-axis. The raising
operator will annihilate the wave vector for such quantum state while the
lowering operator will produce all the eigenvalues lesser than the maximal
value (up to the largest negative) and, naturally, all the eigenfunctions.
The significance of the fundamental weights goes beyond this however.
Indeed, suppose we can expand some root $\alpha _{i}$ according to the rule 
\begin{equation}
\alpha _{i}=\sum\limits_{j}m_{ij}\omega _{j}.  \tag{A.10}
\end{equation}
Then, substitution of such expansion into Eq.(A.2) and use of Eq.(A.4)
produces 
\begin{equation}
<\alpha _{k}^{\vee },\alpha _{i}>=\sum\limits_{j}m_{ij}<\alpha _{k}^{\vee
},\omega _{j}>=m_{ik}  \tag{A.11}
\end{equation}
The expression $<\alpha _{k}^{\vee },\alpha _{i}>$ is known in the
literature as Cartan matrix. It plays the central role in defining both
finite and infinite dimensional semisimple Lie algebras [12]. According to
Eq.s(A.4),(A.10),(A.11), the transpose of the Cartan matrix transforms the
fundamental weights into the fundamental roots.

\bigskip

c) \textit{The} \textit{affine Weyl-Coxeter groups}

Physical significance of the affine Weyl-Coxeter reflection groups comes
from the following proposition

\textbf{Proposition A.1}. \textit{Let W be the Weyl group of any Kac-Moody
algebra. Then W is a Coxeter group for which }$m(\alpha ,\beta )\in
\{2,3,4,6,\infty \}.$\textit{\ Any Coxeter group with such }$m(\alpha ,\beta
)$\textit{\ is crystallographic (e.g. see Eq.(A.2) )}

The proof can be found in Ref.[105], pages 25-26.

To understand better the affine Weyl-Coxeter groups, following Coxeter,
Ref.[22] , we would like to explain in simple terms the origin and the
meaning of these groups. It is being hoped, that such discussion might \
significantly facilitate understanding of the results presented in the main
text.

We begin with quadratic form 
\begin{equation}
\Theta =\sum\limits_{i,j}a_{ij}x_{i}x_{j}  \tag{A.12}
\end{equation}
with symmetric matrix $\left\| a_{ij}\right\| $ whose rank is $\rho .$ Such
form is said to be \textit{positive definite }if it is positive for all
values of $\mathbf{x=\{}x_{1},...,x_{n}\}$ ( $n\geq \rho $ in general !)
except zero. \ It is \textit{positive semidefinite} if it is never negative
but vanishes for some $x_{i}^{\prime }s$ not all zero. The form $\Theta $ is
indefinite if it can be both positive for some $x_{i}^{\prime }s$ and
negative for others.\footnote{%
For the purposes of comparison with existing mathematical physics literature
[12] it is suficient to consider only positive and positive semidefinite
forms.} If positive semidefinite form vanishes for some $x_{i}=z_{i}$ $%
(i=1-n)$, then 
\begin{equation}
\sum\limits_{i}z_{i}a_{ij}=0,\text{ }j=1-n.  \tag{A.13}
\end{equation}
For a given matrix $\left\| a_{ij}\right\| $ Eq.(A.13) can be considered as
system of linear algebraic equations for $z_{i}^{\prime }s.$ Let $\mathcal{N}%
=n-\rho $ be the \textit{nullity} of the form $\Theta $. Then, it is a
simple matter to show that \textit{every positive semidefinite connected }$%
\Theta $\textit{\ form is of nullity 1}. The form is connected if it cannot
be presented as a sum of two forms involving separate sets of variables. The
following two propositions play the key role in causing differences between
the infinite affine Weyl-Coxeter (Kac-Moody) algebras and their finite
counterparts discussed in subsections a) and b).

\bigskip

\textbf{Proposition A.2}. \textit{For any positive semidefinite connected }$%
\Theta $\textit{\ form there exist \ unique (up to multiplication by the
common constant)\textbf{\ positive} numbers z}$_{i}$\textit{\ satisfying
Eq.(A.13).}

\textbf{Proposition A.3. }\textit{If we modify a positive semidefinite
connected \ }$\Theta $ \textit{form by making one of the variables vanish,
the obtained form becomes positive definite.}

\bigskip

Next, we consider the quadratic form $\Theta $ as the norm and the matrix $%
a_{ij}$ as the metric tensor. Then, as usual, we have $\mathbf{x}\cdot 
\mathbf{x=}$ $\Theta =\left| \mathbf{x}\right| ^{2}$ and, in addition, $%
\mathbf{x}\cdot \mathbf{y=}\sum\limits_{i,j}a_{ij}x_{i}y_{j}$ $\equiv
\sum\limits_{i}x^{i}y_{i}=\sum\limits_{i}x_{i}y^{i}$ so that if vectors $%
\mathbf{x}$ and $\mathbf{y}$ are orthogonal we get $\sum%
\limits_{i,j}a_{ij}x_{i}y_{j}=0$ as required. Each vector \textbf{x}
determines a point (\textbf{x}) and a hyperplane [\textbf{x}] with respect
to some reference point \textbf{0} chosen as origin. The distance $\emph{l}$
between a point (\textbf{x}) and a hyperplane [\textbf{y}], measured along
the perpendicular, is the projection of \textbf{x} along the direction of 
\textbf{y}, i.e. 
\begin{equation}
\emph{l}=\frac{\mathbf{x}\cdot \mathbf{y}}{\left| \mathbf{y}\right| }. 
\tag{A.14}
\end{equation}
Let now (\textbf{x}$^{\prime })$ be the image of (\textbf{x}) by reflection
in the hyperplane [\textbf{y}]. Then, $\mathbf{x}-\mathbf{x}^{\prime }$ is a
vector parallel to \textbf{y} of magnitude 2\emph{l.} Thus 
\begin{equation}
\mathbf{x}^{\prime }=\mathbf{x}-2\frac{\mathbf{x}\cdot \mathbf{y}}{\left| 
\mathbf{y}\right| }\mathbf{y}  \tag{A.15}
\end{equation}
in accord with Eq.(A.1). From here, equation for the reflecting hyperplane
is just $\mathbf{x}\cdot \mathbf{y=}0\mathbf{.}$ Let the vector \textbf{y}
be pre assigned then, taking into account Propositions A2 and A3 we conclude
that for the nullity \ $\mathcal{N}$=0 the only solution possible is \textbf{%
x}=\textbf{0. }That is to say in such case $n$ reflecting hyperplanes have
point \textbf{0 }as the only\textbf{\ }common intersection point. Complement
of these hyperplanes in \textbf{R}$^{n}$ forms a chamber system\textbf{\ }%
discussed already in a). For $\mathcal{N}$=1 equation $\mathbf{x}\cdot 
\mathbf{y=}0$ may have many \textit{nonnegative} solutions for \textbf{x. }%
Actually\textbf{, }such reflecting hyperplanes occur in a finite number of
different directions. More accurately, such hyperplanes belong to \textit{%
finite} number of families, each consisting of hyperplanes \textit{parallel
\ to each other}. If we choose a single representative from each family in a
such a way that it passes through \textbf{0}, then complement of such
representatives is going to form a polyhedral cone as before. But now, in
addition, we have a group of translations $\mathit{T}$ for each
representative of hyperplane family so that the total affine Weyl group $%
W_{aff\text{ }}$ is the semidirect product :$W_{aff\text{ }}$ =$T\unlhd W.$
The fundamental region for $W_{aff\text{ }}$ is a simplex (to be precise, an
open simplex, Bourbaki, Ref.[7], Chr.5, Proposition 10) \ called \textbf{%
alcove }bounded by $n+1$ hyperplanes (walls) $n$ of which are reflecting
hyperplanes passing through \textbf{0} while the remaining one serves to
reflect \textbf{0} into another point $\mathbf{0}^{\prime }.$ If one
connects \textbf{0} with $\mathbf{0}^{\prime }$ and reflects this line in
other hyperplanes one obtains a lattice. By analogy with solid state physics
[36] one can construct a dual lattice (just like in a) and b) \ above) \ the
fundamental cell of which is known in physics as the Brilluin zone. For the
alcove the fundamental region of the dual lattice ( the Brilluin zone) is
the polytope having \textbf{0} for its centre of symmetry, i.e. zonotope[22].

d) \textit{The} \textit{pseudo-reflection groups}

Although the pseudo-reflection groups are also described by Bourbaki,Ref.
[7], their geometrical (and potentially physical) meaning is beautifully
explained in the book by McMullen [83]. In articular, all \ earlier
presented reflection groups are isometries of Euclidean space. Their action
preserves some quadratic form which is real. More generally, one can think
of \ reflections in spherical and hyperbolic spaces. From this point of view
earlier described polytopes (polyhedra) represent fundamental regions for
respective isometry groups. Action of these groups on fundamental regions
causes tesselation of these spaces (without gaps). The collection of spaces
can be enlarged by considering reflections in complex n-dimensional space 
\textbf{C}$^{n}.$ In this case the Euclidean quadratic form is replaced by
the positive definite Hermitian form. Since locally \textbf{CP}$^{n}$ is the
same as $\mathbf{C}^{n+1}$ and since \textbf{CP}$^{n}$ is at the same time a
symplectic manifold with well known symplectic two form $\Omega $ [43], this
makes the pseudo-reflection groups (which leave $\Omega $ invariant)
especially attractive for physical applications. This is indeed the case as
the main text indicates. The pseudo-reflections are easily described. By
analogy with Eq.(A.1) (or (A.15)) one writes 
\begin{equation}
r_{\alpha }(v)=v+(\xi -1)<v,\alpha ^{\vee }>\alpha  \tag{A.16}
\end{equation}
where $\xi $ is nontrivial solution of the cyclotomic equation $x^{h}=x$ and 
$\alpha ^{\vee }=\alpha /<\alpha ,\alpha >$ with $<x,y>$ being a positive
definite Hermitian form satisfying as before $<r_{\alpha }(v),r_{\alpha
}(\mu )>=$\ $<\nu ,\mu >$ and $\alpha $ being an eigenvector such that $%
r_{\alpha }(\alpha )=\xi \alpha \footnote{%
According to Bourbaki, Ref. [7], Chr5, paragraph 6, if $\xi $ is an
eigenvalue of psedo-reflection operator, then $\xi ^{-1}$ is also an
eigenvalue with the same multiplicity.}.$ In addition, $\left[ r_{\alpha
}(\alpha )\right] ^{k}=\xi ^{k}\alpha $ for $1\leq k\leq h-1.$ This follows
from the fact that 
\begin{equation}
\left[ r_{\alpha }(\nu )\right] ^{k}=\nu +(1+\xi +\cdot \cdot \cdot +\xi
^{k-1})(\xi -1)<v,\alpha ^{\vee }>\alpha  \tag{A.17}
\end{equation}
and taking into account that $(1+\xi +\cdot \cdot \cdot +\xi ^{k-1})(\xi
-1)=\xi ^{k}-1.$

Finally, the Weyl-Coxeter reflection groups considered earlier can be
treated exactly as pseudo-reflection groups if one replaces a single
Euclidean reflection by the so called Coxeter element [64] \ $\omega $ which
is product of individual reflections belonging to the distinct roots of $%
\Delta .$ Hence, the Euclidean Weyl-Coxeter reflection groups can be
considered as subset of pseudo-reflection groups so that useful information
about these groups can be obtained from considering the same problems for
the pseudo-reflection groups. It can be shown [64,84] that the Coxeter
element $\omega $ has eigenvalues $\xi ^{m_{1}},...,\xi ^{m_{l}}$ with \emph{%
l} being dimension of the vector space $\Delta $ while the exponents $%
m_{1},...,m_{l}$ are positive integers less than $h$ and such that $%
\sum\nolimits_{i=1}^{l}(h-m_{i})=\sum\nolimits_{i=1}^{l}m_{i}$ . This result
implies that the number $\sum\nolimits_{i=1}^{l}m_{i}=N$ - the number of
positive roots in the Weyl-Coxeter group is connected with the Coxeter
number $h$ via relation : $N=\frac{1}{2}lh,$Ref$.[$85$],$page79.

\bigskip

\textbf{Appendix B.} \textbf{Analytical properties of the Veneziano and
Veneziano-like 4- particle amplitudes.}

\bigskip

a) \textit{The} \textit{Veneziano amplitudes}

Using Eq.(4.6) the four-particle Veneziano amplitude is given by 
\begin{equation}
A(s,t,u)=\Gamma (-\alpha (s))\Gamma (-\alpha (t))\Gamma (-\alpha (u))[\sin
\pi (-\alpha (s))+\sin \pi (-\alpha (t))+\sin \pi (-\alpha (u))]  \tag{B.1}
\end{equation}
To analyze the analytical properties of this amplitude we need to use the
following known expansions 
\begin{equation}
\sin \pi z=\pi z\prod\limits_{k=1}^{\infty }(1-(\frac{z}{k}))(1+(\frac{z}{k}%
))  \tag{B.2}
\end{equation}
and 
\begin{equation}
\frac{1}{\Gamma (z)}=ze^{-Cz}\prod\limits_{k=1}^{\infty }(1+(\frac{z}{k}%
))e^{-\dfrac{z}{k}}  \tag{B.3}
\end{equation}
with $C$ being Euler's constant 
\begin{equation*}
C=\lim_{n\rightarrow \infty }(1+\frac{1}{2}+\frac{1}{3}+...+\frac{1}{n}-\ln
n).
\end{equation*}

The above results \ when combined with the Veneziano condition, $\alpha
(s)+\alpha (t)+\alpha (u)=-1,$Eq$.(4.5),$ allow us to write (up to a
constant factor) a typical singular portion of the Veneziano amplitude (the
tachyons are to be considered separately): 
\begin{eqnarray}
A(s,t,u) &=&\frac{1}{nlm(1-\frac{\alpha (s)}{n})}\frac{1}{(1-\frac{\alpha (t)%
}{m})}\frac{1}{(1-\frac{\alpha (u)}{l})}\times  \notag \\
&&[(1-\frac{\alpha (s)}{n})C(n)+(1-\frac{\alpha (t)}{m})C(m)+(1-\frac{\alpha
(u)}{l})C(l)]  \TCItag{B.4}
\end{eqnarray}
where $C(n)$, etc. are known constants and $m,n,l$ as some nonnegative
integers. For actual use of this result the explicit form of these constants
might be important. Looking at Eq.(B.2) we obtain, 
\begin{equation}
C(n,\alpha )=\pi \alpha \frac{1}{(1-\dfrac{\alpha }{n})}\prod\limits_{k=1}^{%
\infty }(1-(\frac{\alpha }{k}))(1+(\frac{\alpha }{k})).  \tag{B.5}
\end{equation}
where $\alpha $ can be $\alpha (s)$, etc. Clearly, this definition leads to
further simplifications, e.g. to the manifestly symmetric form:

\bigskip\ 
\begin{eqnarray}
A(s,t,u) &=&\frac{1}{nlm}[\frac{C(n,\alpha (s))}{(1-\frac{\alpha (t)}{m})}%
\frac{1}{(1-\frac{\alpha (u)}{l})}  \notag \\
&&+\frac{C(m,\alpha (t))}{(1-\frac{\alpha (s)}{n})}\frac{1}{(1-\frac{\alpha
(u)}{l})}  \notag \\
&&+\frac{C(l,\alpha (u))}{(1-\frac{\alpha (s)}{n})}\frac{1}{(1-\frac{\alpha
(t)}{m})}].  \TCItag{B.4.a}
\end{eqnarray}
Consider now special case : $\alpha (s)=\alpha (t)=n.$ In this case we
obtain: 
\begin{eqnarray}
A(s &=&t,u)=\frac{1}{n^{2}m}\frac{1}{(1-\frac{\alpha (s)}{n})^{2}}%
[C(l,\alpha (u))+2C(n,\alpha (s))\frac{(1-\frac{\alpha (s)}{n})}{(1-\frac{%
\alpha (u)}{l})}  \notag \\
&=&\frac{1}{n^{2}m}\frac{1}{(1-\frac{\alpha (s)}{n})^{2}}[\sin \pi \alpha
(u)+2\sin \pi \alpha (s)]\frac{1}{(1-\frac{\alpha (u)}{l})}=0  \TCItag{B.4b}
\end{eqnarray}
This result is in accord with that of Ref. [62] where it was obtained
differently. The tachyonic case is rather easy to consider now. Indeed,
using Eq.s(B.1)-(B.3) and taking into account the Veneziano condition, let
us assume that, say, that $\alpha (s)=0$ then, in view of symmetry of
Eq.s.(B.4a),(B.4b), we need to let $\alpha (t)=0$ as well to check if
Eq(B.4b) holds. This leaves us with the option: $\alpha (u)=-1.$With such
constraint we obtain $($ since $\Gamma (1)=1)$%
\begin{equation*}
A(s,t,u)=\frac{\pi }{\alpha (t)}+\frac{\pi }{\alpha (s)}-\frac{\pi }{\alpha
(t)}-\frac{\pi }{\alpha (s)}=0
\end{equation*}
as required. Hence, indeed, even in the tachyonic case Eq.(B.4b) holds in
accord with earlier results [62]. This means only that one cannot observe
the tachyons in both channels simultaneously. But even to observe them in
one channel is unphysical. Moreover, Eq.(B.4b) implies that only situations
for which $\alpha (s)\neq \alpha (t)\neq \alpha (u)$ could be physically
observable. By combining the Veneziano condition with such constraint leaves
us with the following options:

a) $\alpha (s),\alpha (t)>0,\alpha (u)<0$;

b) $\alpha (s)>0,\alpha (t),\alpha (u)<0$ plus the rest of cyclically
permuted inequalities.

This means that not only the tachyons of the type $\alpha (s)=0$ (or $\alpha
(t)=0$ , or $\alpha (u)=0$ ) could be present but also those for which, for
example, $\alpha (s)<0.$ This is so because according to results of standard
open string theory [55] in 26 space-time dimensions $\alpha (s)=1+\frac{1}{2}%
s$. When $\alpha (s)=0$ such convention produces the only one tachyon: $s=-2$
$=M^{2}$ and the whole spectrum (open string) is given by $%
M^{2}=-2,0,2,...,2n$ where $n$ is non negative integer. Incidentally, for
the closed bosonic string under the same conditions the spectrum is known to
be $M^{2}=-8,0,8,...,8n$. No other masses are permitted. \ The requirements
like those in a) and b) above produce additional complications. For
instance, let $\alpha (s)=1,$then consider the following option : $\alpha
(t)=3$ so that $\alpha (u)=-5.$ This leads us to the tachyon mass : $M^{2}$ $%
=-12$ . It is absent in the spectrum of both types of bosonic strings. The
emerging \ apparent difficulty can actually be bypassed somehow due to the
following chain of arguments. Consider, for example, the amplitude $V(s,t)$
and let both $s$ and $t$ be non tachyonic and, of course, $\alpha (s)\neq
\alpha (t)$ . Then, naturally, $\alpha (u)<0$ \textit{is} tachyonic but,
when we use Eq.(B.4a), we notice at once that this creates no difficulty
since $\alpha <0$ condition simply will eliminate the resonance in the
respective channels, e.g. if $V(s,t)$ will have poles for both $s$ and $t$ 
\textit{then} \textit{the same particle resonances will occur} in $V(s,u)$
and $V(t,u)$ channels so that, \textit{except the case} $\alpha (s)=0$
leading to the pole with mass $M^{2}=-2,$ no other tachyonic states will
show up as resonances and, hence, they cannot be observed. At the same time, 
\textit{effectively}, we have only two types of resonances in the Veneziano
amplitude which we would like to denote symbolically (up to permutation) as$%
\mathcal{V}_{u}(s)$ and $\mathcal{V}_{u}(t)$. The resonances for $\mathcal{V}%
_{u}(s)=$ $V(s,t)+$ $V(s,u)$ and, accordingly, $\mathcal{V}%
_{u}(t)=V(t,s)+V(t,u)$. Such conclusion is valid only if one requires $%
V(s,t)=V(t,s)$. It is surely the case for the Veneziano amplitude,
Eq.(B.4a). Accordingly, should Veneziano amplitude be free of tachyons (e.g. 
$\alpha (s)=0)$ it would be perfectly acceptable. Evidently, in the light of
the results just obtained it can be effectively written as 
\begin{equation}
A(s,t,u)=\mathcal{V}_{u}(s)+\mathcal{V}_{u}(t).  \tag{B.5}
\end{equation}
This result survives when instead of Veneziano we would like to use the
Veneziano-like amplitudes obtained in the main text as demonstrated below.

\bigskip

b) \textit{The} \textit{Veneziano-like amplitudes}

Results of Section 4.4 allow us to write the following result for the 4-
particle Veneziano-like amplitude, e.g. see Eq.s(4.46),(4.47), 
\begin{equation}
A(s,t,u)=V(s,t)+V(s,u)+V(t,u),  \tag{B.6}
\end{equation}
where, for instance, 
\begin{equation}
V(s,t)=(1-\exp (i\frac{\pi }{N}(-\alpha (s))(1-\exp (i\frac{\pi }{N}(-\alpha
(t))B(-\frac{\alpha (s)}{N},-\frac{\alpha (t)}{N}).  \tag{B.7}
\end{equation}
Although this result was obtained by the same analytic continuation as in
the case of Veneziano amplitude, the \ resulting analytical properties of
such Veneziano-like amplitude are markedly different. as we would like to
demonstrate now.

We begin by noticing that in view of Eq.(4.41b) the Veneziano condition in
its simplest form: $a+b+c=1,$ upon analytic continuation leads again to the
requirement: $\alpha (s)+\alpha (t)+\alpha (u)=-1$ if we identify, for
example, $\frac{<c_{1}>}{N}$ $=a_{1}$\ with -$\alpha (s),$etc. This naive
identification leads to some difficulties however. Indeed, since physically
we are interested in poles and zeros of gamma functions as results of
previous subsection indicate, we expect our parameters $a,b$ and $c$ to be
integers. This is possible only if absolute values of $<c_{1}>,<c_{2}>$ and $%
<c_{3}>$ are greater or equal than $N$. By allowing these parameters to
become greater than $N$ we would formally violate the requirements of
Corollary 2.11. by Griffiths, e.g. see Eq.(4.39) and discussion around it.
Fortunately, the occurring difficulty can by resolved. For instance, one can
postulate Eq.s (B.6),(B.7) as \textit{defining equations} for the
Veneziano-like amplitudes as it was done by Veneziano for Veneziano
amplitudes. In this case one is confronted with the problem of finding the
physical model producing such amplitudes. To facilitate search for such
model it is reasonable to impose the same constraints as for the standard
Veneziano amplitudes in the present case. Clearly, if we want to use the
results of the main text, in addition to these constraints, the constraint
coming from Corollary 2.11. should also be imposed. Corollary 2.11. formally
forbids us from consideration of ratios $<c_{i}>/N$ \ whose absolute value
is greater than one as it was discussed in Section 4.4.4. This fact creates
no additional problems however. This can be seen already on example of
Eq.(B.2) for $\sin \pi x$. Indeed, consider the function 
\begin{equation*}
F(x)=\frac{1}{\sin \pi x}.
\end{equation*}
It will have the first order poles for $x=0,\pm 1,\pm 2,...$ If we define
the bracket operator $<...>$ by analogy with that defined in the main text,
e.g. $\ 0<<x>\leq 1$ $\forall x,$ then to reproduce the poles of $F(x)$ it
is sufficient to write 
\begin{equation}
F(x)=\frac{1}{\sin \pi x}=\frac{1}{1-<x>}.  \tag{B.8a}
\end{equation}
Clearly, the above result can be read as well from right to left, i.e.
removal of brackets, is equivalent to unwrapping\footnote{%
That is going to the universal covering space} $S^{1}$ into $R^{1}$. By
looking at Eq.(B.3) for $\Gamma (z)$ and comparing it with Eq.(B.2) we
notice that all singularities of $\Gamma (z)$ are exactly the same as those
for $F(x)$. Hence, the same unwrapping is applicable for this case as well.

These observations lead us to the following set of prescriptions: a) use
Eq.(4.45) in Eq.(4.46) in order to obtain the full Veneziano-like amplitude,
b) remove brackets, c) perform analytic continuation to negative values of $%
c_{i}^{\prime }s$ ,d) identify $-c_{i}/N$ with $-\alpha (i),($ $i=s$, $t$ or 
$u$). After this, let, for instance, $\alpha (s)=a+bs$ with both $a$ and $b$
being positive (or better, non negative) constants. Then, the tachyonic
pole: $\alpha (s)=0,n=0$ is killed by the corresponding phase factor in
Eq.(B.7). The mass spectrum is determined by a) the actual numerical values
of constants $a$ and $b,$ b) by the phase factors and c) by the values of
parameter $N$ (even or odd).

For instance, the condition, Eq.(4.41b), leads to the requirement that the
particle masses satisfying equation $\alpha (s)=2l,l=1,2,...$ cannot be
observed since the emerging pole singularities are being killed by zeroes
coming from the phase factor. In the case of 4-particle amplitude Eq.(4.61)
should be used with $n=1$ thus leading to constraints: $N\geq 3$ and $1\leq
m\leq 3.$ If we choose $m=3$ we obtain similar requirement forbidding
particles with masses coming from equation $\alpha (s)=3l,l=1,2,...$.

Such limitations are not too severe, however. Indeed, let us consider for a
moment the existing bosonic string parameters associated with the Veneziano
amplitude. For the open string the convention is $\alpha (s)=1+\frac{1}{2}s$
so that the tachyon state is determined by the condition: $\alpha (s)=0$
producing $s=-2$ $=M^{2}$. \ If now $1+\frac{1}{2}s=l,$ we obtain: $%
s=2(l-1), $ $l=1,3,5,..$.(for $m=2$) or $l=1,2,4,5,..$.(for $m=3$).Clearly,
these results produce the mass spectrum of open bosonic string (without
tachyons). If we want the graviton to be present in the spectrum we have to
adjust the values of constants $a$ and $b$. For instance, it is known [55]
that for closed string the tachyon occurs at $s=-8$ $=M^{2}$ . This result
can be achieved either if we choose $\alpha (s)=2+\frac{1}{4}s$ or $\alpha
(s)=1+\frac{1}{8}s.$ To decide which of these two expressions fits better
experimental data we recall that the massless graviton should have spin
equal to two. If we want the graviton to be present in the spectrum we must
select the first option. This is so because of the following arguments.
First, we have to take into account that for large $s$ and fixed $t$ the
amplitude $V(s,t)$ can be approximated by [55], page 10, 
\begin{equation}
V(s,t)\simeq \Gamma (-\alpha (t))(-\alpha (s))^{\alpha (t)}  \tag{B.9}
\end{equation}
while the Regge theory predicts [55], pages 3,4, that 
\begin{equation}
V_{J}(s,t)=-\frac{g^{2}(-s)^{J}}{t-M_{J}^{2}}\simeq \frac{-g^{2}(-\alpha
(s))^{J}}{\alpha (t)-J}  \tag{B.10}
\end{equation}
for the particle with spin $J$. This leaves us with the first option.
Second, by selecting this option our task is not complete however since so
far we have ignored the actual value of the Fermat parameter $N$. Such
ignorance causes emergence of the fictitious tachyon coming from the
equation $2+\frac{1}{4}s=l$ for $l=1$.This difficulty is easily resolvable
if we take into account that the ''Shapiro-Virasoro'' condition, Eq.(4.41b),
is reducible to the ''Veneziano condition, Eq.(4.41a), in case if all $%
<c_{i}>$ in Eq.(4.41b) are even. At the same time, if $N$ in Eq.(4.41a) is
even, it can be brought to the form of Eq.(4.41b). Hence, in making
identification of $-c_{i}/N$ with $-\alpha (i)$ we have to consider two
options: a) $N$ is odd, then $c_{i}/N=\alpha (i)$, b) $N$ is even, $N=2\hat{N%
}$ , then $c_{i}/\hat{N}=\alpha (i).$The fictitious tachyon is removed from
the spectrum in case if we choose the option b). Clearly, after this, in
complete analogy with the ''open string'' case we re obtain tachyon-free
spectrum of the ''closed'' bosonic string.

To complete our investigation of the Veneziano-like amplitudes we still need
some discussion related to Eq.s(B.9),(B.10). To this purpose, using the
integral representation of $\Gamma $ given by 
\begin{equation*}
\Gamma (x)=\int\limits_{0}^{\infty }\frac{dt}{t}t^{x}\exp (-t)
\end{equation*}
and assuming that $x$ is large and positive the leading term saddle point
approximation is readily obtained so that we obtain the leading term 
\begin{equation*}
\Gamma (x)=Ax^{x}\exp \left( -x\right) ,
\end{equation*}
where $A$ is some constant. Applying \ with some caution this result to 
\begin{equation*}
V(s,t)=\frac{\Gamma (-\alpha (s))\Gamma (-\alpha (t))}{\Gamma (-\alpha
(s)-\alpha (t))}\equiv B(-\alpha (s),-\alpha (t))
\end{equation*}
we obtain Eq.(B.9)\footnote{%
This result is more carefuly reobtained below.}. Although such arguments
formally explain the origin of the Regge asymptotic law, Eq.(B.9), they do
not illuminate the combinatorial origin of this result essential for its
generalization. To correct this deficiency, following Hirzebruch and Zagier
[90] \ let us consider the identity 
\begin{eqnarray}
\frac{1}{(1-tz_{0})\cdot \cdot \cdot (1-tz_{k})} &=&(1+tz_{0}+\left(
tz_{0}\right) ^{2}+...)\cdot \cdot \cdot (1+tz_{n}+\left( tz_{n}\right)
^{2}+...)  \notag \\
&=&\sum\limits_{n=0}^{\infty
}(\sum\limits_{k_{1}+...+k_{k}=n}z_{0}^{k_{0}}\cdot \cdot \cdot
z_{k}^{k_{k}})t^{n}.  \TCItag{B.11}
\end{eqnarray}
When $z_{0}=...=z_{n}=1,$the inner sum in the last expression yields the
total number of monomials of the type $z_{0}^{k_{0}}\cdot \cdot \cdot
z_{n}^{k_{n}}$ with $k_{0}+...+k_{k}=n$. The total number of such monomials
is given by the binomial coefficient [79] 
\begin{equation}
p(k,n)\equiv \left( 
\begin{array}{c}
n+k \\ 
k
\end{array}
\right) =\frac{(n+1)(n+2)\cdot \cdot \cdot (n+k)}{k!}.  \tag{B.12}
\end{equation}
Hence, for this case Eq.(B11) is converted to useful expansion, 
\begin{equation}
P(k,t)\equiv \frac{1}{\left( 1-t\right) ^{k+1}}=\sum\limits_{n=0}^{\infty
}p(k,n)t^{n}.  \tag{B.13}
\end{equation}
Taking into account the integral presentation of beta function, replacing $%
k+1$ by $\alpha (s)+1$ in Eq.(B.13) and using it in beta function
representation of $V(s,t)$ produces after straightforward integration the
following result 
\begin{equation}
V(s,t)=-\sum\limits_{n=0}^{\infty }p(\alpha (s),n)\frac{1}{\alpha (t)-n} 
\tag{B.14}
\end{equation}
well known in string theory [55].The r.h.s. of Eq.(B.14) can be interpreted
as the Laplace transform of the partition function, Eq.(B.13). Eq.(1.39) of
Section 1, indicates that such interpretation is not merely formal. To see
this, following Vergne [51] consider a region $\Delta _{k}$ of \textbf{R}$%
^{k}$ consisting of all points $\nu =(t_{1},t_{2},...,t_{k}),$ such that
coordinates $t_{i}$ of $\nu $ are non negative and satisfy the equation $%
t_{1}+t_{2}+...+t_{k}\leq 1.$ Clearly, such restriction is characteristic
for the simplex in \textbf{R}$^{k}.$Consider now the \textit{dilated}
simplex $n\Delta _{k}$ for some nonnegative integer $n$. The volume of $%
n\Delta _{k}$ is easily calculated and is known to be 
\begin{equation}
\text{vol(}n\Delta _{k})=\frac{n^{k}}{k!}.  \tag{B.15}
\end{equation}
Next, let us consider points $\nu =(u_{1},u_{2},...,u_{k})$ with \textit{%
integral} coordinates \textit{inside} the dilated simplex $n\Delta _{k}.$
The total number of points with integral coordinates inside $n\Delta _{k}$
is given by $p(k,n)$, Eq.(B.12), i.e. 
\begin{equation}
p(k,n)=\left| n\Delta _{k}\cap \mathbf{Z}^{k}\right| =\frac{(n+1)(n+2)\cdot
\cdot \cdot (n+k)}{k!}.  \tag{B.16}
\end{equation}
The function $p(k,n)$ happen to be the non negative integer. It arises
naturally as the dimension of the quantum Hilbert space associated (through
the coadjoint orbit method ) with symplectic manifold of dimension $2k$
constructed by ''inflating'' $\Delta _{k}.$ \ Physically relevant details
are provided in Section 5 while here we only use these observations to
complete our discussion of the Regge-like result, Eq.(B.9).To this purpose
using Eq.(B.14) and assuming that $\alpha (t)\rightarrow k^{\ast }$ we can
approximate the amplitude $V(s,t)$ by 
\begin{equation}
V(s,t)\simeq -\frac{p_{\alpha (s)}(k^{\ast })}{\alpha (t)-k^{\ast }}\simeq -%
\frac{p_{\alpha (s)}(\alpha (t))}{\alpha (t)-k^{\ast }}  \tag{B.17}
\end{equation}
For large $\alpha (s)\footnote{%
Notice that the negative sign in front of $\alpha (s)$ was already taken
into account}$ by combining Eq.s(B.16) and (B.17) we obtain 
\begin{equation}
V(s,t)\simeq \frac{-1}{\alpha (t)-k^{\ast }}\frac{\alpha (s)^{k^{\ast }}}{%
k^{\ast }!}  \tag{B.18}
\end{equation}
In view of the footnote remark, and taking into account that $k^{\ast
}\simeq \alpha (t),$ this result coincides with Eq.(B.9) as required. In
addition, however, for large $k^{\prime }s$ it can be further rewritten as 
\begin{equation}
V(s,t)\simeq \frac{-1}{\alpha (t)-k^{\ast }}(\frac{\alpha (s)}{k^{\ast }}%
)^{k^{\ast }}\simeq \frac{-1}{\alpha (t)-k^{\ast }}(\frac{\alpha (s)}{\alpha
(t)})^{\alpha (t)}  \tag{B.19}
\end{equation}
Obtained result is manifestly symmetric with respect to exchange $%
s\rightleftharpoons t$ in accord with earlier mentioned requirement $%
V(s,t)=V(t,s)$. Moreover, it explicitly demonstrates that the angular
momentum of the graviton is indeed equal to two as required.

\pagebreak

\bigskip

\bigskip

\textbf{References}

\bigskip

1.Stanley, R.: Combinatorial reciprocity theorems. Adv. Math. \textbf{14},

\ \ 194-253 (1974)

2.Kholodenko, A.: New Veneziano amplitudes from ''old'' Fermat

\ \ \ (hyper)surfaces. In C.Benton (Ed): \textit{Trends in Mathematical
Physics}

\textit{\ \ \ Research. }New York\textit{\ : }Nova Science Publ., 2004.
arXiv: hep-th/0212189

3.Polyakov, A.: \textit{Gauge Fields and Strings}. New York : Harwood
Academic

\ \ \ Publ., 1987

4.Danilov,V.: The geometry of toric varieties. Russ.Math.Surveys \textbf{33},

\ \ \ 97-154 (1978)

5.Audin, M.: \textit{The Topology of Torus Actions on Symplectic Manifolds.}

\ \ \ Boston: Birkh\"{a}user, Inc., 1991

6.Carson, J., Muller-Stach, S., Peters,C.: \textit{Period Mappings and Period%
}

\ \ \textit{\ Domains. }Cambridge,UK: Cambridge University Press, 2003

7. Bourbaki, N.: \textit{Groupes et Algebres de Lie} (Chapitre 4-6).

\ \ \ \ Paris: Hermann, 1968

8. Solomon, L.: Invariants of finite reflection groups.

\ \ \ \ Nagoya Math.Journ.\textbf{22}, 57-64 (1963)

9. Ginzburg,V.: \textit{Representation Theory and Complex Geometry.}

\ \ \ \ Boston: Birkh\"{a}user Verlag, Inc., 1997

10.Wells, R.: \textit{Differential Analysis on Complex Manifolds}.

\ \ \ \ \ Berlin: Springer-Verlag, Inc., 1980

11.Humphreys, J.: \textit{Introduction to Lie Algebras and Representation }

\ \ \ \ \ \textit{Theory. }Berlin: Springer-Verlag, Inc., 1972

12.Kac,V.: \textit{Infinite Dimensional Lie Algebras}.

\ \ \ \ \ Cambridge,UK: Cambridge University Press, 1990

13.Brion, M.: Points entiers dans les polyedres convexes.

\ \ \ \ \ Ann.Sci.Ecole Norm. Sup. \textbf{21}, 653-663 (1988)

14.Stanley, R.: \textit{Enumerative Combinatorics}. Vol.1.

\ \ \ \ \ Cambridge,UK: Cambridge University Press, 1999

15.Atiyah, M., Bott, R.: A Lefschetz fixed point formula for

\ \ \ \ \ elliptic complexes : I . Ann.Math.\textbf{86}, 374-407 (1967), ibid

\ \ \ \ \ A Lefschetz fixed point formula for elliptic complexes :

\ \ \ \ \ II.Applications. Ann.Math. \textbf{88}, 451-491 (1968)

16.Bott, R.: On induced representations. Proc.Symp.Pure

\ \ \ \ \ Math.\textbf{48}, 1-13 (1988)

17.Ruelle, D.: \textit{Dynamical Zeta Functions for Piecevise Monotone}

\ \ \ \ \ \textit{Maps of the Interval. }Providence, RI: AMS, 1994

18.Cvitanovi\v{c}, P.: \textit{Classical and Quantum Chaos}. Unpublished.

\ \ \ \ \ Available at: www.nbi.dk/ChaosBook/

19.Gelfand, I., Kapranov, M., Zelevinsky, A. : \textit{Discriminants,
Resultants}

\ \ \ \ \textit{and Multidimensional Determinats}. Boston: Birkh\"{a}user,
Inc., 1994

20.Guileemin,V., Lerman, E., Sternberg, S.: \textit{Symplectic Fibrations}

\ \ \ \ \ \textit{and Multiplicity Diagrams}. Cambridge,UK: Cambridge

\ \ \ \ \ University Press, 1996

21.Cratier, P.: On Weil's character formula. BAMS \textbf{67}, 228-230 (1961)

22.Coxeter, H.: \textit{Regular Polytopes}.New York: The Macmillan Co. 1963

23.Ziegler,G.: \textit{Lectures on Polytopes}. Berlin: Springer-Verlag,
Inc., 1995

24.Ewald,G.: \textit{Combinatorial convexity and Algebraic Geometry}.

\ \ \ \ \ Berlin: Springer-Verlag, Inc., 1996

25.Brown,K.: \textit{Buildings}. Berlin: Springer-Verlag, Inc., 1989

26.Fulton,W.: \textit{Introduction to Toric Varieties}. Ann.Math.Studies 
\textbf{131}.

\ \ \ \ \ Princeton: Princeton University Press, 1993

27.Borel, A.: \textit{Linear Algebraic Groups}. Berlin: Springer-Verlag,
Inc., 1991

28.Macdonald, I.: \textit{Linear Algebraic Groups}. LMS Student Texts 
\textbf{32}.

\ \ \ \ \ Cambridge,UK: Cambridge University Press, 1999

29.Knapp, A.: \textit{Representation Theory of Semisimple Groups.}

\ \ \ \ \ Princeton: Princeton University Press, 1986

30.Kirillov, A.: \textit{Elements of the Theory of Representations.} (in
Russian)

\ \ \ \ \ Moscow: Nauka, 1972

31.Stanley, R.: Invariants if finite groups and their applications to

\ \ \ \ \ combinatorics. BAMS (New Series) \textbf{1},475-511,1979

32.Kholodenko, A. : Kontsevich-Witten model from 2+1 gravity:

\ \ \ \ \ new exact combinatorial solution. J.Geom.Phys. \textbf{43}, 45-91
(2002)

33.Humphreys, J.: \textit{Linear Algebraic Groups.}

\ \ \ \ \ Berlin: Springer-Verlag, Inc., 1975

34.Hiller, H.: \textit{Geometry of Coxeter Groups}. Boston: Pitman Inc., 1982

35 Knudsen, F., Kempf, G., Mumford, D., Saint-Donat, B.,

\ \ \ \textit{\ \ Toroidal Embeddings} I. LNM\ \textbf{339.}Berlin:
Springer-Verlag, Inc., 1973

36.Ashcroft, N, Mermin, D.: \textit{Solid State Physics.}

\ \ \ \ \ Philadelphia: Saunders College Press, 1976

37.Bernstein, I, Gelfand, I., Gelfand, S.: Schubert cells and cohomology

\ \ \ \ \ of the spaces G/P. Russian Math. Surveys \textbf{28}, 1-26 (1973)

38 Serre, J-P.: \textit{Algebres de Lie Semi-Simples Complexes}.

\ \ \ \ \ New York: Benjamin, Inc., 1966

39.Fomenko, A., Trofimov,V.: \textit{Integrable Systems on Lie Algebras}

\ \ \ \textit{\ \ and Symmetric Spaces}, New York: Gordon and Breach
Publishers,1988

40.Kholodenko, A.: Use of meanders and train tracks for description of

\ \ \ \ \ defects and textures in liquid crystals and 2+1 gravity.
J.Geom.Phys.

\ \ \ \ \ \textbf{33}, 23-58 (2000)

41.Kholodenko, A.: Use of quadratic differentials for description of

\ \ \ \ defects and textures in liquid crystals and 2+1 gravity. J.Geom.Phys.

\ \ \ \ \textbf{33}, 59-102 (2000)

42.Mimura, M., Toda, H.: \textit{Topology of Lie Groups, I and II. }

\ \ \ \ \ Providence, RI: AMS, 1991

43.Atiyah, M.: Angular momentum, convex polyhedra and algebraic

\ \ \ \ \ geometry, Proceedings of the Edinburg Math.Society \textbf{26},
121-138 (1983)

44.Atiyah, M.: Convexity and commuting Hamiltonians.

\ \ \ \ \ Bull.London Math.Soc.\textbf{14}, 1-15 (1982)

45.Schrijver, A.: \textit{Combinatorial Optimization. Polyhedra and
Efficiency.}

\ \ \ \ \ Berlin: Springer-Verlag, Inc., 2003

46.Gass,S.: \textit{Linear Programming}. New York: McGraw Hill Co., 1975

47.Guillemin,V., Sternberg, S.: Convexity properties of the moment

\ \ \ \ \ mapping. Invent.Math. \textbf{67}, 491-513 (1982)

48.Guillemin,V., Ginzburg,V., Karshon,Y.: \textit{Moment Maps, }

\textit{\ \ \ \ \ Cobordisms and Hamiltonian Group Actions. }

\ \ \ \ \ Providence, RI: AMS, 2002.

49.Delzant,T.: Hamiltoniens periodiques et image convexe de

\ \ \ \ \ l'application moment. Bull.Soc.Math.France \textbf{116}, 315-339
(1988)

50.Frankel, T.: Fixed points and torsion on K\"{a}hler manifolds.

\ \ \ \ \ Ann.Math. \textbf{70}, 1-8 (1959)

51.Vergne, M.: Convex polytopes and quanization of symplectic manifolds.

\ \ \ \ \ Proc.Natl.Acad.Sci. \textbf{93}, 14238-14242 (1996)

52.Hopf, H., Samelson, H.: Ein Satz \"{u}ber Wirkungsr\"{a}ume geschlossener

\ \ \ \ \ Liescher Gruppen. Comm.Math.Helv. \textbf{13}, 240-251 (1940)

53.Flaska, H.: Integrable systems and torus actions.

\ \ \ \ \ In :O.Babelon, P.Cartier, Y.Schwarzbach (Eds)

\ \ \ \ \ \textit{Lectures on Integrable} \textit{Systems}, Singapore:

\ \ \ \ \ World Scientific Pub.Co., 1994

54.Veneziano,G.: Construction of crossing symmetric, Regge

\ \ \ \ \ behaved amplitude for linearly rising trajectories. Il Nuovo

\ \ \ \ \ Chim. \textbf{57}A, 190-197 (1968)

55.Green, M., Schwarz, J.,Witten, E.: \textit{Superstring Theory. Vol.1}.

\ \ \ \ \ Cambridge,UK: Cambridge University Press, 1987

56.Stanley, R.: \textit{Combinatorics and Commutative Algebra.}

\ \ \ \ \ Boston: Birkh\"{a}user, Inc., 1996

57.Virasoro, M.: Alternative construction of crossing-symmetric

\ \ \ \ \ amplitudes with Regge behavior, Phys. Rev. \textbf{177}, 2309-2314
(1969)

58.Etingof, P., Frenkel, I., Kirillov, A. Jr.:\textit{\ Lectures on
Representation}

\textit{\ \ \ \ \ Theory and Knizhnik-Zamolodchikov Equations,}

\ \ \ \ \ Providence, RI: AMS, 1998

59.Gelfand, I., Kapranov, M., Zelevinsky, A. : Generalized Euler

\ \ \ \ \ integrals and A-hypergeometric functions. Adv. in Math. \textbf{84}%
,

\ \ \ \ \ 255-271 (1990); ibid \textbf{96}, 226-263 (1992)

60.Orlik, P., Terrao, H. : \textit{Arrangements and Hypergeometric Integrals.%
}

\ \ \ \ \ Math.Soc.Japan Memoirs, Vol.9. Tokyo: Japan Publ.Trading Co., 2001

61.Milnor, J.: On the 3-dimensional Brieskorn manifolds M(p,q,r).

\ \ \ \ \ In: \textit{Knots, Groups, and 3-Manifolds} (Papers dedicated to
the memory

\ \ \ \ of R. H. Fox), pp. 175--225. Ann. of Math. Studies. \textbf{84}.

\ \ \ \ \ Princeton : Princeton Univ. Press,1975

62.De Alfaro, V., Fubini, S., Furlan, G., Rossetti, C.: \textit{Currents in
Hadron}

\ \ \ \ \ \textit{Physics}. Amsterdam, Elsevier Publ.Co.,1973

63.Deligne, P., Mostow, G. : \textit{Commensurabilities Among Lattices}

\ \ \ \ \textit{\ in} PU (1, n).Ann. of Math. Studies. \textbf{132}.

\ \ \ \ \ Princeton : Princeton Univ. Press,1993

64.Carter, R.: \textit{Simple Groups of Lie Type}. New York:

\ \ \ \ \ John Wiley \& Sons Inc., 1972

65.Stanley, R.: Relative invariants of finite groups generated by

\ \ \ \ \ pseudoreflections. J.of Algebra \textbf{49}, 134-148 (1977)

66.Atiyah, M., Bott, R.: The moment map and equivariant cohomology.

\ \ \ \ \ Topology \textbf{23}, 1-28, 1984

67.Guillemin,V., Sternberg, S.: \textit{Supersymmetry and Equivariant}

\ \ \ \textit{\ \ de Rham Theory. }Berlin: Springer-Verlag Inc., 1999

68.Cox, D., Katz, S.: \textit{Mirror Symmetry and Algebraic Geometry}.

\ \ \ \ \ Providence, RI: AMS, 1999

69.Griffiths, P.: On periods of certain rational integrals.

\ \ \ \ \ Ann.of Math. \textbf{90}, 460-495; ibid 495-541 (1969)

70.Manin,Y.: Algebraic curves over fields with differentiation.

\ \ \ \ \ AMS Translations \textbf{206}, 50-78 (1964)

71.Gross, B.:On the periods of Abelian integrals and formula

\ \ \ \ \ of Chowla and Selberg. Inv.Math. \textbf{45}, 193-211 (1978)

72.Leray, J.:Le calcul differential at integral sur une variete

\ \ \ \ \ analytique complexe. Bull.Soc.Math.France \textbf{57}, 81-180
(1959)

73.Hwa, R., Teplitz, V.: \textit{Homology and Feynman Integrals.}

\ \ \ \ \ New York: W.A.Benjamin, Inc., 1966

74.Lang, S.: \textit{Introduction to Algebraic and Abelian Function}s.

\ \ \ \ \ Berlin: Springer-Verag, Inc., 1982

75.Edwards, J.: \textit{Tretease on the Integral Calculus, Vol.2 }

\ \ \ \ \ London: Macmillan Co., 1922

76.Deligne, P.:Hodge cycles and abelian varieties.

\ \ \ \ \ In : Lecture Notes in Math. \textbf{900.}

\ \ \ \ \ Berlin: Springer-Verag, Inc., 1982

77.Yui, N.: Arithmetics of certain Calabi-Yau varieties and mirror symmetry.

\ \ \ \ \ In: B.Conrad, K.Rubin (Eds). \textit{Arithmetic Algebraic Geometry}%
.

\ \ \ \ \ Providence, RI: AMS, 2001

78.Koushnirenko, A.: The Newton polygon and the number of solutions

\ \ \ \ \ of a system of $k$ equations in $k$ unknowns. Uspekhi.Math.Nauk 
\textbf{30},

\ \ \ \ \ 302-303 (1975)

79.Stanley, R.: \textit{Enumerative Combinatorics}.Vol.1.

\ \ \ \ \ Cambridge,UK: Cambridge University Press, 1997

80.Apostol,T.:\textit{Modular Functions and Dirichlet Series in Number Theory%
}.

\ \ \ \ \ Berlin: Springer-Verlag, Inc., 1976

81.Guillemin,V.:\textit{\ Moment Maps and Combinatorial Invariants of}

\ \ \ \ \textit{\ Hamiltonian \ \ }T$^{n}$ \textit{Spaces}. Boston:
Birh\"{a}user, Inc., 1994

82.Shepard, G., Todd, J.: Finite unitary reflection groups.

\ \ \ \ \ Canadian J.of Math. \textbf{6}, 274-304 (1954).

83.McMulllen, P: \textit{Absract Regular Polytopes}.

\ \ \ \ \ Cambridge,UK: Cambridge University Press, 2002

84.Kane, R.: \textit{Reflection Groups and Invariant Theory.}

\ \ \ \ \ Berlin: Springer-Verag, Inc., 2001

85.Humphreys, J.: \textit{Reflection Groups and Coxeter Groups}.

\ \ \ \ \ Cambridge,UK: Cambridge University Press, 1997

86.Orlik, P.Terao, H.: \textit{Arrangements of Hyperplanes.}

\ \ \ \ \ Berlin: Springer-Verlag, Inc., 1992

87.Lerche,W.,Vafa,C.,Warner, N.: Chiral rings in N=2

\ \ \ \ \ superconformal theories. Nucl.Phys.\textbf{B324}, 427-474 (1989)

89.Stanley, R.: \textit{Enumerative Combinatorics}.Vol.2.

\ \ \ \ \ Cambridge, UK: Cambridge University Press, 1999

90.Hirzebruch, F., Zagier, D.: \textit{The Atiyah-Singer Theorem and}

\ \ \ \ \textit{Elemetary Number Theory}. Berkeley, Ca: Publish or Perish
Inc.,1974

91.Khovanskii, A., Pukhlikov, A.: A Riemann-Roch theorem for integrals

\ \ \ and sums of quasipolynomials over virtual polytopes.

\ \ \ St.Petersburg Math.J. \textbf{4}, 789-812 (1992)

92.Witten, E.: Supersymmetry and Morse theory.

\ \ \ \ J.Diff.Geom.\textbf{17},661-692 (1982)

93.Hori,K., Katz, S., Klemm, A., Pandharipande, R., Thomas, R.,

\ \ \ \ Vafa,C.,Vakil, R., Zaslow, E.:

\ \ \ \ \textit{Mirror Symmetry}. Providence, RI: AMS, 2003

94.Guillemin, V.: K\"{a}ehler structures on toric varieties.

\ \ \ \ \ J.Diff.Geom.\textbf{40}, 285-309 (1994)

95.Guillemin,V.: Reduced phase spaces and Riemann-Roch.

\ \ \ \ \ In : \textit{Lie Theory and Geometry}. Boston: Birkh\"{a}user
Inc., 1994

96.Kontsevich, M.: Enumeration of rational curves via torus actions.

\ \ \ \ \ In : \textit{The Moduli Space of Curves}. Boston: Birkh\"{a}user,
Inc., 1995

97.Ellingsrud, G., Stromme, S.: Bott's formula and enumerative geometry.

\ \ \ \ \ J.Am.Math.Soc.\textbf{9},175-193 (1996)

98.Dixmier, J.: \textit{Enveloping Algebras}. Amsterdam: Elsevier Publ.Co.,
1977

99.Goldman,W.: \textit{Complex Hyperbolic Geometry. }

\ \ \ \ \ Oxford: Clarendon Press, 1999

100.Kholodenko, A.: Boundary conformal field theories, limit sets

\ \ \ \ \ \ of Kleinian groups and holography, J.Geom.Phys. \textbf{35},
193-238 (2000)

101.Kholodenko, A.: Statistical mechanics of 2+1 gravity from Riemann

\ \ \ \ \ \ \ zeta function and Alexander polynomial: exact results.

\ \ \ \ \ \ \ J.Geom.Physics. 38, 81-139 (2001)

102.Thurston, W.: \textit{Three Dimensional Geometry and Topology.}

\ \ \ \ \ \ \ Princeton: Princeton U.Press, 1977

103.Chari,V., Pressley, A.: \textit{Quantum Groups.}

\ \ \ \ \ \ \ Cambridge,UK: Cambridge University Press, 1995

104.Kantor,W., Liebler, R., Payne, S., Schult, E. : \textit{Finite Geometries%
},

\ \ \ \ \ \ \ \textit{Buildings and Related Topics}. Oxford: Clarendon
Press, 1990

105. Kumar,S.: \textit{Kac-Moody Groups, Their Flag Varieties}

\ \ \ \ \ \ \textit{\ and Representation} \textit{Theory}. Boston,
Birkhauser Inc., 2002

106.Fulton, W., Harris, J.: \textit{Representation Theory. A first Course.}

\ \ \ \ \ \ \ Berlin: Springer-Verlag, Inc., 1991

\ \ \ \ \ \ \ 

\ \ \ \ 

\ \ \ 

\ \ \ \ \ 

\ \ \ \ 

\ \ \ \ \ 

\ \ \ \ 

\ \ \ \ 

\ \ \ 

\end{document}